\documentclass{aa}  

\usepackage{graphicx}
\usepackage[varg]{txfonts}
\usepackage{xcolor}

\begin{document} 
\title{The rotation period distribution of the rich Pleiades-age Southern open cluster NGC\,2516 
		  \thanks{Based on observations at Cerro Tololo Inter-American Observatory, National Optical Astronomy Observatory under proposal 2008A-0476.}
	      \thanks{The full Tables~\ref{tab:members} and \ref{tab:rotationperiods} are only available in electronic form
		  	at the CDS via anonymous ftp to cdsarc.u-strasbg.fr (130.79.128.5)
		  	or via http://cdsweb.u-strasbg.fr/cgi-bin/qcat?J/A+A/}}

   \subtitle{Existence of a representative zero-age main sequence distribution}

   \author{D. J. Fritzewski\inst{1}
          \and
          S. A. Barnes\inst{1,2}
          \and
          D. J. James\inst{3,4}
          \and
          K. G. Strassmeier\inst{1}
          }

   \institute{Leibniz-Institut f\"ur Astrophysik Potsdam (AIP),
              An der Sternwarte 16, 14482 Potsdam, Germany\\
              \email{dfritzewski@aip.de}
         \and
              Space Science Institute, 4750 Walnut St., Boulder, CO 80301, USA
         \and
              Center for Astrophysics $\vert$ Harvard \& Smithsonian, 60 Garden Street, Cambridge, MA 02138, USA
         \and
	          Black Hole Initiative at Harvard University, 20 Garden Street, Cambridge, MA 02138, USA
             }

\titlerunning{Rotation Periods in NGC 2516}
\authorrunning{D. J. Fritzewski et al.}

   \date{}
\abstract{}
{
        We wish to measure the cool star rotation period distribution for the Pleiades-age rich open cluster NGC\,2516 and use it to determine whether cluster-to-cluster variations exist in otherwise identical open clusters.
}
{
  We obtained 42\,d-long time-series CCD photometry of NGC\,2516 in the $V$ and $I_c$ filters using the Yale 1\,m telescope at CTIO and performed a number of related analyses, including PSF-based time-series photometry.
  Our data are complemented with additional information from several photometric datasets, literature radial velocities, and \emph{Gaia}~DR2 astrometry.
  All available data are used to construct an integrated membership list for NGC\,2516, containing 844 stars in our $\approx 1\degr$ field of view.
}
{
  We derived 308 rotation periods for late-F to mid-M cluster members from our photometry.
  We identified an additional 247 periodic M~dwarf stars from a prior study as cluster members, and used these to construct a 555-star rotation period distribution for NGC\,2516.
  The colour-period diagram (in multiple colours) has almost no outliers and exhibits the anticipated triangular shape, with a diagonal slow rotator sequence that is preferentially occupied by the warmer stars along with a flat fast rotator sequence that is preferentially populated by the cooler cluster members.
  We also find a group of extremely slowly rotating M dwarfs ($10\mathrm{\,d} \lesssim P_\mathrm{rot}\lesssim 23$\,d), forming a branch in the colour-period diagram which we call the `extended slow rotator sequence'.
  This, and other features of the rotational distribution can also be found in the Pleiades, making the colour-period diagrams of the two clusters nearly indistinguishable.
  A comparison with the well-studied (and similarly aged) open cluster M\,35 indicates that the cluster's rotational distribution is also similarly indistinguishable from that of NGC\,2516.
  Those for the open clusters M\,50 and Blanco\,1 are similar, but data issues for those clusters make the comparisons somewhat more ambiguous.
  Nevertheless, we demonstrate the existence of a representative zero-age main sequence (ZAMS) rotational distribution and provide a simple colour-independent way to represent it.
  We perform a detailed comparison of the NGC\,2516 rotation period data with a number of recent rotational evolution models.
  Using X-ray data from the literature, we also construct the first rotation-activity diagram for solar-type stars in NGC\,2516, one that we find is essentially indistinguishable from those for the Pleiades and Blanco\,1.
}
{
  The two clusters NGC\,2516 and Pleiades can be considered twins in terms of stellar rotation and related properties (and M\,35, M\,50, and Blanco\,1 are similar), suggesting that otherwise identical open clusters also have intrinsically similar cool star rotation and activity distributions.
}

\maketitle

%
\section{Introduction}
\label{sec:intro}

Coeval stellar populations within open clusters are widely used to provide snapshots of stellar evolution, fitting them into an age-ranked succession to allow an empirical understanding of the underlying phenomena.
However, very few clusters share an identical age and it is debatable whether cluster-to-cluster variations beyond the compositional one exist.

One particularly sensitive way to probe the existence of any cluster-to-cluster variations between otherwise coeval systems is to measure and compare the rotation rates (preferably periods) of the corresponding late-type stars in the relevant clusters. The dual reasons for this are that rotation periods can be measured with great precision (routinely better than 1\,\%) and that cool star rotation periods themselves change by up to an order of magnitude over a timescale comparable to young open cluster ages \citep{1972ApJ...171..565S, 2003ApJ...586..464B}.

The stellar rotation period is measured by following the light modulation caused by cool starspots rotating into and out of view (\citealt{1947PASP...59..261K, 1987A&AS...67..483V, 2009A&ARv..17..251S}).
Although the number of measured rotation periods in open clusters has increased significantly over the past years
(see references in e. g. \citet{2003ApJ...586..464B, 2007ApJ...669.1167B} and \citealt{2014prpl.conf..433B}), their number is still small in comparison with the general open cluster population, even when restricted to that within 1\,kpc.
Hence, limited work is available when directly comparing two open clusters of similar age.

\cite{2009MNRAS.392.1456I} have directly compared the rotation periods of low-mass (mostly M) stars in NGC\,2516 and the similarly-aged M\,50 cluster, and found no dependence on the cluster environment. 
We confirm and extend their basic conclusion, while noting two issues: 1. their observational baseline was limited, resulting in a number of period aliases, and 2. subsequent membership information also shows that a certain number of their stars are non-members, making detailed comparisons ambiguous. We sort out these issues, and append the corresponding cleaned sample to our own below.

\cite{2014ApJ...782...29C}
compared their 33-star KELT-South-based rotation period distribution for Blanco\,1 with the HATNET-based period distribution for the Pleiades \citep{2010MNRAS.408..475H}, interpreting the measured difference as a small (146\,Myr vs 134\,Myr) age difference between the clusters. Both \cite{2011MNRAS.413.2218D} and \cite{2019ApJ...879..100D} have compared Praesepe with the Hyades and found measurable differences between their rotational distributions.
In fact, they have also interpreted the difference between the rotational distributions as an age difference of 47\,Myr and 57\,Myr respectively between the two open clusters.
Both studies used the stellar rotation period as an age estimator via gyrochronology \citep{2003ApJ...586..464B, 2010ApJ...722..222B}, which itself makes use of the fact that cool stars steadily lose angular momentum in a mass-dependent way as they age, a consequence of their winds \citep{1958ApJ...128..664P, 1967ApJ...148..217W}, resulting in the well-known relationship between the equatorial rotation velocity and age, $t$: $V_\mathrm{eq}\sim t^{-0.5}$, first identified by \cite{1972ApJ...171..565S} for solar-type stars\footnote{See \cite{2016AN....337..810B} for a modern equivalent of the \cite{1972ApJ...171..565S} relationship using rotation periods in open cluster stars.}.
We now know that the relationship can be generalized to rotation periods $P$, as a function of stellar mass, $m$: $P(m) \propto t^{0.5}$, a relationship believed to be roughly true for FGK stars (e.g. \cite{2003ApJ...586..464B}), and often used as a simple approximate implementation of gyrochronology.  

The focus of recent studies has been mostly on the slow rotators whose general spin-down is empirically well-delineated and can be applied to the open clusters of relatively advanced age, as witnessed by results in the $2.5$\,Gyr NGC\,6819 \citep{2015Natur.517..589M} and the $4$\,Gyr-old M\,67 \citep{2016ApJ...823...16B} open clusters.
However, many details of the evolution in young open clusters are unclear, and arguably are more important in understanding the transitions between rapid and slow rotators which observations have shown to exist within the same open cluster. Hence, the best way to test for any possible cluster-to-cluster variations is to compare two (or more) coeval young open clusters, preferably of the same composition.

Here we present new rotation period measurements for the zero-age main sequence (ZAMS) open cluster NGC\,2516, which is in many ways comparable to the Pleiades. It has a similar age and richness\footnote{\cite{2018A&A...618A..93C} list 798 members for NGC\,2516 vs. 992 members for the Pleiades.}, although it is somewhat more distant (409\,pc vs. 136\,pc, \citealt{2018A&A...618A..93C}), the last fact making it less prominent in the night sky.
Given the similarities, NGC\,2516 and the Pleiades are an ideal pair to search for cluster-to-cluster variations in astrophysical properties, in particular stellar rotation.
Rich cluster pairs are crucial because if any such variations exist, we would expect to find them at the level of detail which can only be probed with large numbers of stars spanning the whole mass range.

The Pleiades is probably the best-studied open cluster because of its proximity, richness, and northern location.
Several studies have measured rotation periods therein, beginning with \cite{1987A&AS...67..483V}.
In fact, that work represents the first time rotation periods were derived in any open cluster. It also demonstrated the existence of both slow and fast rotators (including some near break-up speed) in the same open cluster (see also \citealt{1982Msngr..28...15V}).
This was followed by other ground-based work on the cluster, including $v\,\sin\,i$ work by \cite{1987PASP...99..471S}, and more recently, rotation period work by \cite{2010MNRAS.408..475H}.
These ground-based efforts have now been superseded by an even more recent large space-based study using Kepler/K2 satellite data \citep{2016AJ....152..113R}, which now has become our principal source for comparison with NGC\,2516.

The age of NGC\,2516 is believed to be ${\sim}150\pm35$\,Myr \citep{1993A&AS...98..477M, 2002AJ....123..290S}.
The age of the Pleiades has been determined by various authors, with a recent estimate being ${\sim}110 - 160$\,Myr \citep{2018ApJ...863...67G}, in agreement with the lithium depletion age of $125 - 130$\,Myr \citep{1998ApJ...499L.199S}.
The uncertainties in classical isochrone fitting results, especially for young clusters, are large enough for the two clusters to be considered essentially coeval.

The metallicity of NGC\,2516 has been disputed in the past, with photometric estimates finding values as low as [Fe/H]$=-0.3$ \citep{1997MNRAS.287..350J}.
In contrast, spectroscopic observations have found near-solar values of [Fe/H]$=0.01\pm0.03$ \citep{2002ApJ...576..950T} and [Fe/H]$=-0.08\pm0.01$ \citep{2018MNRAS.475.1609B}.
\cite{2002ApJ...576..950T} have directly inter-compared equivalent stars in both clusters and found NGC\,2516 to have $\Delta[\mathrm{Fe/H}] = +0.04\pm0.07$ with respect to the Pleiades, which itself has a measured value of [Fe/H]$=+0.075\pm0.011$ \citep{2009AJ....138.1292S}.
The near-agreement between the spectroscopic metallicity determinations of both clusters makes the Pleiades and NGC\,2516 even more similar.

There are three other comparable open clusters, M\,35, M\,50, and Blanco\,1, which have been studied respectively by \cite{2009ApJ...695..679M}, \cite{2009MNRAS.392.1456I}, and \cite{2020MNRAS.492.1008G}. For various reasons, the relevant data are not as comparable with NGC\,2516 as those for the Pleiades. Nevertheless, we perform the relevant analyses and comparisons below.

A prior rotation period study of NGC\,2516 has been presented by \cite{2007MNRAS.377..741I}, as mentioned earlier. That study focused on the lower-mass, and consequently, pre-main sequence late-K and M-type stars ($0.15 \lesssim M/M_{\odot} \lesssim 0.7$), but did not measure the F, G, and early K stars. We complete the picture here by measuring the important higher-mass range of solar-like stars, including stars from late-F to mid-M (with enough overlap among the M\,stars to compare the results).
In fact, we subsequently supplement our sample with additional rotators from their sample that satisfy our membership criteria to present the most complete picture possible of stellar rotation in NGC\,2516.
Finally, it should be noted that this work subsumes prior work on rotation of NGC\,2516 cool stars in the otherwise unpublished PhD thesis of \cite{1997PhDT.........7B}.

This paper is structured as follows.
In Section~\ref{sec:observations} we present our observations and the photometric reductions.
In Sect.~\ref{sec:membership} we determine the cluster members among our observed stars.
Sect.~\ref{sec:ts} describes our time-series analysis.
We present the colour-period diagram for NGC\,2516 and related analysis in Sect.~\ref{sec:results}.
In Sect.~\ref{sec:models} we compare the observed rotation periods to angular momentum models from the literature and
in Sect.~\ref{sec:comPleiades} we compare the stellar rotation periods of NGC\,2516 with those of the Pleiades and other open clusters.
Finally, in Sect.~\ref{sec:xray}, we construct and present the first rotation-X-ray activity diagram for FGKM stars in NGC\,2516, followed by our conclusions.

\section{Observations and photometry}
\label{sec:observations}
We observed the southern Galactic open cluster NGC\,2516 between 19 February 2008 and 1 April 2008 from the Cerro Tololo Inter-American Observatory (CTIO) with the Yale 1\,m telescope operated by the SMARTS consortium. Within those 42\,d only a 4\,d scheduling gap from 8 March 2008 to 11 March 2008 interrupted the time-series observations, leading to a well-sampled photometric time-series. 
All observing nights had superb seeing conditions, typically better than $\sim$1.3\arcsec.

The CTIO Yale 1\,m telescope was equipped with the \emph{Y4KCam} camera, based on a $4064\,\mathrm{px}\times4064\,\mathrm{px}$ STA CCD detector and with a field of view (FoV) of $19.3\arcmin\times19.3\arcmin$ and a  $0\farcs289\,\mathrm{px}^{-1}$ image scale. The average detector read noise was $4.8\,e^-$\,px$^{-1}$.

\begin{figure}
	\includegraphics[width=\columnwidth]{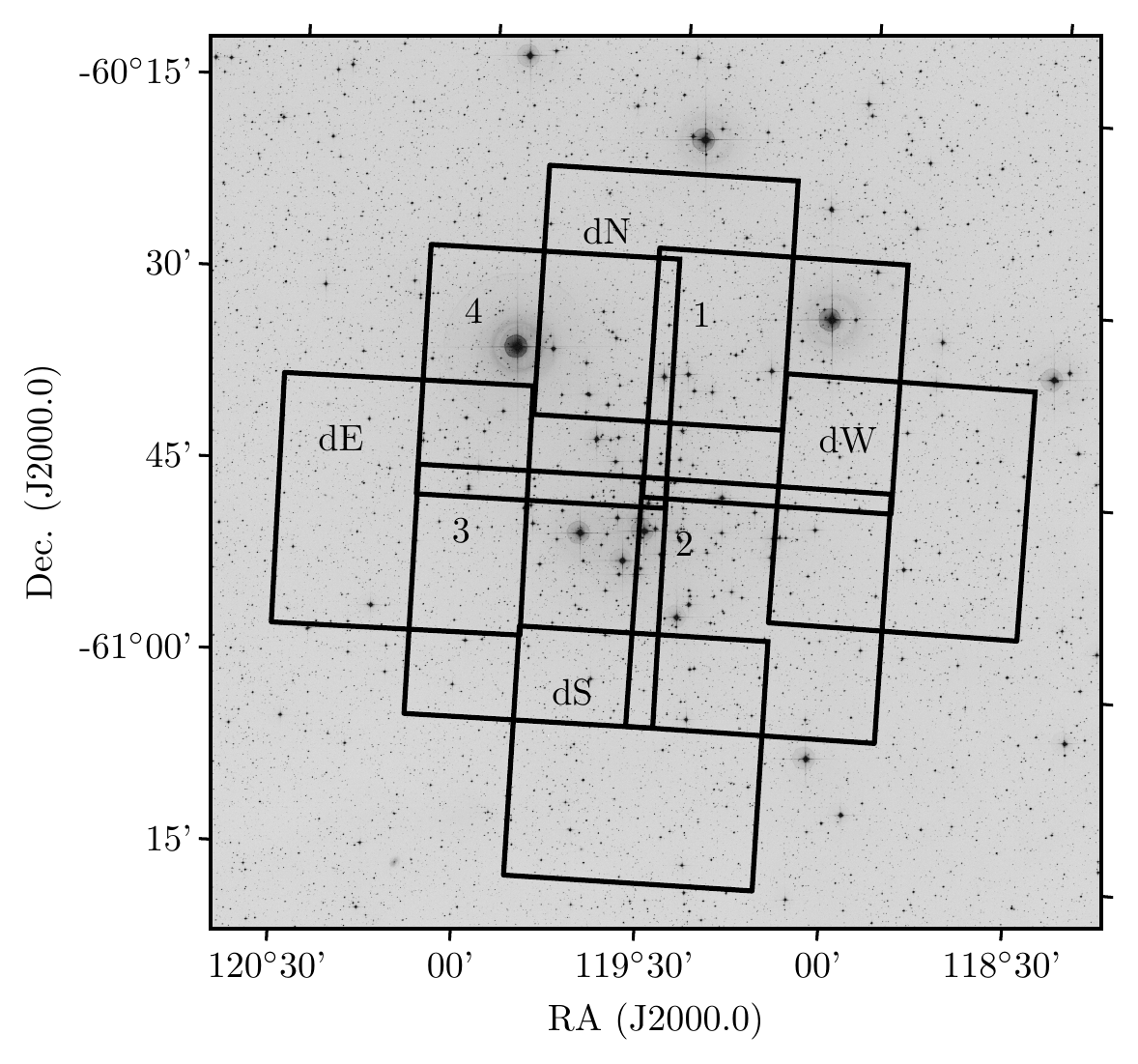}
	\caption{Field of NGC 2516 from the Digitized Sky Survey 2 (red filter), with our time-series observing fields overlaid in black contours.
		Individual fields are square, with each labelled in the North-East corner.
		The inner fields (1 through 4) were observed with exposure times of 60\,s in $V$ and 120\,s in $I_c$ filters, while the outer (deep) fields received 600\,s in the $I_c$ filter.  
		Apart from a small overlap with our outer fields, the region covered by \cite{2007MNRAS.377..741I} lies beyond that of this study, extending the total area covered by time-series observations to >1\,sq. degree.
	}
	\label{fig:fields}
\end{figure}

With an extent of more than 1\,deg$^2$ on the sky, NGC\,2516 is too large to fit into a single FoV. Therefore, we covered the open cluster with eight different fields, as shown in Fig.~\ref{fig:fields}.
We cycled through these fields to obtain a homogeneous time-series. The centre of our observing programme is $\alpha=\mbox{7:58:05}$, $\delta=\mbox{-60:48:47}$ (J2000.0), near the cluster giant CPD-60~980.

In order to provide greater reliability, the four inner fields (numbered F1 to F4) were observed both in the Johnson-Kron-Cousins $V$ and $I_c$ band, with exposure times of 120\,s and 60\,s, respectively. In addition, we observed four outer fields (called deepN/E/S/W). These outer fields partially overlap with the inner fields, allowing consistency checks, but for these we optimized the exposure for the fainter M dwarf members with an exposure time of 600\,s in $I_c$.

In total, we obtained between 82 and 108 observations for each field with up to ten visits (median four) per field per night. A summary of the observations is provided in Table~\ref{tab:obslog}.

\begin{table}  
	\caption{Number of exposures in the different fields and filters.}
	\label{tab:obslog}
	\begin{tabular}{l  l  c c  }
		\hline
		\hline
		Field name & Filter & Exposure time & Number of visits\\
		& & (s) &\\
		\hline
		F1 & $I_c$ & 60 & 106\\
		& $V$ & 120 & 108\\
		F2 & $I_c$ & 60 & 104\\
		& $V$ & 120 & 104\\			
		F3 & $I_c$ & 60 & 106\\
		& $V$ & 120 & 105\\
		F4 & $I_c$ & 60 & 105\\
		& $V$ & 120 & 103\\
		deepN & $I_c$ & 600 & 86\\
		deepE & $I_c$ & 600 & 88\\
		deepS & $I_c$ & 600 & 88\\		
		deepW & $I_c$ & 600 & 82\\
		\hline	 
	\end{tabular}
\end{table}

\subsection{Data reduction and photometry}
\label{sec:photometry}

Supplementary to the science data, zero-second bias images and per-filter sky-flat fields were acquired each night. Using IRAF\footnote{IRAF is distributed by the National Optical Astronomy Observatories, which are operated by the Association of Universities for Research in Astronomy, Inc., under cooperative agreement with the National Science Foundation.}, a median bias frame was subtracted from all calibration and science images, and the science data in each filter were corrected for pixel-to-pixel sensitivity differences using a per-filter balance frame. The dark current of the \emph{Y4Kcam} is sufficiently low ($21\,e^-\,\mathrm{px}^{-1}\,\mathrm{h}^{-1}$) that dark current correction was not applied to the images.

In order to construct the best possible light curves, we used point spread function (PSF) photometry and the \textsc{DaoPhot~II} Software \citep{1987PASP...99..191S, 1994PASP..106..250S, 2003PASP..115..413S}. The key to good PSF subtraction of each star from the obtained images is a clean model PSF that includes as many properties of the data as possible. To construct such a model PSF, we examined (for each field) the frames with the best seeing\footnote{These fiducial images were chosen from the night of UT20080319, with a seeing of $0.9\arcsec$ to $1.2\arcsec$ for the eight different fields} to identify suitable stars.

After several experiments with the data and different set-ups for the PSF, we settled on four criteria for our PSF stars. Our PSF stars are required (1) to have a count number in the middle of the dynamic range of the CCD and (2) each is required to be the only source within at least 25\,px of its centre of light (${\approx} 15\arcsec$ in diameter). 
Furthermore, those stars (3) should have a regular shape on visual inspection, in other words the shape should not differ from nearby stars. Intending to use the same PSF stars for each exposure of a given field, (4) we also avoid stars close to bad columns of the CCD because in some frames these stars could potentially fall onto those columns in cases where the pointing is slightly different. Bad pixels within the fitting area of the PSF would lead to removal of that particular star from the list of PSF stars, lowering the number of available PSF stars for that frame.

For each field, we chose roughly 350 stars to construct the model PSF. Our aim is to exclude as much non-astrophysical variability as possible from the final light curves. Therefore, we used the same PSF stars for all exposures of a given field. Each image was matched to the reference image with \textsc{DaoPhot~II}. This enabled us to translate the ID numbers of the PSF stars in the reference frame to the ID numbers of the working frame and to construct the list of PSF stars for the frame, including only stars with valid data, in particular unsaturated ones.

The workflow in \textsc{DaoPhot~II} starts with source finding, which assigns an ID to each possible point source in the image. Afterwards, aperture photometry is carried out to estimate the brightness of the sources. Those magnitudes are used as an input to create the second-order spatially-varying model PSF, which is obtained by fitting the selected sources with the ``Penny'' function\footnote{\textsc{DaoPhot~II} settings: ANALYTIC MODEL PSF = 5.00, VARIABLE PSF = 2.0}. Subsequently, the actual PSF fitting of all stars is performed in \textsc{AllStar} and delivers the final instrumental magnitude of the sources in the frame.

After performing the PSF photometry with \textsc{DaoPhot~II}, we used the software \textsc{DaoMaster} to match the images to within 0.3\,px to obtain accurate positions of the stars for the creation of the light curves. In the next step, \textsc{DaoMaster} adjusts the instrumental magnitudes of the individual frames to a common system and combines all data points of a star into a light curve. We chose to construct light curves only for stars for which at least 60 data points remained (${\sim}50$ percent of the available frames). In total \textsc{DaoMaster} created ${\sim}24\,000$ light curves which correspond, due to the overlap of the normal and the deep frames and the usage of two filters, to ${\sim}14\,000$ distinct sources.

In the following, we use the IDs assigned by \textsc{DaoPhot~II} preceded by the field as the main identifiers in the text. In the tables we additionally give the global identifier from \emph{Gaia}~DR2. For example, a star from field F2 in $I_c$ has the name \emph{2i516} while a star from the deep $I_c$ south field is named \emph{dS2516}.

\subsection{Photometric uncertainties}

\begin{figure}
	\includegraphics[width=\columnwidth]{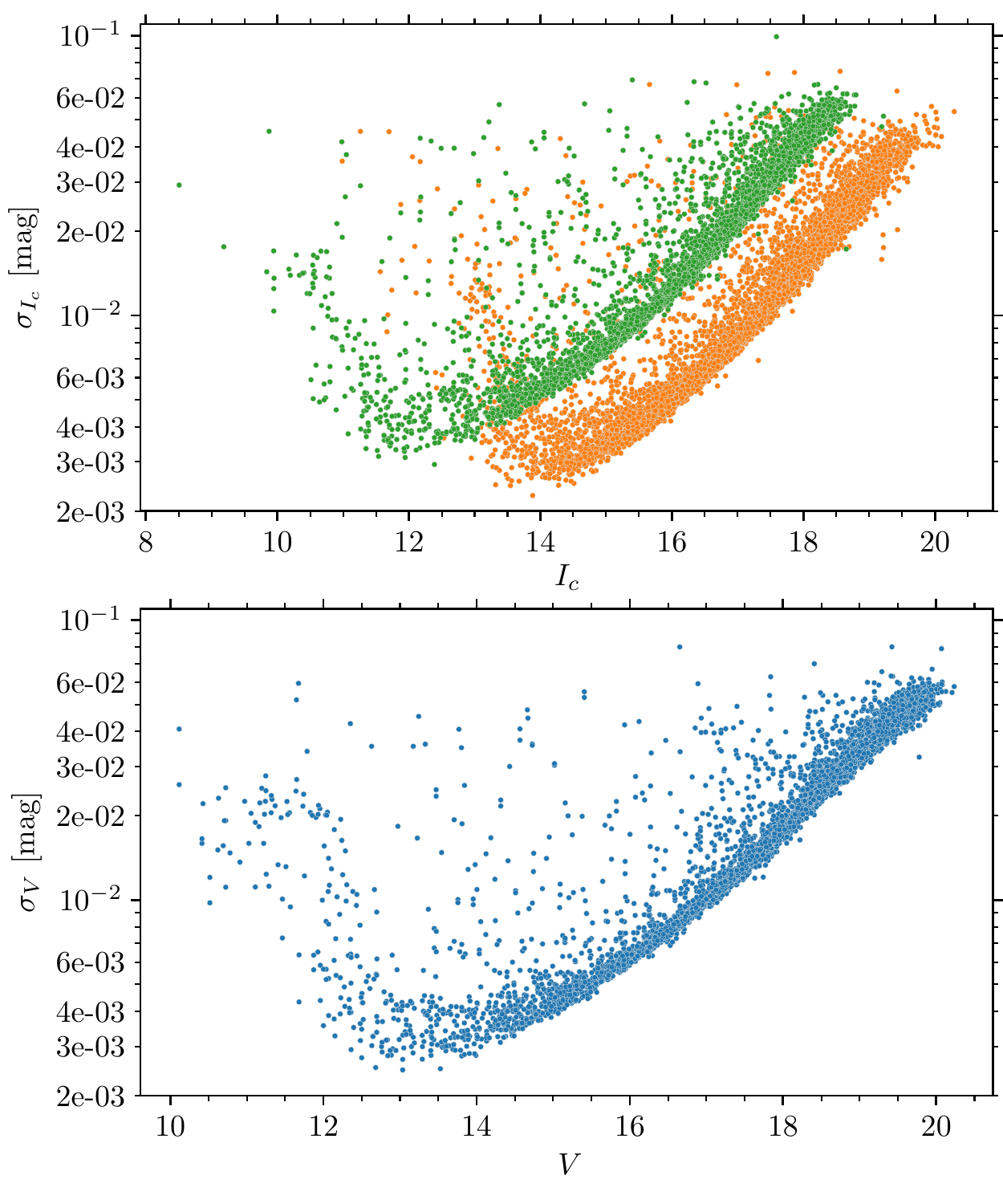}
	\caption{Mean photometric uncertainties of the light curves ($\sigma$) as a function of the magnitude. \emph{Top:} In green we show values for $I_c$ (60\,s) and in orange the deep $I_c$ (600\,s). \emph{Bottom:} Same as the upper panel but for $V$ (120\,s) data.}
	\label{fig:photuncert}
\end{figure}

In Fig.~\ref{fig:photuncert}, we show the mean uncertainties for all light curves over the magnitude range in their corresponding filters. The data are separated by kind of exposure. The deep 600\,s $I_c$ filter data exhibits the lowest noise level, as one would expect. For the 60\,s $I_c$ data a comparable uncertainty is reached for stars 2\,mag brighter. This also means that those light curves deliver significantly better results for the brighter stars (which are either overexposed or saturated in the deep $I_c$ filter frames). In combination, we achieved a precision of 4\,mmag or better for the whole range from $I_c=11.0-16.8$\,mag. This corresponds to a range from solar-type to late-K stars, the core mass range for our study.

The 120\,s $V$ band observations are comparable to the short $I_c$ ones but generally deliver a lower noise level. With the inclusion of the deep $I_c$ frames we are able to probe stars down to $I_c=20$, corresponding to $V\approx23$ for stars on the cluster sequence.

\begin{figure}
	\includegraphics[width=\columnwidth]{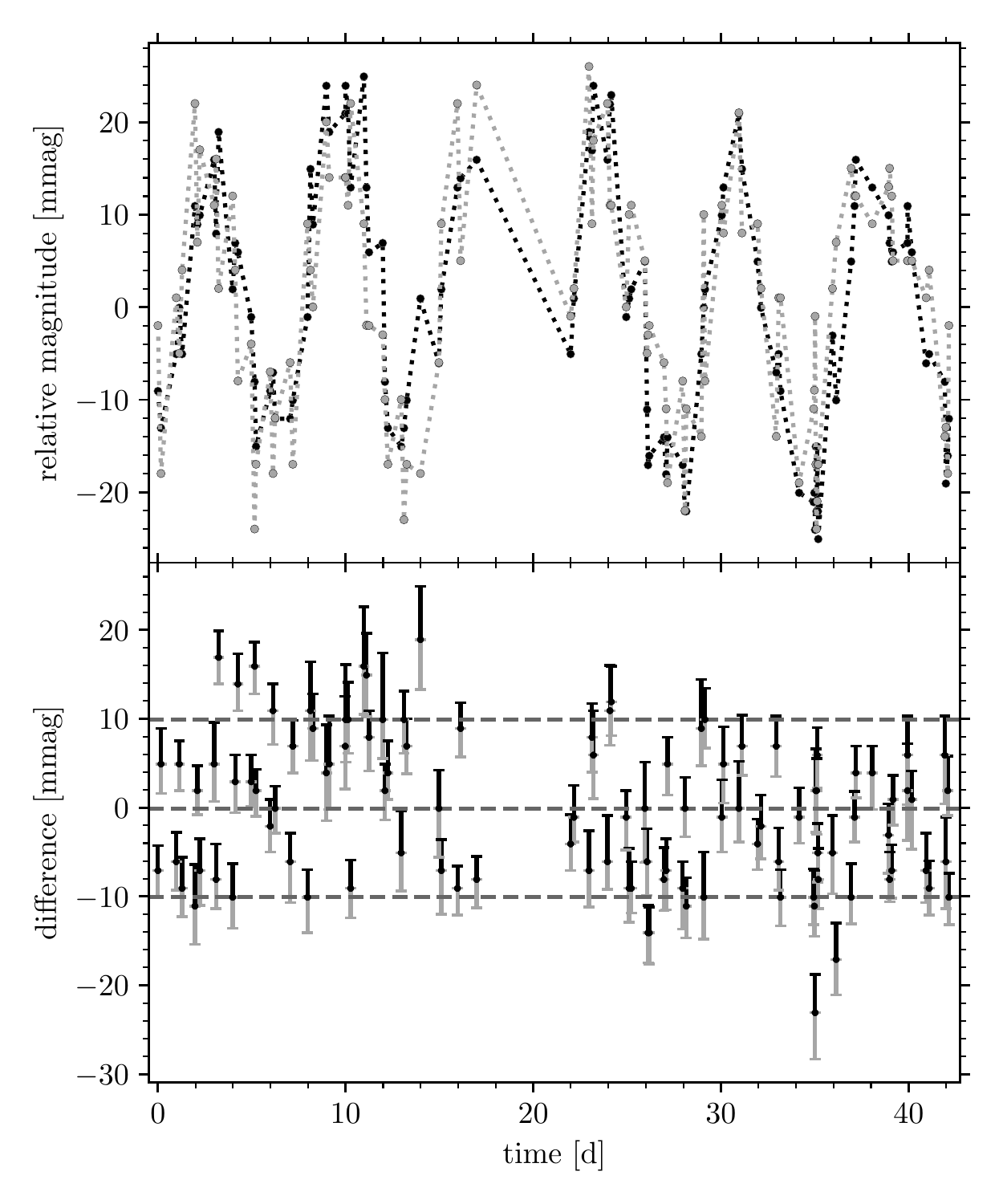}
	\caption{Two $V$ band light curves of the same star Gaia~DR2~5290833258325773312 from fields 1 and 2 (\emph{upper} panel) and their differences (\emph{lower} panel), demonstrating the extent of (dis)agreement.
		The uncertainties shown in the \emph{lower} panel are those estimated by \textsc{DaoPhot~II} for each data point. The upper part of the error bar (black) corresponds to the black light curve while the lower one corresponds to the grey light curve. The three dashed lines, from top to bottom, are the ninetieth, the fiftieth (median), and the tenth percentile of the light curve differences.}
	\label{fig:lcdiff}
\end{figure}

\subsection{Testing the uncertainties with duplicate light curves}
Our observed fields on the sky were designed to overlap, significantly for the deep fields, resulting in multiple light curves from different fields for many stars.
See Fig.~\ref{fig:lcdiff} for an example.
Our observing strategy was to cycle through the fields; consequently two photometric data points for one star in two different fields are often separated by less than one hour. Over this interval, the variability of most of the observed stars is small; hence we can assume that both photometric data points should agree to within the uncertainty given by \textsc{DaoPhot~II}. Here we test this assumption and hence the reliability of the error estimate.

For the comparison we used all light curves of stars found in multiple fields. In total, we have 1260 pairs\footnote{Each filter is counted independently. A star found in two fields with photometry in both $V$ and $I_\mathrm{c}$ is counted as two pairs.} of light curves for 658 individual stars. We calculated the differences of the light curves (DLC) from the mean-subtracted light curves rather than from zero point-adjusted photometry\footnote{This might introduce some additional uncertainties, increasing the differences between the light curves. In this case the mean of the DLC is non-zero. However, the offsets observed in the DLCs are much smaller than the scatter and can be neglected.}. We use the data points closest in time for the calculation of the differences and set an upper limit of six hours (i.e. half a night) after which two data points are no longer considered as a pair.


One example of our analysis is shown in Fig.~\ref{fig:lcdiff}. This particular star (Gaia~DR2~5290833258325773312)\footnote{The light curves are for the star with IDs \emph{1v1471} and \emph{4v46}.} is a slow rotator, with $P_\mathrm{rot} = 7.1$\,d. We chose one of our rotators to illustrate the differences in the light curves because it is much easier to follow a structured light curve rather than random stellar and instrumental variability. The upper panel of Fig.~\ref{fig:lcdiff} shows the two light curves from the 120\,s $V$ band exposures. In general, both light curves follow each other closely. Some outliers can be seen but the starspot induced variability is the main contribution to the flux variations.

The lower panel of Fig.~\ref{fig:lcdiff} shows the differences between the two light curves at sufficiently close time stamps. The median value of the differences is very close to zero. If the uncertainties from \textsc{DaoPhot~II} account for all the variance, we would expect all differences to agree with zero within the uncertainties. However, this is not the case, with the median value of the absolute differences being twice the median value of the uncertainties. Nevertheless, (because of what we show immediately below using all our light curves) we believe the uncertainties to be correctly estimated by \textsc{DaoPhot~II} and ascribe the deviations to instrumental characteristics, short term atmospheric-, and maybe even stellar variations.

\begin{figure}
	\includegraphics[width=\columnwidth]{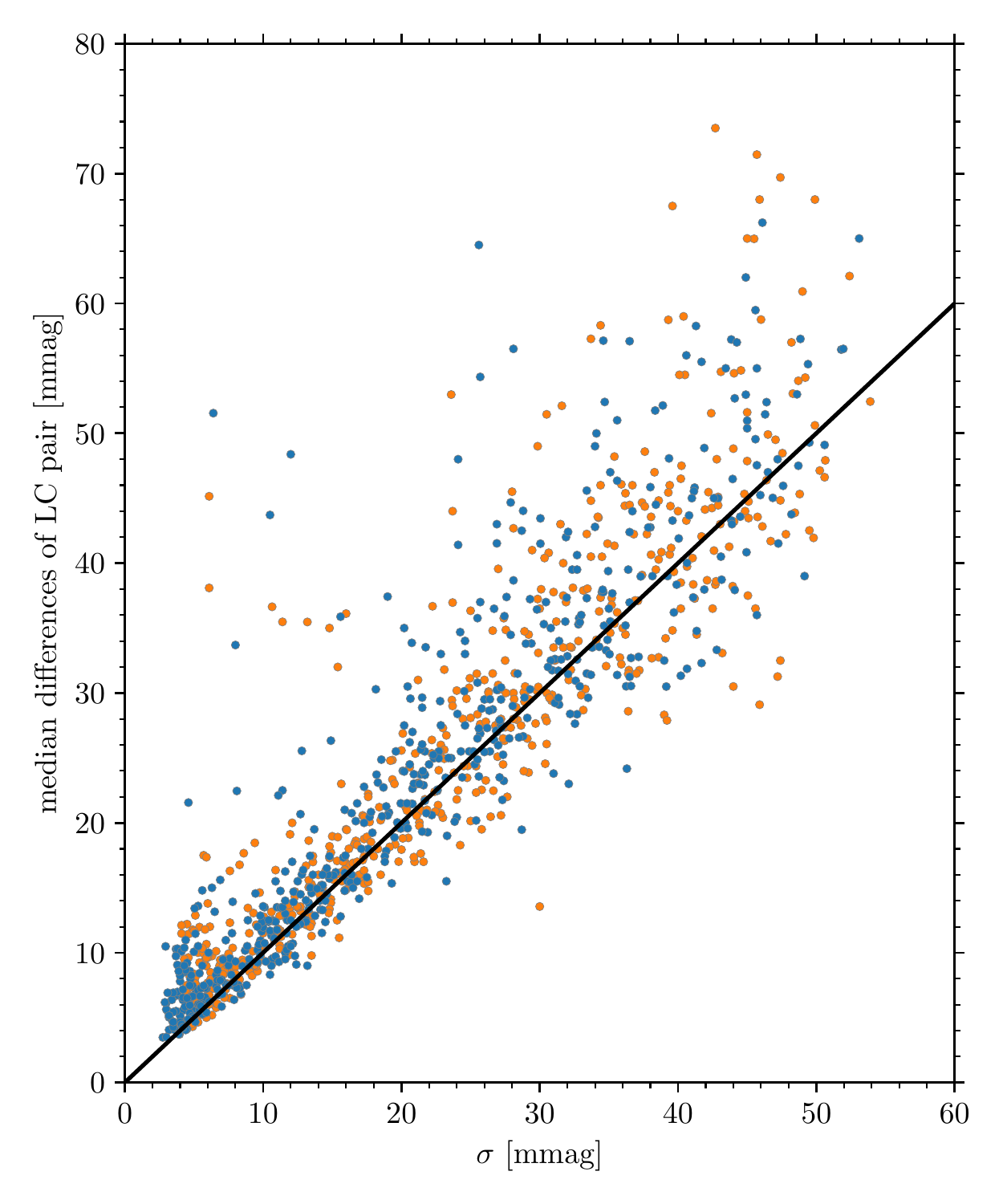}
	\caption{Median measurement error of a light curve as determined by \textsc{DaoPhot~II} ($\sigma$) against the median of the absolute differences of a light curve pair. The two colours indicate the two filters ($V$ blue, $I_c$ orange). In black we mark the line of equality. The majority of the differences are only marginally larger than the estimated photometric uncertainties.}
	\label{fig:MADuncert}
\end{figure}

Beyond that particular example in Fig.~\ref{fig:lcdiff}, we find the median of the absolute differences between the light curves to be slightly larger than the \textsc{DaoPhot~II} errors for most stars.
This relation is shown in Fig.~\ref{fig:MADuncert}. As this effect is of the same order, independent of the light curve uncertainties, we feel certain that the larger differences are not caused by underestimation of the uncertainties but ought to be interpreted as short-term stellar and atmospheric variability. Therefore, we use the uncertainties as is in the analysis.

%

\subsection{Literature photometry}
\label{sec:litphot}

For this work, we use standardized photometry available from prior work. \cite{2001A&A...375..863J} (hereafter J01) presented a comprehensive photometric study of the NGC\,2516 with essentially the same spatial coverage as our time-series photometry. Additional photo\-metry of sub-fields of NGC\,2516 was published by \cite{2002AJ....123..290S} (hereafter S02) and \cite{2006A&A...453..101L}. Although only J01 cover the whole area, we chose S02 as our preferred source for $BVI_c$ photometry and supplemented that coverage with J01 because we find a certain number of erroneous magnitudes in J01 (see Sect.~\ref{sec:photdiff}).
We occasionally also supplement these with the \emph{Gaia}~DR2 photometry \citep{2018A&A...616A...4E} because we have a small number of stars with light curves for which no standardized ground-based photometry is available.
(Thus, we also include \emph{Gaia}~DR2 photometry for all stars in our discussion, and display our data against $(G_\mathrm{BP} - G_\mathrm{RP})$ when appropriate.)

For near-infrared photometry, we use data from the Vista Hemisphere Survey (VHS, \citealt{2013Msngr.154...35M}). This survey is deeper than 2MASS and provides
precise magnitudes even for the low-mass members of the cluster. The $K_s$ filter used in VHS is very similar to the 2MASS filter\footnote{http://casu.ast.cam.ac.uk/surveys-projects/vista/technical/photometric-properties} and the small differences are negligible for our purposes, centred on the colour-period diagram; hence it can be used instead of 2MASS photometry without further adjustments.

\section{Membership and colour-magnitude diagram}
\label{sec:membership}

\begin{figure*}
	\includegraphics[width=\textwidth]{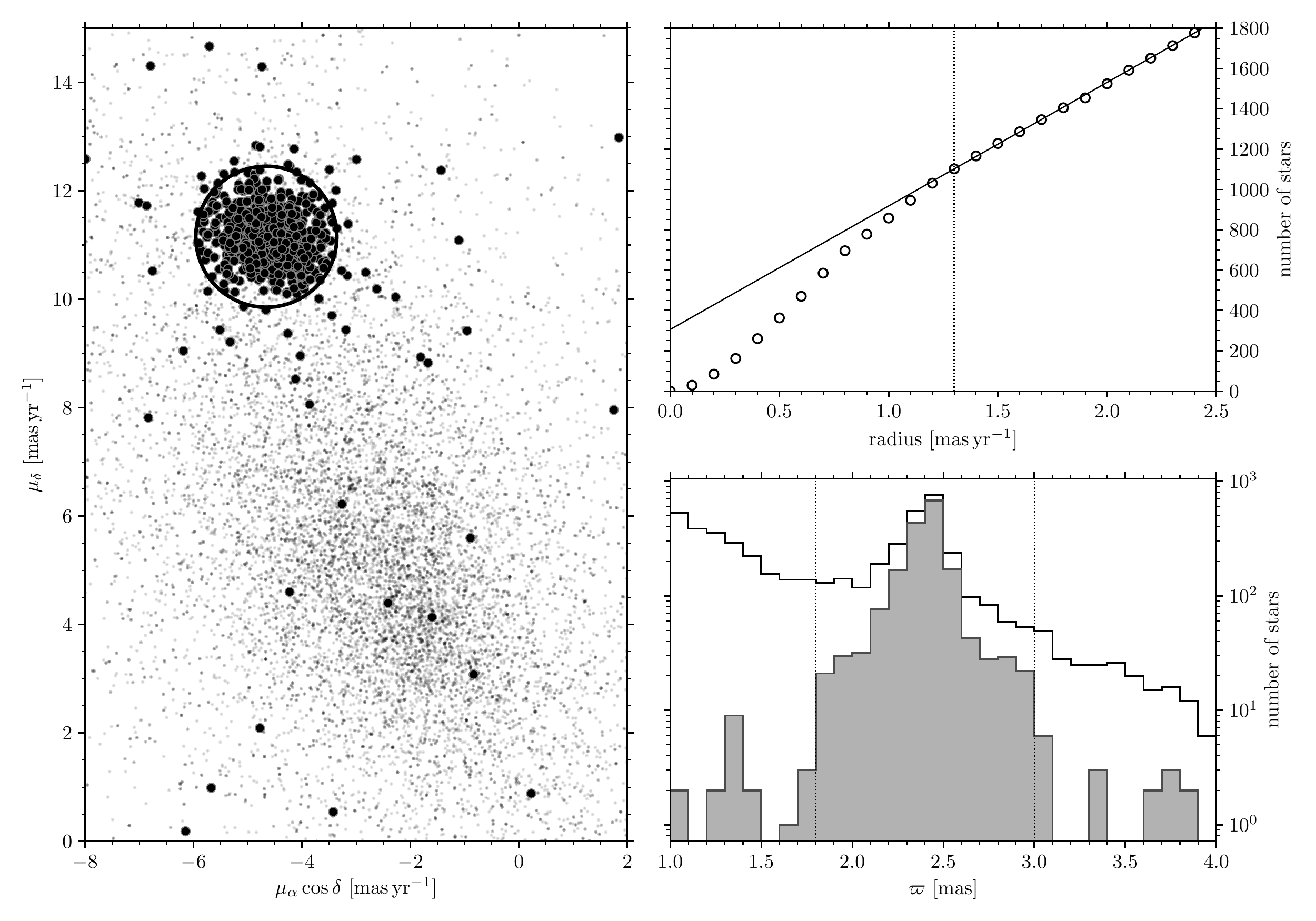}
	\caption{Selection of cluster members based on data from \emph{Gaia}~DR2.
		\emph{Left panel:} Proper motion diagram of the field of NGC\,2516. The members finally selected and satisfying at least two membership criteria are highlighted in black.
		All stars within the black circle (with radius of 1.3\,mas\,yr$^{-1}$; see \emph{upper right} panel) are considered proper motion members.
		\emph{Top right panel:} Number of stars with proper motions against distance from the mean cluster proper motion. The vertical dotted line at 1.3\,mas\,yr$^{-1}$ marks the transition from non-linear to linear growth (indicated by the solid line), i.e. the transition from open cluster members to field stars, correspondingly used as the radius for the circle on the \emph{left} plot.
		\emph{Bottom right panel:} Histogram of \emph{Gaia}~DR2 parallaxes of the field (upper line) and the finally selected members (filled histogram). Stars between the vertical dotted lines are accepted as parallax-based cluster members.}
	\label{fig:gaiamem}
\end{figure*}

A well-defined membership list is the basis for every open cluster study. For NGC\,2516 no such membership (incorporating several criteria) for the whole cluster was available in the literature until the day before the final submission of this manuscript\footnote{The results of a comparison with that work \citep{2020arXiv200609423J} have been included in an Appendix to this paper.}. However, a large number of photometric and radial velocity studies are available, each covering different fields of the cluster. Combined with the \emph{Gaia}~DR2 proper motions and parallaxes, we are able to identify the cluster members among the stars for which we have obtained light curves.
We use the four criteria discussed below, first individually, and then in combination.
We then construct a clean colour-magnitude diagram (CMD) for the cluster.

\subsection{Individual membership criteria}
\label{sec:individMem}
Previously, J01 presented a photometric membership study which coincidently covers the same area as our photometry.
We use this study to provide our basic photometry\footnote{In addition, we use the excellent photometry from the \emph{Gaia}~DR2 as a quality check after defining the membership, but the selection of cluster members is not based on those data.} and rely on the photometric membership presented therein (supplementing this membership as described below).

For the second element of our membership definition we use proper motions from \emph{Gaia}~DR2 \citep{2018A&A...616A...2L}. The mean proper motion of NGC\,2516 is $\mu_\alpha\cos\delta = -4.6579$\,mas\,yr$^{-1}$, $\mu_\delta = 11.1517$\,mas\,yr$^{-1}$ \citep{2018A&A...616A..10G}.
We selected all stars within 1.3\,mas\,yr$^{-1}$ of this mean proper motion as proper motion members (Fig.~\ref{fig:gaiamem} left panel).
The radius was decided by plotting star counts for different radii, and marks the transition from a non-linear to a linear growth rate. See Fig.~\ref{fig:gaiamem} top right panel and \cite{N3532RV} for details.

In addition to the proper motions, we include the parallax in our membership determination. Since we do not know the radial extent of the open cluster we chose to include stars within 0.6\,mas of the mean cluster parallax as parallax members. With $\varpi=2.4$\,mas \citep{2018A&A...616A..10G} this means we include stars between 330\,pc and 550\,pc as parallax members (Fig.~\ref{fig:gaiamem}, lower right panel). The aim is to remove mainly background giants and nearby low-mass stars. Among the finally selected members 96\,\% are parallax members\footnote{For our purposes (selecting candidate members to derive their rotation periods) this is an adequate membership selection. We note that the later derived rotation periods act as an additional membership proxy and we find 95\,\% of the rotators even more concentrated between 380\,pc and 470\,pc.}.

Radial velocity measurements for NGC\,2516 are available from multiple sources including both large surveys and dedicated work on NGC\,2516.
The spectroscopic surveys with radial velocity measurements for NGC\,2516 are RAVE \citep{2006AJ....132.1645S, 2017AJ....153...75K}, \emph{Gaia} ESO \citep{2012Msngr.147...25G}, GALAH\footnote{The GALAH~DR2 excludes most of NGC\,2516 because the analysis of open clusters will be published elsewhere.} \citep{2015MNRAS.449.2604D, 2018MNRAS.478.4513B}, and \emph{Gaia}~DR2 \citep{2018A&A...616A...5C}. Dedicated studies of NGC\,2516 include the first spectroscopic studies of main sequence stars in NGC\,2516 \citep{1998MNRAS.300..550J} and \cite{2002ApJ...576..950T}. \cite{2016A&A...586A..52J} have subsequently measured the radial velocities of low-mass stars and \cite{2018MNRAS.475.1609B} have recently published the results of a multiplicity survey for solar-mass stars.

In order to establish a radial velocity membership list, we combined all measurements available for a given star and calculated its mean radial velocity. The zero-points of the literature data could potentially be in mild disagreement, and hence bias the mean and widen the radial velocity distribution.
Placing these on the same system is beyond the scope of our work, so we
have simply assigned radial velocity membership to all stars with mean radial velocities within 5\,km\,s$^{-1}$ of the cluster radial velocity ($v_\mathrm{N2516}=23.6$\,km\,s$^{-1}$, \citealt{2016A&A...586A..52J}).
We are aware that this procedure is likely to include a small number of non-members into our selection (as do each of the others).
However, the radial velocity is only one of four input criteria for our final membership, which we expect to be much cleaner.

\subsection{Combined membership}
\label{sec:kinnon}
With the individual membership lists from the different methods at hand, we can combine them to construct a membership list that will be the basis for our further study of the open cluster. We have decided to classify a star as a member if it satisfies at least two of the membership criteria (photometry, proper motion, parallax, and radial velocity). We do not require stars to satisfy all available membership criteria because the imprecision in the photometry of the fainter stars could unjustifiably remove them from the sample. In fact, our membership criteria result in retaining many of the M dwarfs that are important for our rotational work below, and also, in contrast to many similar studies, the binary cluster members to the extent possible.
An additional justification for weakening our criterion is that some stars which were classified by J01 as photometric non-members were subsequently shown to be photometric members in the \emph{Gaia}~DR2 photometry. We suspect that for those stars either the $B$ or $V$ magnitude has an incorrect value, placing the star off the cluster sequence in a [$B-V$, $V$] colour-magnitude diagram. In Sect.~\ref{sec:photdiff} we describe our solution to this problem.

As a consequence of the above choices, we include a small number (28) of kinematic non-members in our overall membership sample (a subset of which we will discuss individually later in the rotational analysis).
By this, we mean objects that are radial velocity and proper motion non-members, but photometric and parallax members.
Such kinematic non-members are likely field main sequence stars which happen to be crossing the open cluster at the observational epoch.
For the time being we retain those 28 stars ($= 3\,\%$) in our membership list, but mark them for later removal from the sample, as necessary.
Among these 28 stars, only 10 have multiple radial velocity measurements. Hence, the remaining 18 stars could potentially be binary cluster members (which we would of course prefer to retain).

\subsection{Binaries}
\label{sec:binaries}

\begin{table*}
	\caption{Potential radial velocity binaries in NGC\,2516.}
	\label{tab:binaries}
	
	\begin{tabular}{lllrrrll}
		\hline
		\hline
		ID & \emph{Gaia} DR2 designation & J01ID & RV & nRV & $\Delta_\mathrm{RV}$ & B18 type & Ref.\\
		&& & (km\,s$^{-1}$) &  & (km\,s$^{-1}$) &  & \\
		\hline
		2v1067, 2i1133 & 5290671630116696448 &   6465 & 74.56 & 2 & 10.15 & B & 4, 3\\
		3i356, 3v336 & 5290672867067372672 &   8967 & 23.96 & 2 & 14.48 & B & 4, 2\\
		2v430, 2i441 & 5290716881891963520 &   4125 & 29.34 & 2 & 10.69 & \dots & 3, 5\\
		2v1097, 2i1164 & 5290719836829549056 &   6570 & 25.66 & 3 & 5.78 & B & 4, 3, 2\\
		3i139, 3v140, 2i1656, 2v1524 & 5290720867621791488 &   8172 & 24.34 & 2 & 8.03 & S & 4, 5\\
		3i84, 3v83, 2v1473, 2i1595 & 5290721142499618816 &   7962 & 24.38 & 4 & 5.25 & S & 4, 3, 5\\
		2v1372, 2i1487 & 5290721864054120320 &   7585 & 27.75 & 3 & 5.07 & B & 4, 3, 2\\
		2v1447, 4v62, 3v53, 4i61, 2i1568, 3i53 & 5290721898413863424 &   7864 & 32.54 & 2 & 10.28 & SB2 & 4, 2\\
		4i128, 1v1543, 1i1679, 4v127 & 5290722310730727040 &   8099 & 22.97 & 3 & 7.39 & B & 4, 3, 5\\
		1i1319, 1v1209 & 5290726262100476928 &   6880 & 16.32 & 3 & 11.96 & B & 4, 3, 2\\
		1v581, 2i563, 2v548, 1i619, deepW2579 & 5290735779747993088 &   4560 & 31.94 & 2 & 16.11 & \dots & 3, 5\\
		1i1014, 1v939 & 5290738047490662016 &   5887 & 23.43 & 3 & 7.40 & \dots & 3, 2, 5\\
		1v213 & 5290740040355374848 &   3208 & 18.30 & 4 & 22.20 & \dots & 3, 1, 2, 5\\
		1v857, 1i917 & 5290742857853957504 &   5586 & 47.75 & 2 & 10.50 & \dots & 3, 5\\
		3v687, 3i711 & 5290763095740848128 &  10301 & 27.37 & 3 & 5.32 & \dots & 3, 2, 5\\
		3v649, 3i665 & 5290763160163095936 &  10152 & -28.02 & 2 & 5.48 & B & 4, 3\\
		3i1254, 3v1199 & 5290765397843334912 &  12005 & 15.60 & 2 & 53.60 & \dots & 3, 5\\
		3v1116, 3i1168 & 5290765565341001984 &  11713 & 17.09 & 3 & 36.95 & SB2 & 4, 3, 5\\
		4v1274, 3v1369, 4i1360, 3i1446 & 5290766252535812224 &  12649 & 18.71 & 3 & 6.97 & B & 4, 3, 5\\
		3i1017, 3v977 & 5290767115830162560 &  11233 & 7.71 & 3 & 36.06 & B & 4, 3, 1\\
		3v384, 3i405 & 5290767562506658176 &   9175 & 27.63 & 3 & 7.13 & S & 4, 3, 5\\
		3i901, 4i845, 4v789, 3v874 & 5290768215341699200 &  10863 & 25.84 & 2 & 9.20 & B & 4, 3\\
		4v902, 4i966 & 5290769898968884096 &  11307 & 37.85 & 2 & 44.81 & SB2 & 4, 3\\
		4v321, 3v324, 4i341, 3i345 & 5290814875866306432 &   8920 & 25.12 & 2 & 5.67 & C & 3, 5\\
		4i237, 4v225 & 5290816353334936576 &   8529 & 25.39 & 4 & 5.42 & S & 4, 3, 2, 5\\
		4v773, 4i826 & 5290818964675068288 &  10817 & 25.67 & 3 & 5.01 & B & 4, 3, 5\\
		4i268, deepN1774, 4v254 & 5290819342632176512 &   8634 & 22.27 & 4 & 9.05 & B & 4, 3, 1, 2\\
		4i273, 4v259 & 5290820064186534144 &   8660 & 25.63 & 3 & 7.40 & B & 4, 3, 5\\
		4i796, 4v746 & 5290821919612568064 &  10736 & 33.80 & 3 & 12.25 & B & 4, 3, 5\\
		4v457, 4i490 & 5290822572447442176 &   9486 & 24.98 & 4 & 11.72 & B & 4, 3, 2, 5\\
		1i1447, 1v1335 & 5290833705002367104 &   7328 & 16.56 & 2 & 6.24 & \dots & 3, 5\\
		\hline
	\end{tabular}
	\tablebib{(1) \cite{1998MNRAS.300..550J}, (2) \cite{2002ApJ...576..950T}, (3) \cite{2016A&A...586A..52J} (\emph{Gaia ESO survey}), (4) \cite{2018MNRAS.475.1609B}, (5) \emph{Gaia}~DR2}
	\tablefoot{\emph{RV} is the mean radial velocity based on the \emph{nRV} different studies. $\Delta_\mathrm{RV}$ gives the scatter around the mean radial velocity. The classification from \cite{2018MNRAS.475.1609B} is given in the column \emph{B18 type}. Object IDs are, in column order, our own (upto 6 detections), from Gaia DR2, and from J01.}
\end{table*}

With multiple radial velocities for a number of stars, we are able to search for radial velocity variability, as arising from stellar multiplicity. For each star with multiple measurements, we calculate $\Delta_\mathrm{RV}$ as the peak-to-peak difference of the set of measurements. This way we obtain well-defined values even for a sample size of only two measurements. We chose to call stars with $\Delta_\mathrm{RV} > 5\,\mathrm{km\,s}^{-1}$ likely binaries. This threshold was selected because certain radial velocity measurements have large uncertainties and possibly different zero-points. With this definition, we find 32 likely binaries (listed in Table~\ref{tab:binaries}) with 29 of them being members of NGC\,2516. This large fraction is expected because the target selection of most radial velocity studies focused on the cluster main sequence.

Out of those 32 stars, 19 were also labelled as binaries by \cite{2018MNRAS.475.1609B}. Four of them  are marked as single by \cite{2018MNRAS.475.1609B} although the literature data show evidence for binarity\footnote{This could potentially be explained by the differing baselines of the studies and also differing zero points between them.}.
Stars identified as binaries by \cite{2018MNRAS.475.1609B} but not by our criterion usually have $\Delta_\mathrm{RV} < 3\,\mathrm{km\,s}^{-1}$ which is well within the uncertainty of the combined radial velocities
(and reasonable in the context of a single RV study, where relative precision is better maintained.).
Unfortunately, analysing the binary population further is beyond the scope of this work and we do not investigate this discrepancy here.
However, for the purposes of this work, we include all possible photometric (from J01) and radial velocity binaries from both this study and that of \cite{2018MNRAS.475.1609B} in the later analysis.
We call out and identify the corresponding rotators when we discuss them exhaustively in Sect.~\ref{sec:results}.

\subsection{Photometric anomalies}
\label{sec:photdiff}

\begin{figure}
	\includegraphics[width=\columnwidth]{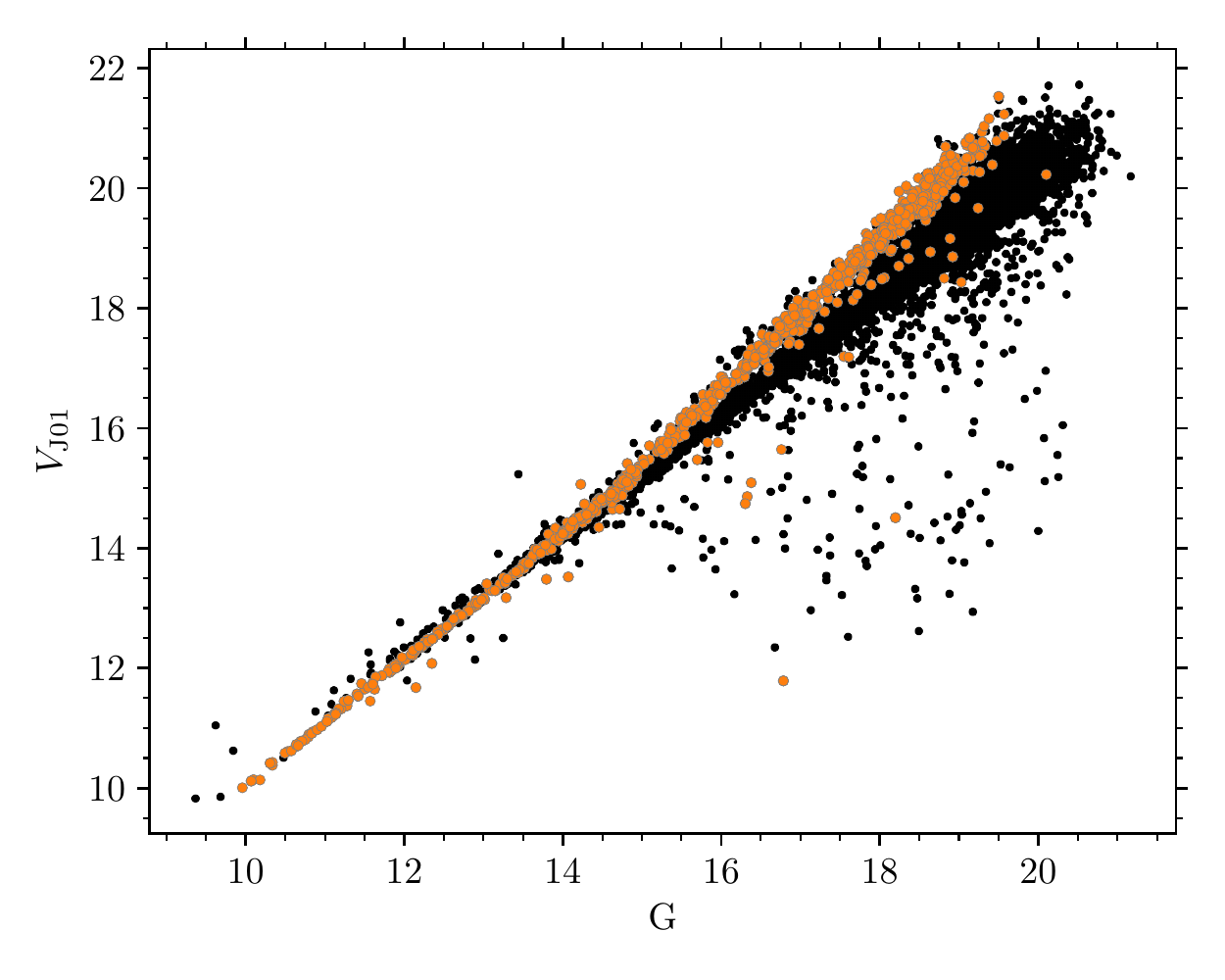}
	\caption{$V$ magnitudes from J01 plotted against \emph{Gaia}~DR2 $G$ magnitudes, with cluster members (satisfying two criteria) marked in orange.
            With the exception of the lower mass stars that are embedded in the background field, the member outliers below the sequences are considered for correction (as described in the text).
		}
	\label{fig:photdeviation}
\end{figure}

Upon inspection of our selected members in various photometric plots, we found that a small number of the measured magnitudes in J01 deviate significantly from \cite{2002AJ....123..290S} and \emph{Gaia}~DR2.
A comparison between \cite{2002AJ....123..290S} and J01 $V$ suggests that there is an offset of up to several magnitudes for some stars.
Correspondingly, in Fig.~\ref{fig:photdeviation}, displaying a comparison between the \emph{Gaia}~DR2 and J01 magnitude, some stars are found well below (= brighter in $V$) the otherwise tight correlation between $V$ and $G$.
It is unusual for even highly reddened stars to be four magnitudes fainter in $G$ than in $V$.

All stars with offsets are found to be brighter in the J01 photometry compared to other sources, indicating that additional flux from nearby stars likely entered the aperture in J01, who performed aperture photometry, rather than PSF fitting. Indeed, all the affected stars lie in the vicinity of brighter stars in our images, where both stars are correctly identified.

Consequently, we prefer the S02 photometry over J01 when available. Unfortunately, a number of rotators in the open cluster lie outside the FoV of S02.
This motivated us to include our own photometry as a third independent data set for stars with otherwise discrepant photometry and simply to transform our instrumental values to place them on the S02 system.

Also, certain stars appear to be obvious outliers in the $V$ vs. $G$ diagram but we do not have a $V$ magnitude measurement other than that from J01.
For those stars, we estimated a corrected magnitude by shifting the star to the S02 system.
Those magnitudes are of course relatively imprecise, but still preferable to a potentially anomalous value.
All corrected magnitudes are marked as such in the membership table.

Finally, although identified as inconsistent, we retain some stellar magnitudes as published.
One star (\emph{3v580}) has neither a proper motion nor a parallax measurement from \emph{Gaia}.
This star is a binary whose second component of similar brightness was also detected by \emph{Gaia} DR2.
While both are heavily blended in our images and form a single image on the CCD, they are resolved in the \emph{Gaia} CMD and fall onto the cluster sequence.
  One component appears later on the low-mass fast rotator sequence in the CPD, which fortunately, given that stars of a wide range of colour (or mass) have similar rotation periods, means that a different magnitude would not influence the shape of the rotational distribution.

We have not shifted the remaining low-mass outliers because we do not have $V$ light curves for them.
Using an estimate based on $G$ is not useful here as even the \emph{Gaia} cluster sequence broadens for the low mass stars and we cannot be sure about the true position in the $V$ vs. $G$ diagram.  

\subsection{Final membership}
\label{sec:finalmem}

\begin{table}
	\caption{Description of the columns of the online Table containing the membership information for NGC\,2516 constructed in this study. 
}
	\label{tab:members}
	
	\begin{tabular}{lcl}
		\hline
		\hline
		Name & unit & description\\
		\hline
		ID & - &  ID in this work\\
		designation & - & \emph{Gaia} DR2 ID \\
		J01 & - & ID from J01\\
		RA & deg & Right ascension from \emph{Gaia} DR2$^1$\\
		Dec & deg & Declination from \emph{Gaia} DR2$^1$\\
		Vmag & mag & $V$ magnitude from S02 or J01\\
		VCorr & - & Indicates $V$ correction (Sect. \ref{sec:photdiff})\\
		B\_V & mag & $(B-V)$ from S02 or J01\\
		bp\_rp & mag & $(G_\mathrm{BP}-G_\mathrm{RP})$ from \emph{Gaia} DR2\\
		pmra & mas\,yr$^{-1}$ & $\mu_\alpha \cos\delta$ from \emph{Gaia} DR2\\
		pmdec & mas\,yr$^{-1}$ & $\mu_\delta$ \emph{Gaia} DR2\\
		parallax & mas& Parallax from \emph{Gaia} DR2\\
		RV& km\,s$^{-1}$ & Mean radial velocity from lit. \\
		PMmem& - & Proper motion member (0/1) \\
		PLXmem& - & Parallax membership (0/1) \\
		RVmem& - & Radial Velocity membership (0/1) \\
		PHmem& - & J01 Photometric membership (0/1) \\
		KNM & - & Kinematic non-member (Sect. \ref{sec:kinnon})\\
		\hline
	\end{tabular}
	\tablebib{J01: \cite{2001A&A...375..863J}, S02: \cite{2002AJ....123..290S}}
	\tablefoot{$^1$ \emph{Gaia} DR2 epoch is J2015.5.}
\end{table}

We have constructed a final list of cluster members from the stars which have survived our filtering as described above. It contains 844 stars (hereafter called members) and is provided as an online table with the data columns described in Table~\ref{tab:members}.
We note that additional members, both brighter and fainter than our photometry, will undoubtedly be found in NGC 2516 in the future.
As already seen from the results in \cite{2018A&A...616A..10G}, NGC\,2516 is one of the three richest nearby open clusters, comparable to the Pleiades \citep{2015A&A...577A.148B} and to NGC\,3532 \citep{N3532RV}.
In summary, NGC 2516 is a very rich open cluster with 844 probable members to date within our ${\sim}1\degr{}$ FoV.

\subsection{Colour-magnitude diagram}

%
\begin{figure*}
	\includegraphics[width=\textwidth]{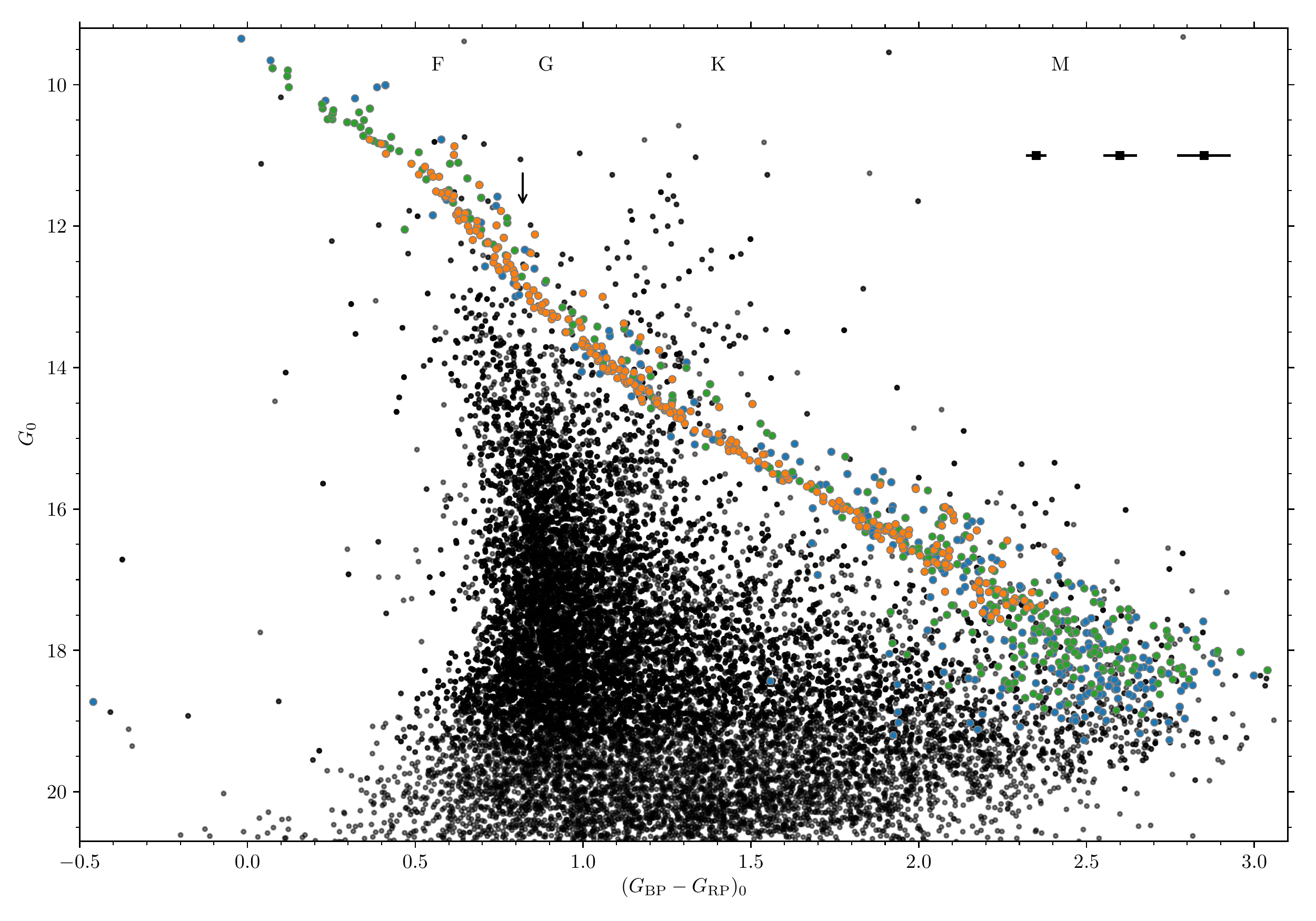}
	\caption{\emph{Gaia}~DR2 colour-magnitude diagram for all stars for which we constructed light curves.
	The best-defined members (satisfying all four criteria) are marked in orange, members according to three criteria are marked in green, and those fulfilling only two in blue. Non-members are marked in black. We note that photometric binaries are retained. The arrow indicates the colour of the Sun \citep{2018MNRAS.479L.102C} for context. The spectral classifications at the top are according to \cite{2013ApJS..208....9P}. In the upper right we display representative uncertainties for stars on the cluster sequence. Uncertainties for $(G_\mathrm{BP}-G_\mathrm{RP})_0 < 2.3$ and for $G$ are within the symbol size and are not shown here.}
	\label{fig:memberCMD}
\end{figure*}

We plot the cluster colour-magnitude diagram (CMD) with the membership information in Fig.~\ref{fig:memberCMD} and find that our criteria successfully retain photometric binaries. From the CMD we find that members satisfying the most criteria are usually photometrically single stars close to the cluster sequence. Stars which fulfil all four membership criteria can be found in the range from $G=11$ to $18$. For brighter and fainter stars, there is usually no radial velocity information available. We note that one of the cluster white dwarfs has found its way into our membership list despite our not explicitly including white dwarfs. One could potentially include constraints from dedicated white dwarf studies (e.g. \citealt{1996A&A...313..810K}) in the cluster; however our focus here is on the main sequence population of NGC\,2516.

Certain stars appear to be \emph{Gaia} photometric non-members in Fig.~\ref{fig:memberCMD} but are otherwise classified as members.
(Recall that \emph{Gaia} photometry was intentionally excluded as a membership constraint.)
Hence this figure provides an immediate visual impression of the false-positive rate in our membership list.
However, we caution that \emph{Gaia} DR2 photometry should not be treated as gospel truth because those data also contain misplaced stars, especially in the vicinity of brighter stars, as we will show later in Sect.~\ref{sec:photnonmem} below by comparison with $(V-K_s)$ photometry for those which are found to be rotators.

%
%

\section{Time-series analysis}
\label{sec:ts}

Several period search algorithms are available in the literature for the determination of the rotation period ($P_\mathrm{rot}$) from the light curve. Extensive comparisons have been published (e.g. \citealt{1999ApJ...516..315S} or \citealt{2013MNRAS.434.3423G}) showing that all methods have their advantages and disadvantages depending on the type of input data or the computing time. For our study, we used an improved version of the procedure applied in our earlier work \citep{2016MNRAS.462.2396F}. In the present paper we apply five different algorithms to derive the rotation period. Additionally, we filter the periods at an intermediate stage with a signal-to-noise criterion.

Our selection of algorithms includes the widely-used and robust Lomb-Scargle (LS, \citealt{1976Ap&SS..39..447L, 1982ApJ...263..835S}) which is the work horse for period determination. Here we apply the generalized Lomb-Scargle (GLS, \citealt{2009A&A...496..577Z}) which takes the photometric errors of the light curve into account in the period determination. The (G)LS fits a combination of sine and cosine waves to the data with periods from a grid and returns the spectral power density of the light curve for each input period.

Our second method is the \textsc{clean} periodogram \citep{1987AJ.....93..968R, 2001SoPh..203..381C}, which is designed to suppress the window function from the periodogram and reduce the power from alias frequencies. For our ground-based observations this mainly means suppressing the one-day alias period imposed on the data by the observational frequency.

The remaining three of the five applied algorithms share the method of period search through phase-folding; nonetheless they differ from each other substantially.
From this class of algorithms we chose the phase dispersion minimization (PDM, \citealt{1978ApJ...224..953S}), string-length (SL, \citealt{1983MNRAS.203..917D}), and the Gregory-Loredo Bayesian periodogram (GL \citealt{1992ApJ...398..146G, 1999ApJ...520..361G}).
PDM uses the phase space to fold the data into a light curve with the minimal dispersion.
The SL algorithm, which folds the data in the time domain and calculates the distance between consecutive points, is very similar.
In addition to phase-folding, the GL periodogram uses Bayesian statistics to find the periodicity of the data.
An advantageous feature of these three algorithms is that, in contrast to the sinusoidal variation used in the LS analysis, the variability of the target is not assumed to follow a specific shape.

We also experimented with the minimum entropy method (MEM, \citealt{1999MNRAS.307..941C, 2013MNRAS.434.2629G}) but concluded that it is not suitable for our data structure, with its median sampling of four data points per night.
In most cases, even when a clear periodicity is visible in our data, the algorithm locks down on the observational frequency of 1\,d and its multiples.
Therefore, we have not considered this method further for our study and have limited ourselves to the five methods described above.

%
\begin{figure}
	\includegraphics[width=\columnwidth]{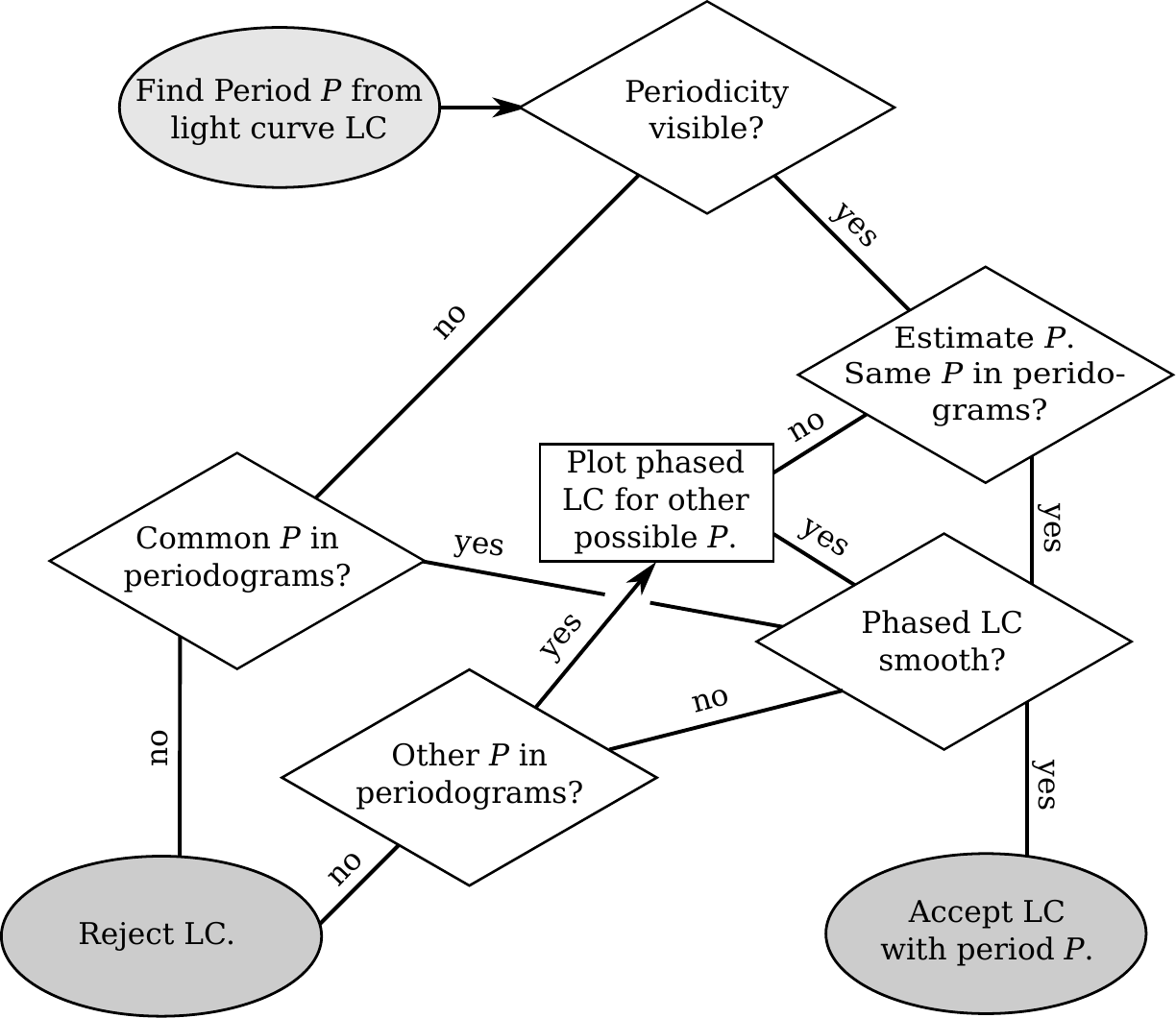}
	\caption{Flowchart for the determination of the preliminary period $P$ of a given light curve (LC). All flows, with the exception of the two explicitly marked with arrows, go downwards.}
	\label{fig:decisiontree}
\end{figure}

\subsection{Signal-to-noise criterion}
\label{sec:snr}

Each of the five algorithms provides a period for a light curve but does not necessarily inform us about its reliability. Although, the individual algorithms include certain measures for the spectral power of a given period which one could in principle, use to define a threshold value, we have decided not to use the power of the corresponding periodograms. Instead, we use a signal-to-noise ratio (S/N) criterion for each determined period.

For a given light curve, we obtain a set of five periods, one from each method, for which we can calculate the S/N. We then fit a sine wave for each period to the light curve. The initial phase is set to zero, and the amplitude and zero point offset are estimated from the light curve. The fitted sine is then subtracted from the light curve to estimate the noise term of the S/N. The peak-to-peak amplitude of the fitted curve is defined as our signal, and the peak-to-peak value of the residuals as the noise. The latter can be used safely because all light curves have been sigma-clipped before applying any algorithm.

The assumption of a sinusoidal shape for the light curve is appropriate because we are searching for starspot induced periodic variability. All our rotators fall into the class of sinusoidal or pseudo-sinusoidal light curves, a selection based on this criterion is not biased towards a single class.

For each light curve, we select the period with the highest S/N as our initial period for the manual inspection and manipulation. (Further details are located in Sect.~\ref{sec:finalP} below.) A comparison of the different S/N values shows that for $\mathrm{S/N}>0.7$ the values are mostly correlated, meaning that the periods of the different methods agree well. Therefore, we have chosen 0.7 as our threshold value and have manually inspected all light curves for which at least one period with $\mathrm{S/N} > 0.7$ was obtained.

\subsection{Rotation period}
\label{sec:finalP}

\begin{figure*}
	\includegraphics[width=\textwidth]{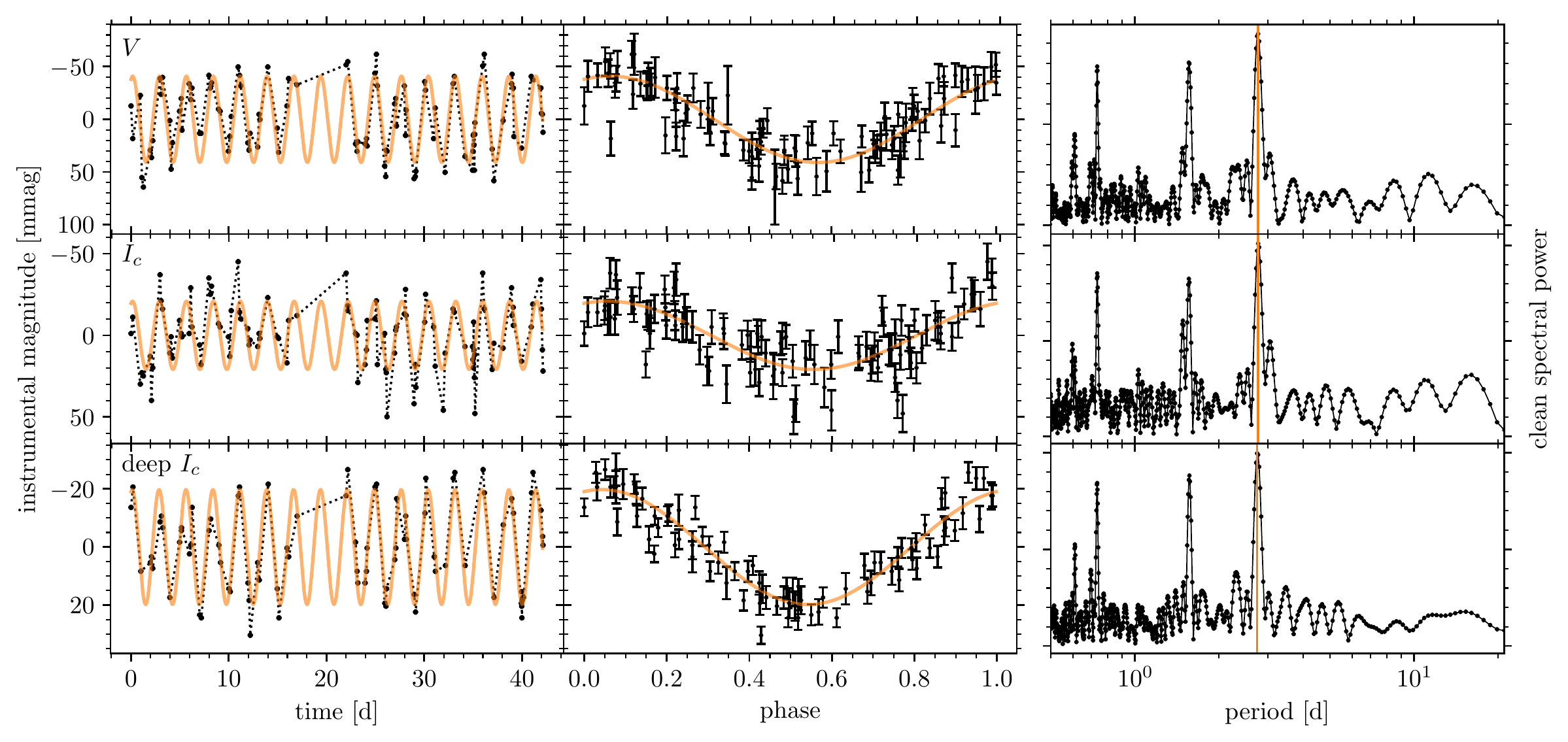}
	\caption{Multiple light curves for the same spectral type M1 star (Gaia~DR2~5290816800011541504, $V=17.7$, $(V-K_s)_0=3.9$) in $V$, $I_c$, and deep $I_c$, demonstrating both similarities and differences. From left to right, we show the light curve in the time (\emph{left}) and phase domains 	 (\emph{centre}), and the \textsc{clean} periodograms (\emph{right}).
          This illustrates how a period of 2.75\,d is confirmed within our data set. The rotation period is marked in the \textsc{clean} periodogram with an orange line at the highest peak. The three highest secondary peaks in the periodogram are beat frequencies of the rotation period with the 1\,d alias. This figure also shows the behaviour of measurement uncertainties in the different bands. The $I_c$ band (\emph{middle} row) shows  especially large uncertainties relative to the light curve amplitude whereas $V$ (\emph{top}) and in particular the longer-exposed deep $I_c$ (\emph{bottom}) are more tightly correlated.}
        \label{fig:multLC}
\end{figure*}

The initial period from the S/N criterion is preliminary, and we require a further check to ensure that it is indeed the stellar rotation period. We followed the decision-making process outlined in the flowchart displayed in Fig.~\ref{fig:decisiontree}. This manual procedure was only applied to our previously-determined members of NGC\,2516 for two reasons: 1. it is labour-intensive, and 2. we are currently interested only in the rotation periods of the cluster members.

We inspected all available information about the light curve to evaluate whether a star exhibits convincing periodic behaviour. Apart from the light curve itself, we use all five periodograms (both in frequency and period space), and the light curve, phase-folded on the period with the highest S/N. We overplotted a sine wave with this period on the light curves in both the time domain and in phase space to guide the eye.
An example for a first class rotator with light curves in multiple filters is displayed in Fig.~\ref{fig:multLC}.

Thus, our flowchart begins with visual classification of the light curve to decide whether periodic behaviour is visible at all. If so, we estimate the period by eye and examine whether this is confirmed by the periodograms. If these show evidence for the same (or similar) period as found manually, we use the phase-folded light curve. It is required to show a tight sequence of data points. In this case, we accept the period right away for further treatment (Sect.~\ref{sec:meanPer} below).

In certain cases, the positions of the highest peaks in the periodograms are not equal to the manually estimated period. In those cases we need to decide whether this algorithmic period is reasonable by comparing the phase-folded representation of both periods. If the phase-folded light curve is smooth we accept the algorithmic period.

In cases where no obvious periodic behaviour of the light curve is visible, we used the periodograms to seek out a common period among the different methods. If a convincing signal is present, we apply the techniques given above; otherwise the light curve is rejected as non-periodic.

Light curves with detected periods that we subsequently classify as likely aliases are treated as special cases. Those light curves are initially preferentially classified as slow rotators (i.e. the longer rotation period is preferred) with periods of a few days, together with some photometric noise.
This preference is mostly supported by the periodograms themselves which confirm the longer periods.
The \textsc{clean} periodogram is especially useful in providing valuable support in these cases because it suppresses multiples of the observational frequency and the beating associated with it ($1/P_\mathrm{beat} = 1/P_\mathrm{rot}\pm 1\,\mathrm{d}^{-1}$), enabling the identification of the underlying rotation period.

Another group of stars affected by aliasing are fast rotators whose periods and corresponding aliases are both below one day. We identified several of these cases in our analysis.
It is not possible to distinguish the true period from an alias without careful visual inspection.
Hence, we closely inspected the light curves themselves, concentrating on the dense coverage of data points on the night of 26 March 2008, where we have up to eight images of NGC\,2516 (inner fields). From this ensemble of data points, we estimated the slope of the variation, which then allows us to favour one of the possible periods. Unfortunately, the coverage is still occasionally too sparse to break the alias completely.
These stars are marked as possible alias periods in our results. 

Among our measured rotation periods some are of course multiples of 1\,d,
but for those light curves even the \textsc{clean} periodogram confirms those periods with high significance.
No signs of possible aliases can be found in those with a thorough analysis of the periodograms. And indeed, such periods are expected to occur naturally.
Therefore, we have included them in our data set without further consideration. 

After following the flowchart (Fig.~\ref{fig:decisiontree}) for all cluster members with light curves (844 stars with 1810 light curves),
we identified 308 rotators (with 530 light curves).
All other members are classified as showing no evidence for rotational
modulation in our study.
The field stars (non-members) are not considered further in this study. 

\subsection{Further treatment}
\label{sec:meanPer}

The periods found above from both the light curve and the periodograms are not yet considered final. Even within a narrow range around this value each of the applied algorithms finds a slightly different period, giving us a small range of possible estimated periods.

In order to provide a mean period, we return to the periodograms and locate for each of the five the positions of the highest peak near the estimated period. For the detection, we chose a window of $\pm 20$ percent around the given period for $P_\mathrm{rot} > 2\,$d and $\pm 0.2\,\mathrm{d}^{-1}$ for faster rotators. From the five periods derived above, we report the mean value as our period measurement and half of the range as the uncertainty.

For some light curves, the periodograms contain alias periods which are very close to the true period. Hence, they might fall into our 20 percent window and be the highest peak for one or two methods, leading to a bias of the mean period and inflation of the uncertainty. In those cases we shrank the window manually to exclude those alias peaks.

This method can lead to rather small uncertainties for those light curves with very coherent periodicity mostly fast rotators with a constant phase. Conversely, for stars with evolving spots the different methods find more diverging periods because the peaks are broader and the centre is not always at the same position.

From visual inspection we classified the light curves into three classes of confidence: (1) Stars of the first class show a clear rotational signal in both the light curve and the periodogram\footnote{For stars in this class one could in principal determine the rotation period even without a periodogram.}.
(2) Stars of the second class show a noisy signal in the light curve but a clear peak in the periodograms.
(3) The third class consists of the possible alias periods discussed above.
These classes are marked accordingly in the Figures and Tables. In Appendix Fig.~\ref{fig:exampleLCs}, we show examples of light curves of the different classes.

\subsection{Periods from multiple light curves}

Some stars appear in multiple fields. Indeed, almost all stars in the inner fields have light curves both in $V$ and $I_c$. One example with three light curves from each of the different exposure configurations is shown in Fig.~\ref{fig:multLC}. The light curves (left panels) show slightly different features that might be explained by the different observation times of the images, photometric noise and stellar activity. In general, the rotation period is visible well in all three light curves.
The central column of the figure shows the phase folded light curve. 
It is obvious that the deep $I_c$ light curve shows both the tightest correlation, and the smallest uncertainties.
The \textsc{clean} periodograms on the right illustrate the very good agreement of the periods for all three independent light curves.
We note the presence of the clearly visible beat periods, approximately symmetric with the 1\,d observing cadence.

In the most extreme case, we are in possession of eight light curves in two filters from the overlap region of all four inner fields.
In such cases, even for the four light curves in the same filter, large differences in signal-to-noise are visible, showing that the detection threshold is not only a function of brightness and period, but also position on the detector.
For both filters the light curve from field\,1 displays the clearest variability. The poorer performance of the $I_c$ light curves due to the lower contrast between stellar spot-group and the photosphere is also apparent.

Light curves of the same star obviously should exhibit the same period.
In order to test this, we checked whether multiple periods for the same star agree within our determined uncertainties. Indeed, this is usually the case.
In our whole sample of 308 stars, we found only eight stars that do not satisfy this criterion. All of them show differences in their determined periods slightly larger than the estimated uncertainties. This can be seen as a test of the reliability of our uncertainty estimate. Because 2.5 percent of the stars have larger uncertainties, we conclude that the estimated uncertainties are in fact 2$\sigma$ errors. This result additionally demonstrates the high reliability with which we can determine the rotation periods with our combined methods.

After having verified the agreement of all periods from different light curves, we simply average the values to determine the final period.
For the uncertainty of the average period, we chose either the largest uncertainty from the various light curves, or, in the case of period differences larger than this, the amplitude of the dispersion.
This procedure provides the final set of 308 rotation periods and corresponding uncertainties for NGC\,2516 listed online, described in Table~\ref{tab:rotationperiods}.

%
\begin{table}
	\caption{Description of the 555 unique-object online table containing our new 308-star stellar rotation period sample, supplemented with the 247-star sample from \cite{2007MNRAS.377..741I}. 
} 
	\label{tab:rotationperiods}
	
	\begin{tabular}{lll}
		\hline
		\hline
		Name & Unit & Description\\
		\hline
		ID & - & ID in this work\\
		designation & - & ID from \emph{Gaia} DR2$^1$\\
		J01 & - & ID from J01\\
		RA & deg & Right ascension from \emph{Gaia} DR2$^1$\\
		Dec & deg & Declination from \emph{Gaia} DR2\\
		Vmag & mag & $V$ magnitude from S02 or J01 or corr. \\
		B\_V0 & mag & $(B-V)_0$ colour from S02 or J01\\
		V\_K0 & mag & $(V-K_s)_0$ colour\\
		bp\_rp0 & mag & $(G_\mathrm{BP} - G_\mathrm{RP})_0$ colour from \emph{Gaia}~DR2\\
		Prot & d & Rotation period\\
		dProt & d & Uncertainty in rotation period\\
		P90 & mag & Light curve amplitude\\
		Class & - & Classification of period\\
		ProtI07 & d & Rotation period from I07\\
		logLxLbol & - & $\log L_\mathrm{X}/L_\mathrm{bol}$ (P06)\\
		Ro & - & Rossby number\\
		PMmem & - & Proper motion membership (0/1)\\
		PLXmem & - & Parallax membership (0/1)\\
		RVmem & - & Radial velocity membership (0/1)\\
		PHmem & - & J01 Photometric membership (0/1)\\
	\hline
	\end{tabular}
	\tablebib{J01: \cite{2001A&A...375..863J}, S02: \cite{2002AJ....123..290S}, I07: \cite{2007MNRAS.377..741I}, P06: \cite{2006A&A...450..993P}
	\tablefoot{$^1$ The \emph{Gaia} DR2 epoch is J2015.5.}. 
    The membership information provided is for both our rotators from Table~\ref{tab:members}, and for the additional ones from I07.}
\end{table}

\subsection{Variability amplitudes}
\label{sec:varamp}

%
\begin{figure}
	\includegraphics[width=\columnwidth]{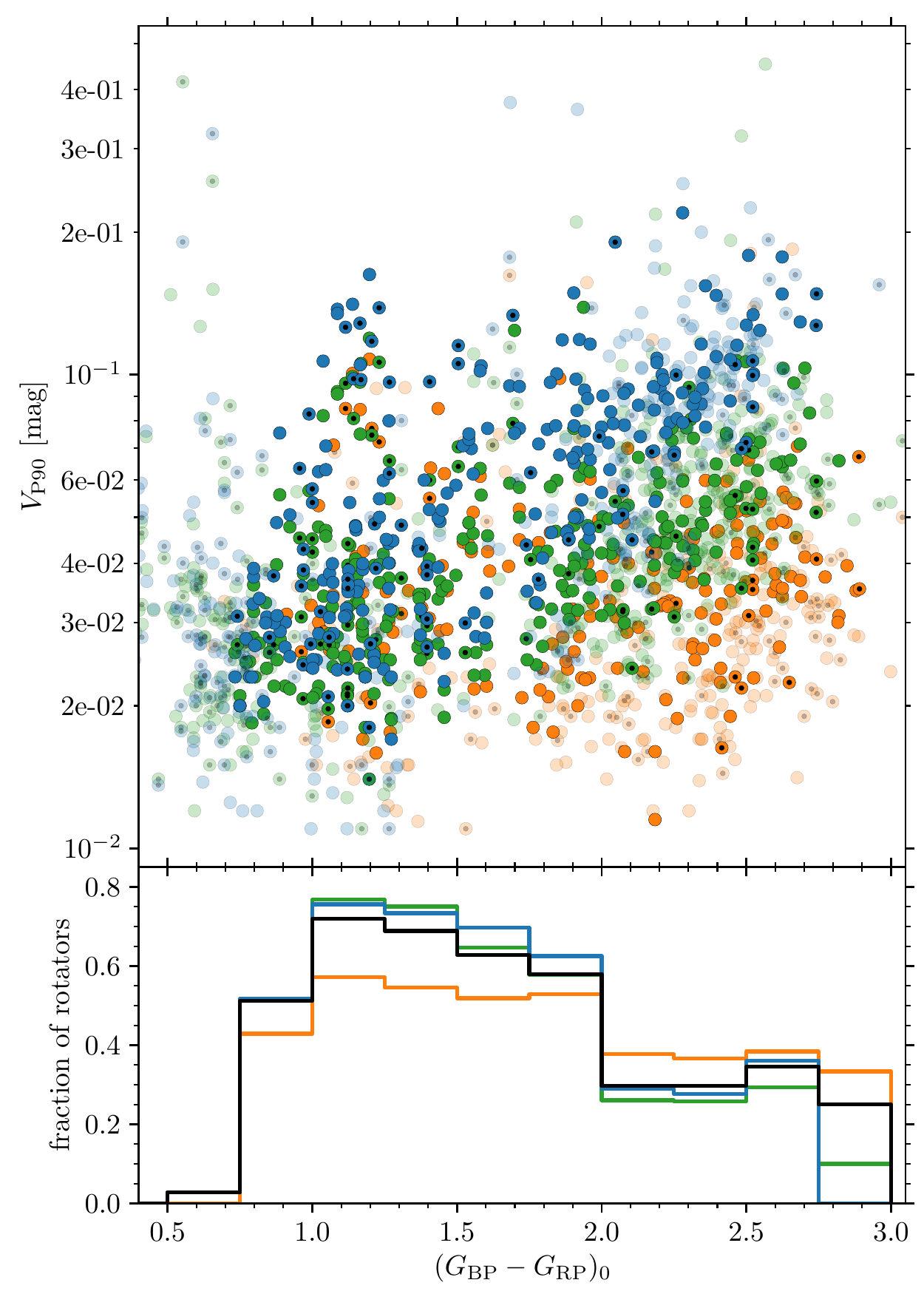}
	\caption{\emph{Top:} Variability amplitude $V_\mathrm{P90}$ against the intrinsic colour $(G_\mathrm{BP}-G_\mathrm{RP})_0$. The colour coding is according to the exposure and filter combination (blue: $V$ 120\,s, green: $I_c$ 60\,s, orange: $I_c$ 600\,s, same as in Fig.~\ref{fig:photuncert}). Stars with derived periods are highlighted and additional cluster members are shown in the background. Possible binaries (radial velocity and photometric) are indicated with additional small black dots. \emph{Bottom:} Fraction of periodic stars as a function of colour for the different exposure and filter settings. The black line marks the overall detection rate. (It is lower for certain bins because of the additional members introduced by the deep outer fields.)}
	\label{fig:amps}
\end{figure}

An easily obtainable measure from the light curve is the variability amplitude. We define it here as the difference between the tenth and the ninetieth percentile of the light curve \citep{2011AJ....141...20B}. This measure ($V_\mathrm{P90}$) is robust against outliers but at the same time captures non-periodic variations. Therefore, we can calculate it for all light curves and compare the rotators to the non-rotators.
	
Accordingly, in the upper panel of Fig.~\ref{fig:amps} we display $V_\mathrm{P90}$ against $(G_\mathrm{BP}-G_\mathrm{RP})_0$ for all members of NGC\,2516, colour-coded by the filter. It is immediately visible that both the rotators and non-rotators have very similar variability amplitudes. It follows that stars without a measured rotation period are likely have unfavourable spot configurations or evolution which do not allow the period to be measured. We have marked the possible binaries (from radial velocity and photometry) among the members and find that they too follow the same distribution.

The trend towards higher variability amplitudes with redder intrinsic colour also relates to the higher photometric uncertainty. The longest exposed light curves exhibit the lowest level of variability. Due to this trend, the detection fraction of the reddest stars is lower.

We calculate this fraction of the rotators among the cluster members and display it in the lower panel of Fig.~\ref{fig:amps}. For the G and early K stars, we were able to derive rotation periods for up to 80 percent of the stars, mostly as a result of the ideal exposure settings. Among the M stars, the deep $I_c$ (600\,s) exposures also appear to have revealed a large number of rotators which we would have missed without these frames.

\subsection{Comparison with prior rotation work in NGC\,2516}
\label{sec:compI07}

%
\begin{figure}
	\includegraphics[width=\columnwidth]{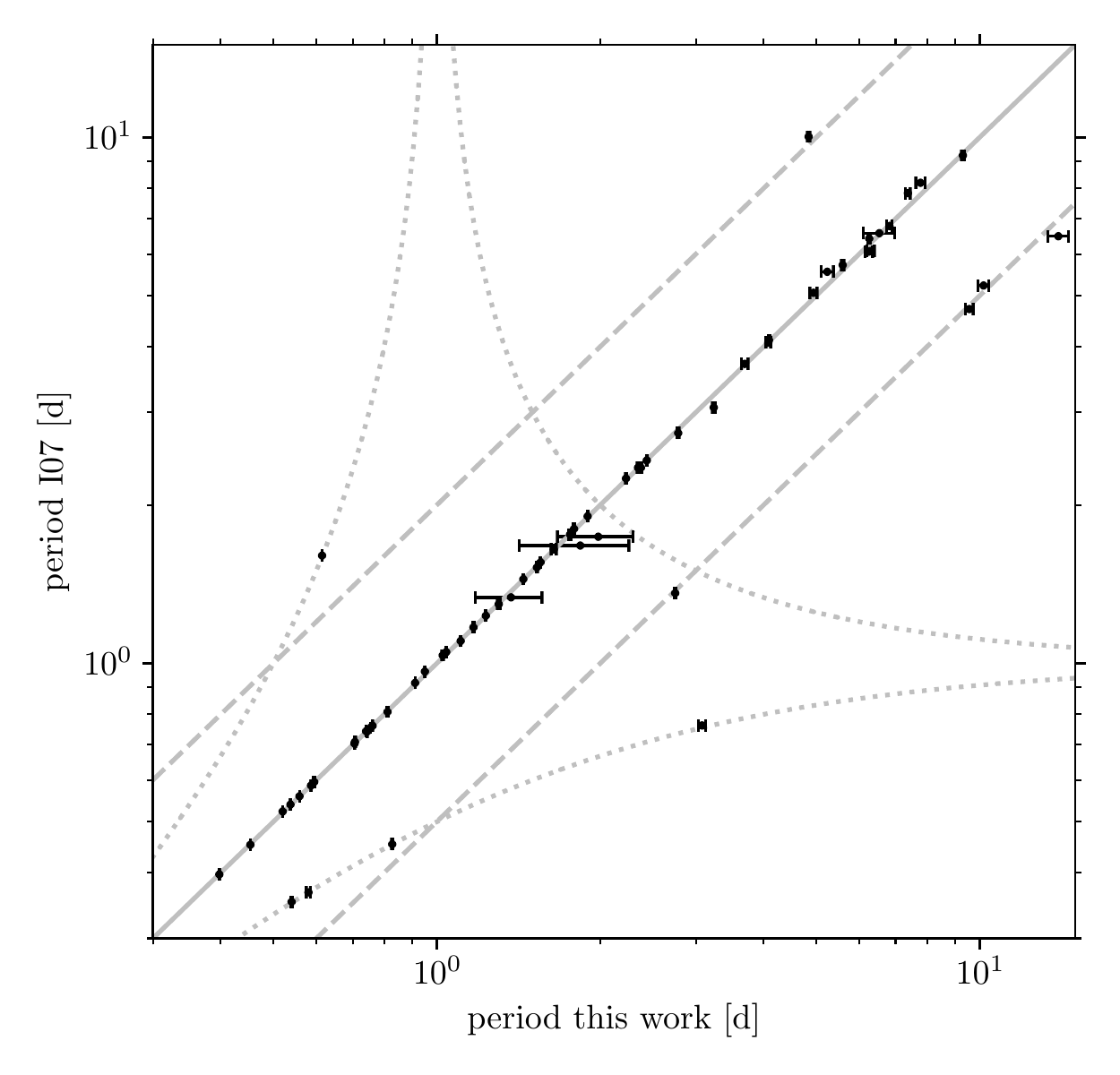}
	\caption{Comparison of our rotation periods with those measured by I07. The line of equality is shown as the solid grey line and the dashed lines indicate the half and double periods. The dotted lines show the beat periods of the observational period. The outliers are discussed in the text.}
	\label{fig:compI07}
\end{figure}

Stellar rotation in NGC\,2516 was previously studied by \cite{2007MNRAS.377..741I} (hereafter I07). In Fig.~\ref{fig:compI07}, we show the comparison between their periods and ours for the set of 62 common (mostly M) stars. In general the periods agree well; the data points scatter around the line of equality. However, the periods of ten stars are in disagreement, all of which are close to or exactly equal to the alias and beating periods. Hence we double-checked our periods relative to those from I07.

Among the ten stars we find five with periods in a 1:2 ratio.
We have re-examined those light curves and periodograms but we cannot confirm the periods provided by I07 in our data.
Hence, we conclude that our periods are preferred, especially because we are in a better position to break the aliases with our much longer baseline.
Whereas I07 had two four-day windows separated by one week, we observed for 42\,d with only a four day scheduling gap.

The remaining five discrepant periods are affected by beating with the observing cadence.
Because both studies could suffer from this alias, it is not easy to chose between the two alternative periods.
Among these five stars, two have already been classified by us as possible alias periods.
For all these five stars we also detect the additional component related to the I07 period in our periodograms but for none can we unambiguously
favour one or the other rotation period.
Hence, we simply retain the period values we derived. All of the five affected stars are early M dwarfs where the range of observed rotation periods are large. A handful of indefinite periods will not influence the later analysis of the whole distribution as seen below.

%
%
%
%

\section{Colour-period diagram for NGC\,2516}
\label{sec:results}

%
\begin{figure*}
	\includegraphics[width=\textwidth]{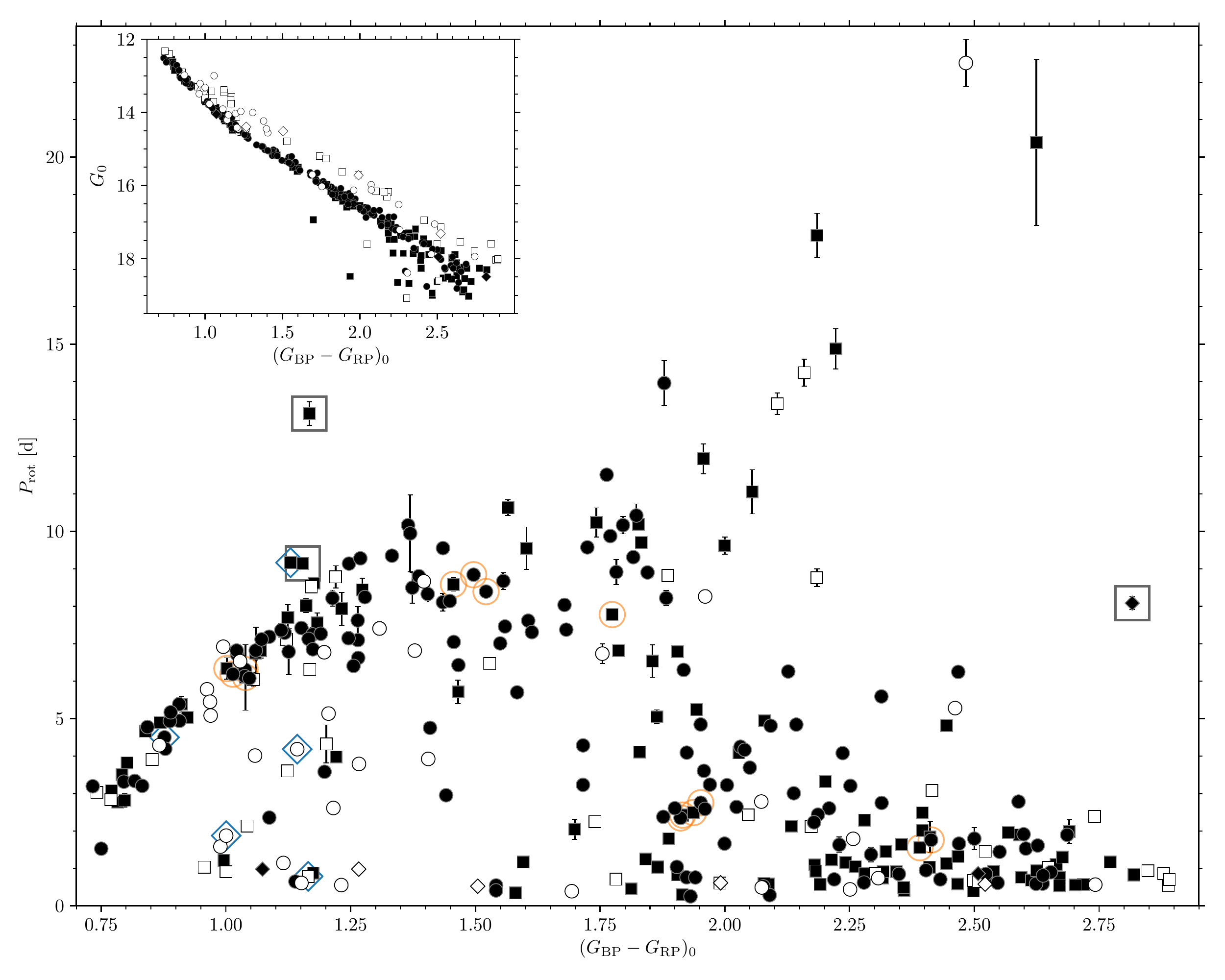}
	\caption{
            Colour-period diagram (CPD) for NGC\,2516 (large figure) using $(G_\mathrm{BP}-G_\mathrm{RP})_0$ colour, with an inset displaying the corresponding colour-magnitude diagram (CMD) for the rotators. Symbols have identical meaning in both plots.
            A well-defined slow rotator sequence (upper boundary) is visible, as well as a fast-rotation boundary that is increasingly well-populated among cooler stars. The triangular region between these is also occupied, and the paucity of outliers is remarkable. 
            Obvious rotation periods are marked with circles (both filled and unfilled) and less evident (algorithmic) periods with squares. (We note that the distributions are identical.)
            Small diamonds indicate the few possibly aliased periods.
            Period uncertainties are only visible when they exceed the symbol sizes.
            Filled symbols denote single stars (really, those without known companions to date), while open symbols indicate known or suspected binaries (of all separations), either from photometry or spectroscopy or both.
            Five symbols enclosed within large blue diamonds indicate those with $P_\mathrm{orb}<10$\,d as determined by \cite{2018MNRAS.475.1609B}, while those enclosed by large black squares are discussed individually in the text.
            Stars enclosed within orange circles are those whose light curves are displayed for illustrative purposes in the Appendix (Fig.~\ref{fig:exampleLCs}).
        }
	\label{fig:CPD}
\end{figure*}

A standard way of displaying and discussing open cluster rotation period measurements is in a colour-period diagram (CPD), with periods plotted against any convenient colour.
We therefore plot our new sample of 308 rotation periods for stars in NGC\,2516 in Fig.~\ref{fig:CPD} in a CPD using \emph{Gaia} $(G_\mathrm{BP}-G_\mathrm{RP})_0$ colour, partially because of its relative precision.
(Equivalent colours, $(V-K_s)_0$ and $(B-V)_0$, are also discussed below.)
As expected for an open cluster of this age, we find that the distribution has a roughly triangular shape, identified by \cite{2003ApJ...586..464B} as a key unifying feature of several ZAMS and post-ZAMS open clusters \footnote{See also \cite{1989ApJ...343L..65K} in connection with the Hyades open cluster.}.
This shape has also been observed in a steadily-increasing series of open clusters since then.
However, the relative absence of outliers in this member-only CPD is remarkable.

The upper boundary of the CPD is delineated by a well-defined slow rotator sequence that stretches from early G-type solar-mass stars all the way to M dwarfs.
The CPD also has a well-defined lower boundary, consisting of the fastest rotators at each colour, ranging from slightly sub-solar-mass stars at $(G_\mathrm{BP}-G_\mathrm{RP})_0 \sim 1$ to the lowest-mass stars in the sample.
Note especially that this fast rotator sample even contains Class\,1 rotation periods (circular symbols, both filled and unfilled in Fig.~\ref{fig:CPD}), whose rotation periods can simply be read off the light curves without the necessity for periodogram analysis.
All indicators point towards the correctness of these rotation periods, and to their being bona-fide cluster members (all plotted symbols).
They should emphatically not be discarded, as in certain prior studies.
As expected from prior work, the fast rotator sample is certainly less populous than the slow rotator sequence among the warmer stars, but accounts for an increasingly-large fraction of the cooler cluster member stars.
The region between these boundaries, the interior of the triangular region, the so-called `gap' region of \citealt{2003ApJ...586..464B}, interpreted there as stars undergoing the uni-directional transition from fast- to slow rotation, is also occupied with both obvious (circular symbols) and algorithmic cluster member rotation periods (square symbols, both filled and unfilled).
Again, the occupancy of this region represents astrophysical reality.

There are four obvious outlier rotation periods (all algorithmic), three at $(G_\mathrm{BP}-G_\mathrm{RP})_0 \sim 1.15$, and one possibly-aliased rotator (diamond) among the reddest stars, at $(G_\mathrm{BP}-G_\mathrm{RP})_0 \sim 2.8$.
Careful manual inspection (Section~\ref{sec:outliers} below) of the relevant data does not permit us to relocate them in any way, and we conclude that these objects are also real.
We note further that these four objects are all classified as single stars to date (solid symbols of various shapes), as opposed to objects that are either photometric or radial velocity binary members (equivalent unfilled symbols).

For the CPD in Fig.~\ref{fig:CPD}, we use \emph{Gaia} photometry because those are the only passbands where photometry is available for all stars with measured rotation periods.
Later we also use $(V-K_s)_0$ as our mass proxy for two reasons. Firstly, the $G_\mathrm{BP}$ magnitude is not reliable for very low mass stars \citep{2019MNRAS.485.4423S}.
Secondly, $(V-K_s)_0$ is increasingly being used in studies of open clusters, particularly as greater numbers of lower-mass stars are being measured.
For instance, it is the main colour used in the recent Pleiades study \citep{2016AJ....152..113R}, our principal comparison to NGC\,2516.
We de-redden the photometry by rescaling appropriately from $\mathrm{E}_{(B-V)}=0.11$\,mag \citep{2002AJ....123..290S}. The optical photometry is mainly from S02 and J01, and the infrared photometry is from the Vista Hemisphere Survey (see Sect.~\ref{sec:litphot}).
The data presented in this figure are available online, and Table~\ref{tab:rotationperiods} describes the columns provided.
  
\subsection{Binarity}
\label{sec:binariesRotation}
Although we do not study the binary population in NGC\,2516 in detail, the literature offers significant information on multiplicity status (see Sect.~\ref{sec:binaries}) which we include in our analysis.
We adopt the photometric binary classification from J01 and supplement this with radial velocity binaries from our analysis (Sect.~\ref{sec:binaries}) and from \cite{2018MNRAS.475.1609B} without considering orbital period or separation. These stars are marked with open symbols in Fig.~\ref{fig:CPD}. We note that our current classification almost certainly includes both false-positives and false-negatives. For instance, with few exceptions, available data do not yet permit us to distinguish between those binaries that are truly interacting rotationally and those that are distant enough to be effectively single stars.

Given the above-mentioned caveats, we find that the fraction of photometric binaries among stars with ${(G_\mathrm{BP}-G_\mathrm{RP})_0}\leq1.8$ is
60.0 percent (12/20) for the fast rotators and a substantially-lower
18.4 percent (18/98) for the slow rotators. 
The gap stars have a binary fraction of 52.9 percent (9/17).

This represents a significant difference, and because it would be absurd for
binaries to become single stars in the course of transitioning from fast-
to slow rotation, we must conclude that this difference represents a real
difference in initial conditions, that is that the rotational evolution is such that
binaries preferentially emerge on the ZAMS as fast rotators, while singles
preferentially appear as slow rotators.

The higher fraction of binary stars among the fast rotators and the gap stars (i.e. presumably somewhat rotationally evolved fast rotators) could point to the influence of multiplicity on the discs of young stars.
For example, a (nearby) binary companion might truncate the disc, which then could dissipate faster than otherwise \citep{1994ApJ...421..651A,1994ARA&A..32..465M}.
In this scenario, stellar rotation would not be able to couple to the disc and the star would not have the corresponding brake on its pre-main sequence spin-up \citep{2019A&A...627A..97M}.
Such a mechanism would, of course, work in addition to the usual tidal interactions that are known to occur (and to influence rotation) in close binary systems.
Our photometric binaries are mostly too far apart for tidal interaction to be relevant.

However, \emph{3i1017/3v977} with $(G_\mathrm{BP}-G_\mathrm{RP})_0=1.0$, $P_\mathrm{rot}=1.89$\,d, marked with a blue diamond in Fig.~\ref{fig:CPD} is one particularly interesting case. For this fast rotator\footnote{UCAC4 ID 146-012601 in \cite{2018MNRAS.475.1609B}} \cite{2018MNRAS.475.1609B} have obtained a radial velocity time series, allowing them to estimate the orbital period. They find a periodicity in the radial velocity data of 1.9\,d. The near-exact match between the orbital and rotational periods indicates a tidally-locked system, or at least the presence of a large spot in phase with the orbital period\footnote{The radial velocity amplitude is too large to originate from stellar activity.}. The shape of the light curve shows a very clean rotational signal, with no sign of an eclipse.

We identified four additional stars with orbital periods below 10\,d in the intersection of our data and that of \cite{2018MNRAS.475.1609B} (marked in Fig.~\ref{fig:CPD}). However, these orbital periods are larger than two days, all unequal to the rotation periods (which have a large range), and show no sign of tidal-locking at the age of NGC\,2516.
Among these four stars, two are classified as single by \cite{2018MNRAS.475.1609B}, but with a small radial velocity amplitude, likely indicating the presence of a sub-stellar companion.

Any remaining binaries, with orbital periods ${\gtrsim}10$\,d can, in any case, be considered effectively single stars for rotational studies of young open clusters \citep{2005ApJ...620..970M,2019ApJ...881...88F}.

\subsection{Outliers}
\label{sec:outliers}
Certain stars in both the CPD (Fig.~\ref{fig:CPD}) and CMD (Fig.~\ref{fig:memberCMD}) can be considered outliers with respect to the observed sequences. 
We have emboxed certain obvious rotational outliers in the CPD in Fig.~\ref{fig:CPD}. The photometric outliers were defined in Sect.~\ref{sec:finalmem}, and are discussed now.
We note that small numbers of outliers are a standard feature of open cluster CPDs, even when the rotation periods derived are completely secure.
For instance, equivalents are also visible in the Pleiades and M\,35 clusters. 
Nor do any of these particular stars display overt signs of being binaries.
However, we show below that the most egregious outlier is a non-member.

\subsubsection{Outliers in the colour-period diagram}

Star \emph{dS1490} with $P_\mathrm{rot}=13.15$\,d at $(G_\mathrm{BP}-G_\mathrm{RP})_0=1.15$ is an obvious outlier well above the slow rotator sequence.
From our membership, we can easily see that this star is a kinematic non-member because both the radial velocity and the proper motion are not consistent with cluster membership.
However, it has the same distance as the cluster and is located on the main sequence, leading to classification as a photometric member because our threshold requirement that two criteria be matched for membership leads to its being included as a member (see Sect.~\ref{sec:kinnon}).
This star is apparently an older field star that is passing through the cluster. 
From its rotation period, we estimate its age
to be $870$\,Myr\footnote{We have used the Barnes (2010) formula for age here and assumed that the star is not otherwise pathological.}, which is older than even the Hyades cluster. 
Therefore, we have removed this star, and the three other kinematic non-members in our sample of rotation periods, from further analysis, and from the figures\footnote{The other identified kinematic non-members among our rotational sample have only one radial velocity measurement each. 
Additionally, all of them fall onto the sequences in the CPD. 
Hence, they could potentially be binary cluster members, and thus have not been removed.}.

The two stars \emph{3v516} and \emph{3v924} ($(G_\mathrm{BP}-G_\mathrm{RP})_0=1.15$, $P_\mathrm{rot}=9$\,d) are cluster members fulfilling all four membership criteria.
However, they appear to be slightly above the slow rotator sequence. When comparing the NGC\,2516 CPD to that of the Pleiades in Sect.~\ref{sec:Plslowrot} (below), we find additional stars in this region of the CPD ($(V-K_s)_0\approx2.2$) which also seem to be outliers.
A discussion of possible reasons and consequences is postponed to Sect.~\ref{sec:Plslowrot} below.

The star \emph{dE2352} can only be found on the CPD
with \emph{Gaia} photometry at $(G_\mathrm{BP}-G_\mathrm{RP})_0=2.8$, $P_\mathrm{rot}=8.09$\,d but not in the other figures because we do not possess either visual and infrared photometry for this star, despite its being well-exposed in our images.
We find no other star in this area of the CPD, but note that our (determined) period of 8.28\,d has a beat period with the 1\,d observing period at 1.14\,d.
That would put it in the same location in the CPD as other fast rotators in NGC\,2516.
The \textsc{clean} periodogram has a peak at that period but it is not as high as the main peak. We were unable to confirm the shorter period although it would fit the overall picture in the CPD. As the highest peak is found for 8.28\,d in the periodograms, we do provide this period in the data table, but flag it as a possible alias.

\subsubsection{Photometric non-members}
\label{sec:photnonmem}

In Sect.~\ref{sec:finalmem}, we found a few photometric non-members in our membership list shown in Fig.~\ref{fig:memberCMD}. Among the rotators, the photometric outliers are found mostly at the faint end of the cluster sequence. In the CMD inset in Fig.~\ref{fig:CPD}, three somewhat bluer stars appear below the main sequence (with $(G_\mathrm{BP}-G_\mathrm{RP})_0\lesssim2.1$). A detailed analysis of the available photometry from the various publications reveals that the bluer the filter the more off-set the stars are from the cluster sequence.
Visual identification on images shows much brighter stars located in the vicinity of the affected cluster members.
Hence, we infer that scattered light influences the determination of the flux in the bluest pass bands\footnote{This shows that even the relatively precise \emph{Gaia} photometry can occasionally be erroneous.} and redder colour indices such as $(V-K_s)$ are barely influenced. As a result we do not flag or remove any of those stars but use $(V-K_s)$ colour instead.

\subsection{Inclusion of rotation periods from \cite{2007MNRAS.377..741I}}
\label{sec:AddI07}

The prior study of I07 used a larger telescope in search of exoplanets among the fainter stars of NGC\,2516.
Although their baseline coverage (of two 4-night runs, with a one week gap) was optimized for planet discovery, they also derived a large number of rotation periods for the low-mass M\,dwarfs in NGC\,2516.
In contrast, our sensitivity (long and baseline) is optimized for the solar-mass, K, and early-M stars, making the two studies complementary.
  
In order to obtain the richest possible distribution of rotation periods for NGC\,2516, we append the periods for the low-mass stars determined by I07 to those from our sample.
However, with both the improved astrometry from the \emph{Gaia} mission and the radial velocity study by \cite{2010MNRAS.407..465J} we are now in a position to remove non-member contaminants.
To clean the I07 sample, we therefore applied the same membership criteria as we applied to our own data.

We found 41 of the 350 rotators in I07 to be non-members. Notably, these are often the outliers described by I07 in their Section 5.1.2, explaining their abnormal behaviour. Additionally, the two bluest slow rotators and the overall slowest rotating star in their sample are also non-members, despite their agreement with the general distribution of stars in the CPD. The remaining non-members are among the fast-rotating stars, hidden within the distribution of members.

\begin{figure}
	\includegraphics[width=\columnwidth]{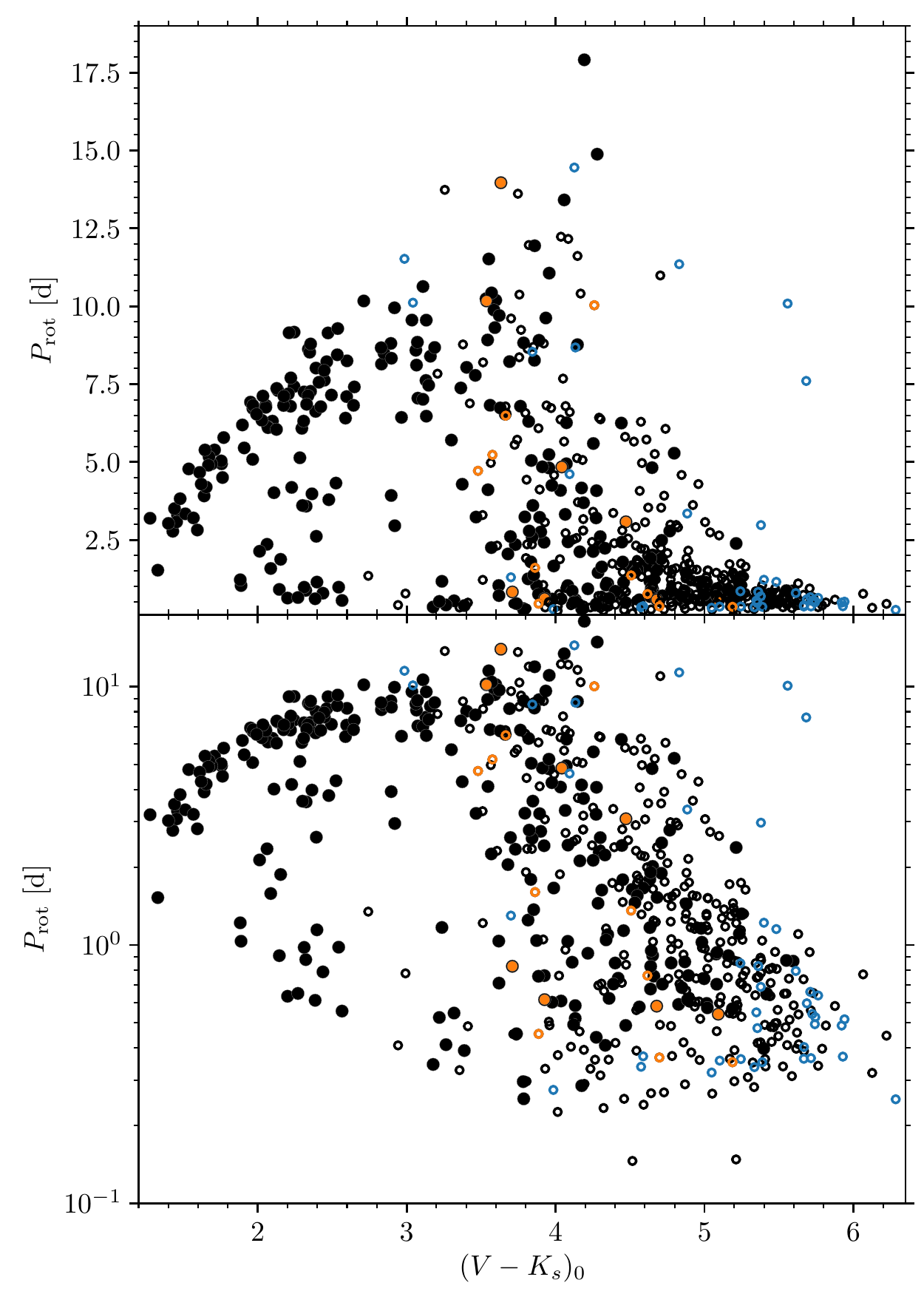}
	\caption{Colour-period diagram for NGC\,2516 with our data (solid) and the data from I07 (open) on both linear and logarithmic scales. Periods in disagreement between the two data sets are marked in orange (with solid circles for this work, open circles for I07).
            Blue symbols indicate non-members from I07 (subsequently suppressed), identified as such  by us using the improved astrometry and new radial velocity measurements.}
	\label{fig:I07CPD}
\end{figure}

In Fig.~\ref{fig:I07CPD}, we display both datasets in a CPD\footnote{2MASS data are used here for the stars of I07 because the VHS did not cover the whole area observed by I07.}. 
After removing the overlap between the two samples, we finally compiled a list of 555 unique individual rotators, which we subsequently use for comparison with the Pleiades distribution\footnote{Ten of these 555 rotators have recently been classified as non-members by \cite{2020arXiv200609423J} using somewhat stricter membership criteria. We list the IDs of these stars in Appendix~\ref{app:nonmembers}.}.

%

The additional low-mass stars contribute in particular to the connecting region in the CPD between the slow rotator sequence and the M dwarf fast rotator sequence, and to the fast rotating tail of very late-type M\,dwarfs.
Thus, the union of the two data sets provides what we consider to be a relatively  comprehensive picture of the rotation period distribution of FGKM stars in NGC\,2516.

%
%

\section{Comparison with models of stellar rotation}
\label{sec:models}

A comparison of rotation period measurements with theoretical models can help to constrain them and provide insight into stellar activity physics.
Here, we display a comparison of the NGC\,2516 data with five such models from the literature in order to provide an overview of the current state\footnote{To a greater or lesser extent, the models presented here include insights from prior theoretical models, including { \cite{1991ApJ...376..204M}, \cite{1989ApJ...338..424P}, \cite{1988ApJ...333..236K}, extending to \cite{1979ApJ...232..531E}}.}.
We present the selected models chronologically and include comments on their differences and distinctive features.

\subsection{Model of \cite{2010ApJ...721..675B} and \cite{2010ApJ...722..222B}}

The oldest of the models considered is the  \cite{2010ApJ...721..675B}, \cite{2010ApJ...722..222B} model, developed as an improvement upon the original gyrochronology formulation presented in \cite{2003ApJ...586..464B}.
This model consists of an analytic overlay on regular stellar structure models for the main sequence\footnote{There is little change in these quantities on the main sequence, but the model can of course be extended to make the underlying quantities dynamic.}, from which it extracts and uses in particular the convective turnover timescale.
Thereafter, the model evolves an assumed initial (ZAMS) distribution of rotation periods forward in time, as required by the age of the cluster being discussed.
Guided by observations then available in the youngest ZAMS clusters, this initial distribution was specified by \cite{2010ApJ...722..222B} to range from 0.12\,d to 3.4\,d, although this can in principle be modified as necessary.
As in prior work by our group, we here retain all features exactly as in that decade-old model to maintain consistency.

The model contains only two parameters, namely the two dimensionless constants $k_c = 0.646$\,d\,Myr$^{-1}$ and $k_I = 452$\,Myr\,d$^{-1}$, which together with the extracted convective turnover timescale, $\tau_c$, of the stellar model, govern the rotational evolution timescales in the fast- and slow-rotator regimes respectively.
These particular parameter choices were specified by comparison with all relevant observations available at that time, and of course, with respect to the convective turnover timescales tabulated in \cite{2010ApJ...721..675B}.
These timescales are included mathematically symmetrically in the model for the rotation period evolution and the formulation is such that the transition from fast- to slow rotation occurs automatically at the fixed Rossby Number,
$Ro = P/\tau_c = \sqrt(k_I k_c) = 0.06$,
where the numerical value in the final equality is specified with respect to the convective turnover timescales in \cite{2010ApJ...721..675B}\footnote{Other published timescales may also be used because timescales from differing publications tend to differ only by a scaling constant, but then the dimensionless constants $k_c$ and $k_I$ should be adjusted accordingly.}.
A slow rotator sequence develops naturally in this model, with an asymptotic shape specified by the convective turnover timescale on the main sequence.
In this model that asymptotic shape is $P \propto \sqrt{\tau_c}$.

Figure~\ref{fig:models} displays this model (with two green curves) for the fast and slow rotator boundaries of the allowed distribution calculated for the isochrone age of 150\,Myr, against the NGC\,2516 rotation period measurements.
(We note that a G2 star has $(V-K_s)_0 = 1.56$ \citep{2013ApJS..208....9P}, which means that the warm end of our rotation period distribution is at F9 \citep{2013ApJS..208....9P}; the Pleiades rotation periods (see Sect.~\ref{sec:comPleiades}) appear to extend the distribution to approximately the convective-radiative boundary.)
Inspection of the lower boundary of the rotation period distribution shows that its fast edge is described adequately by this model.
In fact, the uplift of the curve for the warmer stars with $(V-K_s)_0 \sim 1.3$ reasonably describes the shorter spin-down timescale for these stars and the corresponding early development of the slow rotator sequence for stars warmer than this, at which point spindown is heavily curtailed as the star transitions into the slow rotator regime.

On the other hand, the location of the slow rotators in NGC\,2516 is problematical in this model (upper green curve).
For stars warmer than solar, the initial rotation period of 3.4\,d is clearly too long because that is already significantly above the NGC\,2516 data.
The cluster being at the ZAMS, it is clear that not enough time has elapsed for the initial rotation period distribution to have been forgotten.
This is again an issue for the coolest M-type stars (with $(V-K_s)_0 \gtrsim 4$) in that the initial rotation period of 3.4\,d is clearly an overestimate\footnote{The stated reason for this in \cite{2010ApJ...722..222B} was the desire to retain a simple flat initial rotation period distribution at the ZAMS, and then to show that the shape of the slow rotator sequence would develop automatically.}, and in any case, probably meaningless in a context where the stars are still on the pre-main sequence.
As regards the K-M\,type stars with $2 < (V-K_s)_0 < 4$, because the NGC\,2516 data are significantly above the model, it is correspondingly also clear that either the assumed 3.4\,d initial period is too short, or this model is not sufficiently aggressive in spinning down the slow rotators.
However, this model does have the very desirable feature of retaining fast rotation for the lower-mass stars, as seen in the fact that the isochrone dips towards faster rotation for $(V-K_s)_0 \gtrsim 3.5$.


Finally, we check whether the \cite{2010ApJ...721..675B} and \cite{2010ApJ...722..222B} model is able to separate the fast- and slow rotators adequately.
The separation between these groups (which will become more obvious in Sect.~\ref{sec:comPleiades}) was called the `rotational gap' and proposed in \cite{2010ApJ...722..222B} to be a line of constant Rossby Number (Ro), located at $Ro = P/\tau_c = 1 / \sqrt{k_I k_C}$, i.\,e. $P = 0.06 \, \tau_c$\,, using the numerical constants $k_C$ and $k_I$ in that model.
Accordingly, we plot the line $P = 0.06 \, \tau_c$ in Fig.\,14, using the values of $\tau_c$ in \cite{2010ApJ...721..675B}\footnote{The constants $k_I$ and $k_C$ are particular to the \cite{2010ApJ...721..675B} convective turnover timescales and will require slight recalibration if other timescales are used.}.
We observe that this line is able to separate these two groups of stars relatively effectively in the F,\,G,\,K spectral range. 
One could argue that this line is located at overly-long rotation periods for M-type stars.
For the present, we are uncertain whether to implicate the model itself, the turnover timescales, the intermediate colour transformation required, or even the behaviour of the stars themselves, given that they are still in the pre-main sequence phase of evolution.

\subsection{Model of \cite{2015ApJ...799L..23M}}
The model presented in \cite{2015ApJ...799L..23M} is similar to the \cite{2010ApJ...722..222B} model in that it also expresses the torque on an (assumed) solid-body stellar model as a function of the convective turnover timescale, albeit a higher (second) power.
The remaining dependencies apart from the rotation rate in the (separable) torque formulation are absorbed into a function $T_0 = T_0\,(T_{\odot},M_\star, R_\star, Q, m)$ which itself includes adjustable powers, $m$, of the mass ($M_\star$) and radius ($R_\star$) of the star, a tunable parameter $Q$, and a constant, $T_{\odot}$, that is scaled to reproduce the torque on the current Sun.
It also parametrizes the stellar magnetic field and mass loss rate jointly as a separate power, $p$, of the Rossby Number $Ro$.
Consequently, it has a few more degrees of freedom than the \cite{2010ApJ...722..222B} model.
The resulting spin-down timescales for both the fast and slow rotator regimes have similar mass dependencies to those in \cite{2010ApJ...722..222B}, as can be seen explicitly in Fig.\,1 in \cite{2015ApJ...799L..23M}, where a number of other older models are also displayed for comparison.

\begin{figure}
	\includegraphics[width=\columnwidth]{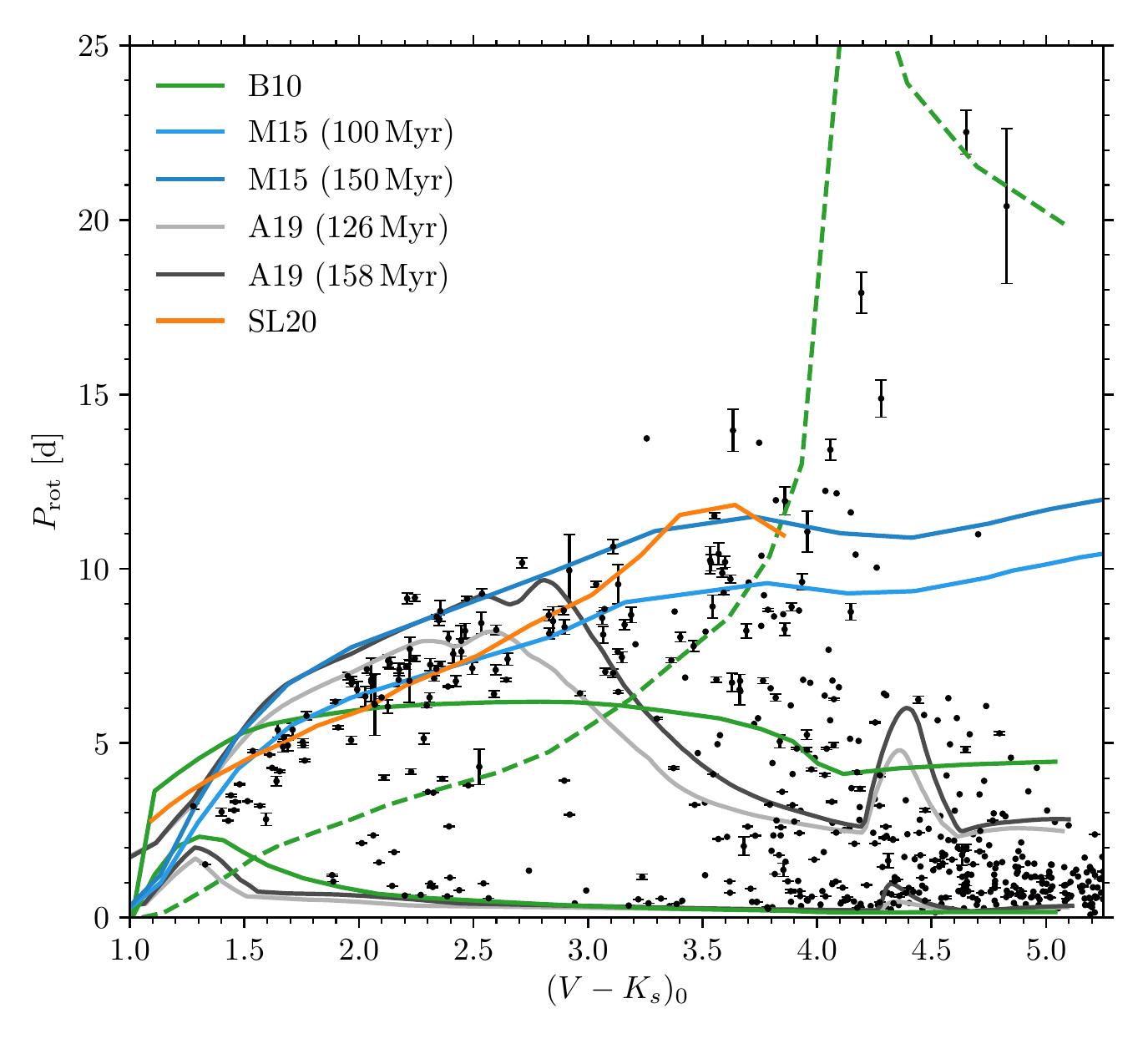}
	\caption{Colour-period diagram for NGC\,2516 comparing multiple rotational isochrones (chronological by date of publication top to bottom in the legend).
          Green curves show the upper and lower boundaries of models from \cite{2010ApJ...722..222B} (B10) with corresponding initial periods of 3.4\,d and 0.12\,d respectively. The dashed line indicates the transition ($Ro = 0.06$) between fast- and slow rotation in the B10 model. 
          The blue curves show the asymptotic spin rate from \cite{2015ApJ...799L..23M} (M15) for 150\,Myr (upper blue) and 100\,Myr (lower blue).
          The grey curves show  the models of \cite{2019A&A...631A..77A} (A19) for 158\,Myr and 126\,Myr, each for both the slow and fast rotators.
          The orange curve shows the two-zone model of \cite{SL20} (SL20) for 150\,Myr. (See the text for a detailed comparison of each model with the data.)
          } 
	\label{fig:models}
\end{figure}

In Fig.~\ref{fig:models}, we show the asymptotic spin rates (in blue) from the \cite{2015ApJ...799L..23M} models for ages of 150\,Myr (as with other models) and 100\,Myr, together with our NGC\,2516 rotation periods.
The reason for displaying a younger 100\,Myr age in addition to the 150\,Myr isochronal age of NGC\,2516 is that the younger model is significantly closer to the NGC\,2516 data than the older one.
In this connection, we note that a difference between the model age and another age estimate, such as an isochronal age, is not necessarily a problem unless the rotation periods are used in deriving ages via gyrochronology, as opposed to generally understanding angular momentum evolution.
As already pointed out by \cite{2015ApJ...799L..23M}, the asymptotic model plotted is not descriptive of the behaviour of the fast rotating M dwarfs at the low-mass end of the distribution.
We conclude from the comparison that the spin-down in this model is somewhat more aggressive than warranted by the NGC\,2516 data.

%

\subsection{Model of \cite{2018ApJ...862...90G}}

\cite{2018ApJ...862...90G} have proposed a model using scaling laws from \cite{2016A&A...595A.110G} for angular momentum loss rates as a function of magnetic morphology.
In this model, the magnetic morphology of cool stars is proposed to be a smoothly-changing function of the Rossby number, $Ro$.
This functional dependence is analogous to the period derivative dependence in \cite{2010ApJ...722..222B} in that here the dependence is also on the Rossby number and its inverse, and includes two related dimensionless constants $a$ and $b$ (see their Eq.~5).
This form for $n$ permits drastic changes in the angular momentum loss rates, as noted in \cite{1988ApJ...333..236K}, \cite{1996ApJ...462..746B}, and reaching back to \cite{1968MNRAS.138..359M}.
Ignoring any possibly additional parameters in the derivation of the scaling laws, one finds that 
$\dot{P} = \dot{P}\,(c,\, a,\,b,\,I,\,B,\,P,\,\tau)$, where $I$ is the star's moment of inertia, assumed to be constant.
This is therefore a seven parameter model in general, although the term with the magnetic field strength, $B$, is apparently allowed to be suppressed under certain conditions, as in their paper.

\cite{2018ApJ...862...90G} were able to produce a desirable bifurcation in $P$ because of the features described above. 
However, their models have very significant departures from the shape of the open cluster rotation period distributions at various ages (including Pleiades and NGC\,2516 age), as one can see in the comparisons shown in Fig.~4 of their paper, and as can be seen by calculating asymptotic solutions.
Unfortunately, their models are not available in a form that we can readily reproduce in Fig.~\ref{fig:models} to compare directly with our data.

\subsection{Model of \cite{2019A&A...631A..77A}}
\cite{2019A&A...631A..77A} have largely implemented the \cite{2015ApJ...799L..23M} prescription of angular momentum loss into the Geneva/Montpellier stellar models\footnote{These also permit radial differential rotation in radiative regions.}, together with additional parameters and initial conditions tuned to reproduce various open cluster rotation observations. (For instance, a disk-locking timescale is included on the pre-main sequence.)
The computed mass dependence of rotation for the slower and warmer stars is almost identical to that of \cite{2015ApJ...799L..23M}, as one might expect from the adoption of that prescription here and as can be seen from the grey curves corresponding to this model in Fig.~\ref{fig:models}.
However, \cite{2019A&A...631A..77A} have modified the ($Ro$-based) threshold for saturation significantly to retain fast rotation for longer timescales among lower mass stars, making it similar to the one from \cite{2010ApJ...722..222B}.
Consequently, this model has even more degrees of freedom than that of \cite{2015ApJ...799L..23M}.

We display the \cite{2019A&A...631A..77A} models for two different ages, 126\,Myr, the age they assign to the Pleiades, and 158\,Myr, the model they provide that is closest in age to the isochrone age of NGC\,2516.
The linear rotation period scale in our plot, together with the improved dataset, allows a more detailed view of the comparison than the logarithmic plot in \cite{2019A&A...631A..77A}.

The slow rotator curves for both ages, especially the older one, in the \cite{2019A&A...631A..77A} models can be observed to be located at significantly longer periods than allowed by the NGC\,2516 data for $(V-K_s)_0 \lesssim 2-3$.
The spindown is clearly over-aggressive.
However, the modification of the threshold for saturation noted above allows the isochrones (displayed in Fig.~\ref{fig:models}) to deviate away from the slow rotator sequence at $(V-K_s)_0 > 2.7$ and approach the fast-rotating region for M dwarfs.
The turndown of the cooler slow rotator models towards shorter rotation periods is clearly a positive feature of the model, even if it is not quite happening at the correct stellar mass.
We hesitate to speculate about the clearly-visible kinks in the \cite{2019A&A...631A..77A} models at $(V-K_s)_0 \sim 2.5$ and $(V-K_s)_0 \sim 4.5$, presuming them to be isochrone-related artefacts.

\cite{2019A&A...631A..77A} also include a model for the fast rotators in their paper, enabling us to display the corresponding comparison in Fig.~\ref{fig:models}.
These models are quite similar to the model from \cite{2010ApJ...722..222B} across the entire mass range, particularly with respect to being able also to produce a noticeable spindown among the warmer stars $(V-K_s)_0 \sim 1.3$.
(This was first successfully modelled by \cite{2010ApJ...722..222B} (green curve), invoking a Rossby Number-based transition from fast- to slow-rotation.)
However, the NGC\,2516 observations, because they are significantly above the fast rotator model curves, appear to inform us here that the \cite{2019A&A...631A..77A} models should be spun down somewhat more aggressively here than they currently are.
    
\subsection{Model of \cite{SL20}}
Our final model comparison is to one devoted exclusively to the slow rotator sequence that is presented in \cite{SL20}, further developing the two-zone model presented in \cite{2015A&A...584A..30L}, where the radiative core is permitted to decouple from the convective envelope.
This model is overlaid on the Yale stellar models and the basic angular momentum loss formulation is taken from \cite{2010ApJ...721..675B} and \cite{2010ApJ...722..222B}.
Consequently, it has an additional degree of freedom above that of \cite{2010ApJ...722..222B} in that it implements one additional (strongly mass dependent) coupling timescale between the two zones, applicable to the slow rotator sequence itself.
This timescale is presumably independent of the (also strongly mass-dependent) fast- to slow-rotator transition that is observed in young open clusters.
The fast rotators are not addressed in this model.

In Fig.~\ref{fig:models}, we show the 150\,Myr model from \cite{SL20} in orange.
This curve generally follows the slow rotator sequence of NGC\,2516 reasonably well for stars of lower mass than the Sun ($(V-K_s)_0 \gtrsim 1.5$).
This is perhaps unsurprising, given that the Pleiades (assumed to be 120\,Myr) were used as a reference cluster (although not an exclusive one) for the model.
There appears, however, to be some deviation towards slower-than-observed periods for the solar- and higher-mass stars, as one can see especially well when we include the Pleiades rotation periods (in the next section).
The spindown is also perhaps somewhat too aggressive for the slow rotators cooler than $(V-K_s)_0 \sim 3$.
Finally, since this model is formulated exclusively for the slow rotators, it does not model the lowest mass fast rotators.

\subsection{Suggestions for the applications of angular momentum models}

In summary, each of these models has certain advantages, and is able to reproduce particular aspects of the CPD, all at the expense of having increased the number of degrees of freedom over \cite{2010ApJ...722..222B},
with \cite{SL20}, \cite{2018ApJ...862...90G}, \cite{2015ApJ...799L..23M}, and \cite{2019A&A...631A..77A} in order of increasing number of parameters.
Our recommendation, therefore, is instead to consider the data presented here as specifying both the initial (near-ZAMS) distribution for rotational evolution models aimed at older main sequence populations and as the target distribution for pre-main sequence models.

%
%

\section{Comparison with other similar open clusters, especially the Pleiades}
\label{sec:comPleiades}

We noted in Sect.~\ref{sec:intro} that the Pleiades is the best open cluster for comparison with NGC\,2516. In addition to the similar age and richness, the Pleiades is one of the best-studied open clusters and is often seen as the archetype ZAMS open cluster.
One might therefore ask whether there is such a thing as an archetypical open cluster for rotation, or whether cluster-to-cluster variations actually exist. 
After a detailed comparison with the Pleiades, we briefly investigate other similarly-aged open clusters with respect to those questions.

\subsection{Comparison between the Pleiades \& NGC\,2516}

We use the very rich and most complete K2 dataset presented in \cite{2016AJ....152..113R} for our detailed comparison with the Pleiades rotational distribution.
We also consider questions raised by \cite{2016AJ....152..115S} about specific features in the rotational distribution of the Pleiades.
For NGC\,2516 we used the above-presented rotation periods together with the cleaned period sample from I07 (Sect.~\ref{sec:AddI07}).

We note that biases exist in the data, and can influence the comparison; therefore, awareness of limitations is important.
For instance, the variability amplitude imposes a significant detection limit
on our ground-based data for NGC\,2516, whereas this is largely irrelevant for the space-based (Kepler/K2) Pleiades data.
Therefore, any possibly-identified voids in our rotational distribution should be treated with caution, and detectability ought to be considered\footnote{Readers might wish to refer back to Sec~\ref{sec:varamp} and Fig.~\ref{fig:amps} at this point.}.
However, the overall completeness of the membership information is excellent for both open clusters, and should not be a limiting factor.

\subsubsection{Pleiades \& NGC\,2516: Full distribution of rotation periods}
\begin{figure*}
	\includegraphics[width=\textwidth]{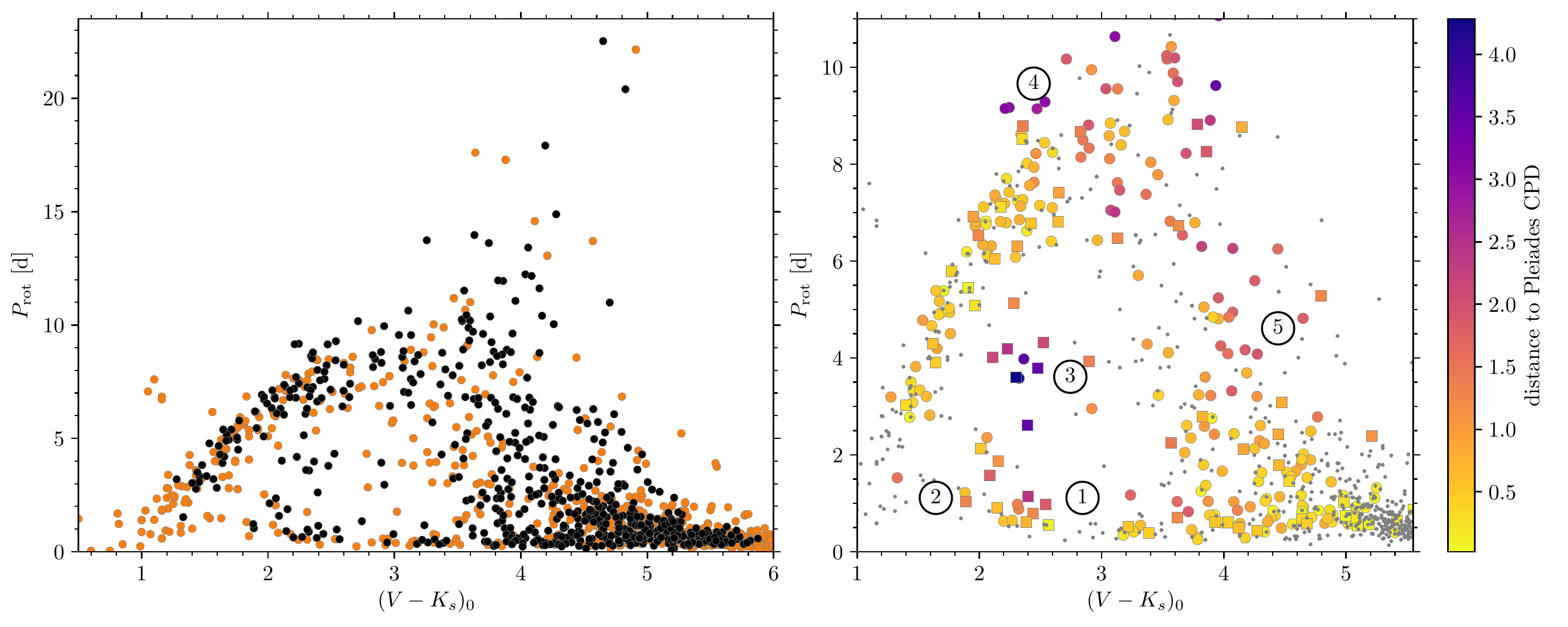}
	\caption{            \emph{Left:} Colour-period diagram for both NGC\,2516 (black, union of data from this work and I07) and the Pleiades (orange).
            The similarity of the two distributions across all sampled regions of the colour-period diagram is noticeable.
            \emph{Right:} Distance of a given NGC\,2516 star (coloured symbols) in the CPD to the corresponding Pleiades star.
            Distance is measured in standard deviations (see text), and colours range from yellow (very near) to deep purple (most distant; see scale).
			The fast- and slow sequences, representing the edges of the rotation distributions, are mostly very close (yellow), while the intermediate and sparsely populated regions of the CPD are quite distant (deep purple.) 
            The Pleiades data are displayed with grey dots in the background.
            Circular symbols indicate single stars, while square symbols indicate possible binaries.
            The encircled numerals from 1 to 5 indicate specific regions discussed in detail in the text. (This panel has a restricted period axis range compared to the left panel.)
        } 
	\label{fig:plcomp}
\end{figure*}

We display the Pleiades rotation periods from \cite{2016AJ....152..113R} in the left panel of Fig.~\ref{fig:plcomp} (orange) together with our rotation periods for NGC\,2516 (black) in a $(V-K_s)_0$ CPD.
The two clusters present the same morphology across the entire colour-period diagram, making the visual appearance of the two clusters almost indistinguishable.
In particular, the tight slow rotator sequences follow each other closely at the warm end of the distribution, as do the fast rotators at the cool end.
Despite the larger (and apparently intrinsic) rotational scatter among the intermediate-mass (K - M) stars ($(V-K_s)_0 \sim 3-4$), the two distributions are also essentially visually indistinguishable.
Finally, both clusters contain sequences of early-M stars with very long (up to ${\sim}23$\,d) rotation periods.
In fact, we find that almost every group of stars in one CPD, with the exception of the brightest ones in NGC\,2516 (which are saturated in our images), has a corresponding group in the other CPD.

%

Nevertheless, certain minor differences between the two distributions are present, which we wish to highlight quantitatively.
For each star in the NGC\,2516 CPD, we  measure the distance to the nearest star in the Pleiades CPD.
As our difference metric, we simply chose the Euclidean distance (hereafter simply distance) in the normalized CPD.
We normalized both the $(V-K_s)_0$ and $P_\mathrm{rot}$ axes to the interval $[0, 1]$ with the end points chosen such that the values from NGC\,2516 fill the whole range.
In order to avoid being distracted by the visually-obvious larger separations among the long-period rotators, we limited the comparison region to periods $P_\mathrm{rot} < 10.5$\,d\footnote{
This approach was chosen because CPDs are usually displayed with near-equal aspect ratios, despite the fact that the rotation periods span over one order of magnitude. The normalization ensures a result that simply quantifies the qualitative results obtained through visual inspection.}.
Correspondingly, the right panel of Fig.~\ref{fig:plcomp} shows the CPD of NGC\,2516 with each star colour-coded by its distance from the closest Pleiades star. (The Pleiades stars are also displayed unobtrusively with grey dots in the background.)
The distance is normalized to the standard deviation of the distribution.

The smallest distances are found for the stars on the two clear fast- and slow rotational sequences. (These stars are yellow in the colour map chosen.)
Stars towards the edges of those sequences are somewhat sparser in the CPD, and consequently are more distant from the corresponding stars in the Pleiades, as can be visually appreciated through their darker colours in Fig.~\ref{fig:plcomp}.
A number of stars are several standard deviations away from their nearest counterparts in the Pleiades.
Those groups of stars are labelled with encircled numerals in the right panel of Fig.~\ref{fig:plcomp} to enable us to discuss them individually below.

\subsubsection{NGC\,2516 vs. Pleiades: Fast rotators}
The fastest rotators (marked with an encircled numeral \emph{1} in  Fig.~\ref{fig:plcomp})
in both clusters are very similar.
In particular, their appearance as distinct sequences with $P_\mathrm{rot}\lesssim 1$\,d and a mostly empty region above it in the CPD is notable. This sequence extends from the warmest stars in the CPD, where it meets the slow rotator sequence, to the coolest M\,dwarfs.

We observe a slight lift of the fast rotator sequence in both clusters (\emph{2}) for $(V-K_s)_0\sim1-2$. In particular, when combining the two samples the rise of the fast rotator sequence is visible best in Fig.~\ref{fig:plcomp}. All stars in this colour range seem to have spun down by a factor of two or so, from their initial ZAMS rotation period, hence this feature could indicate the evolution of both clusters beyond the rotational ZAMS, suggesting a very similar age.
However, the exact counterparts for the higher-mass stars in NGC\,2516 are missing. Such stars have very low variability amplitudes (c.f. Fig.~\ref{fig:amps}), which means that we likely do not detect them with our ground-based photometry, despite their likely existence in NGC\,2516.

\subsubsection{NGC\,2516 vs. Pleiades: Rotational gap}
As stars spin down from the fast to the slow rotator sequence, they eventually have to cross the gap region (\emph{3}) between the two sequences. Indeed, we find a group of intermediate ${\sim}$4\,d rotators in NGC\,2516 that are relatively isolated in the gap region.
These stars have no equivalent in the Pleiades and are among the stars with the largest distance to any Pleiades star.
Curiously, their measured periods are approximately half that of the slow rotator sequence at this colour.
We have therefore re-checked their light curves carefully to exclude a half or double period alias and to identify systematic trends like peaks of alternating height, none of which we see at the level of precision of our light curves.
We therefore consider these period measurements to be genuine and list them as such.
This view is supported by their X-ray activity (see Sect.~\ref{sec:xray} below), which is also noticeably lower than for the fast rotators but higher than for slow rotators of the same colour.
Hence, we conclude that this group constitutes an intrinsic feature of the NGC\,2516 CPD.

Such a group of gap stars has also been found in the nearby stellar stream Meingast~1 \citep{2019A&A...622L..13M, 2020arXiv200205728R} by \cite{2019AJ....158...77C}.
A comparison of the rotation periods between this stream and the Pleiades revealed that the two objects have similar ages.
With the knowledge that the gap stars found in this stream are also present in the slightly older NGC\,2516, we conclude that they are not a sign of youth relative to the Pleiades, as proposed by \cite{2019AJ....158...77C}.

In conclusion, the existence of gap stars in NGC\,2516 and Meingast~1 suggests that they are an intrinsic feature.
The fact that only a fraction of the fast rotators have evolved into the gap stars provides evidence that the transition from fast to slow rotators is not occurring at the same age even for all stars of the same mass (or colour) but with a degree of randomness, or perhaps `metastability' in the language of \cite{2014ApJ...789..101B}.

\subsubsection{NGC\,2516 vs. Pleiades: Slow rotators}
\label{sec:Plslowrot}

The slow rotator sequences of both clusters are very similar and correspondingly we find the shortest distances ($\lesssim $1 S.D.; yellow) here.
Although this result appears to be trivial it is very important for the validity of gyrochronology, which can only be used effectively if otherwise similar (nearly-)coeval open clusters show (near-)identical slow rotator sequences.
However, it has to be noted that we find a number of stars in NGC\,2516 that rotate slightly slower than the corresponding Pleiades stars towards the red end of the slow rotator sequence (\emph{4}, $(V-K_s)_0\approx2.2-3$).
These stars also appear to be above the rest of the slow rotator sequence of NGC\,2516.
Among them are not only the stars \emph{3v516} and \emph{3v924}, pointed out in Sect.~\ref{sec:outliers}, but additional stars with large distance to the Pleiades as seen in Fig.~\ref{fig:plcomp}.

All of these stars are cluster members according to all four membership criteria.
This makes it very unlikely that they are non-members.
All of the light curves show some scatter around the maxima, possibly indicating two spot groups on the stellar surface.
The proximity of the two spots could bias the period determination towards a longer period. However, the noise level of these light curves does not permit a definitive statement of whether two spots are present. Hence, those apparently slower rotating stars might not form a genuine feature of NGC\,2516.

We note that \emph{3v516} (marked with a blue diamond in Fig.~{\ref{fig:CPD} at $(G_\mathrm{BP}-G_\mathrm{RP})_0=1.15$, $P_\mathrm{rot}=9$\,d at, and here located at $(V-K_s)_0 = 2.1$}) potentially hosts a sub-stellar companion (\citealt{2018MNRAS.475.1609B} and discussion in Sect.~\ref{sec:binariesRotation}) but we do not believe it to have had an impact on the rotation period. In particular the large number of stars above the sequence are unlikely to all host low-mass companions.
Finally, with respect to the slow rotators, a paucity at $(V-K_s)_0 \sim 2.6$ was noted among the Pleiades stars on the slow rotator sequence, and was called the `kink' by \cite{2016AJ....152..115S}. This is separately discussed in Sect.~\ref{sec:kink} below.

\subsubsection{NGC\,2516 vs. Pleiades: M stars - fast and intermediate rotation}

So far we have only considered the FGK stars. Those stars have arrived on the main sequence. In contrast most M stars at the age of NGC\,2516 are still in the pre-main sequence phase. The fast rotators are in very similar positions in both clusters, as indicated by the small distances (yellow), especially for the reddest objects in our study. At intermediate rotation periods ($P_\mathrm{rot}\sim 5$\,d), we observe somewhat larger distances between the two clusters.

In Fig.~\ref{fig:plcomp} the M stars with intermediate rotation periods (\emph{5}) seem to be of slightly redder colour than in the Pleiades. We are not certain whether this is an expression of an age difference (in the sense of NGC\,2516 being marginally older) or simply occurring by chance because of the low numbers (${\sim}10$) of stars.

\subsubsection{NGC\,2516 vs. Pleiades: M stars - extended slow rotator sequence}

In the uppermost region of the CPD (left panel of Fig.~\ref{fig:plcomp}) at $P_\mathrm{rot}\gtrsim 10$\,d, a group of slowly rotating M stars is present\footnote{We include their light curves in Fig.~\ref{fig:lcslowM} so that readers can appreciate the confidence with which these periods are determined.}.
These are well above the usual locations of the majority of the M-type stars discussed in the previous section. Furthermore, they appear to be an extension of the slow rotator sequence.
\cite{2016AJ....152..115S} labelled the very slow rotating M dwarfs in the Pleiades `abnormal' and questioned their membership.
We, however, find such stars in NGC\,2516 in similar locations in the CPD.
Consequently, we propose that those stars are not abnormal but instead are the continuation of the slow rotator sequence in the low-mass regime.
We call this the `extended slow rotator sequence'.

In the case of NGC\,2516 we are quite certain about the membership of these stars. Five of the eight slowest rotators are members satisfying all four membership criteria (Sect.~\ref{sec:membership}).
One additional star lacks radial velocity measurements and therefore satisfies only three criteria.
Among the remaining two stars, one star is technically a photometric non-member, located slightly below the main sequence, but is a kinematic member at the cluster distance.
The final one has no proper motion or parallax measurements from \emph{Gaia}~DR2 because it is a (photometric) binary.
Consequently, none of these can be considered as non-members, and we conclude that all eight stars are indeed cluster members.

The stars on the extended slow rotator sequence are all obviously pre-main sequence (PMS) stars at the age of NGC\,2516 and they are still contracting, hence would spin up in the absence of angular momentum loss.
However, their long rotation periods suggest either very efficient loss mechanisms (meaning early spindown), or else a greatly delayed spinup.

We note that very slowly rotating M\,dwarfs have been discovered among field stars \citep{2016ApJ...821...93N, 2018AJ....156..217N}, and that it is possible that we are observing the young equivalent of those stars.
\cite{2016ApJ...821...93N, 2018AJ....156..217N} proposed that these stars spin down strongly only after their PMS phase to reach the long periods.
It is difficult to say how our observations of extremely slow rotators in NGC\,2516 relate to this assertion.

An extended slow rotator sequence has likely not yet been observed in older open clusters because either the time baseline of the observations is too short (typical for ground-based monitoring) or this small group of stars was simply overlooked in prior studies as such stars appear to be similar to noise from non-members in the CPD\footnote{Indeed, slowly rotating M dwarfs well above the continuation of the slow rotator sequence are observed in Praesepe \citep{2017ApJ...842...83D}.}.

With the evidence that those slow rotators are indeed members of young open clusters, more questions arise. In particular it is unclear where those stars come from. Why do those M dwarfs show such low rotation rates in the first place? NGC\,2516 is only the second young cluster (after the Pleiades) where M dwarfs of such slow rotation are observed; hence their evolution might not yet be described adequately with the current wind-braking models.
This leaves their true rotational evolution as an open question.

\subsubsection{NGC\,2516 vs. Pleiades: The kink (or not) in the Pleiades slow rotator sequence}
\label{sec:kink}
As a final detail, we inquire whether the kink observed in the Pleiades period distribution is also present in NGC\,2516. \cite{2016AJ....152..115S} noted a paucity in the slow rotator sequence at $(V-K_s)_0 \sim 2.6$.
At this colour the slowest rotators are faster than expected from the overall shape of the slow rotator sequence. In NGC\,2516 this kink is simply absent (Fig.~\ref{fig:plcomp}). Indeed, we find five stars which perfectly fill the gap observed in the Pleiades. The slow rotator sequence in NGC\,2516 is flattening around this colour and hence is not following the slope of the bluer slow rotator sequence. No obvious kink in the data is observed.

\cite{2018A&A...619A..80G} suggest that the kink is caused by tidal interaction with planets of those stars in the Pleiades.
If this were the case, then the absence of such a feature in NGC\,2516 would mean that the corresponding stars there do not have tidally-interacting planets.
Now, recall that in both the Pleiades and NGC\,2516 five slow rotators are observed in the range $2.6 < (V-K_s)_0 < 3.0$.
If one followed the logic of their suggestion to its obvious conclusion, one would have to find some way of designing planetary architectures in such a way that stars in that narrow mass range in the Pleiades would host tidally-interacting planets, but not the corresponding stars in NGC\,2516.   
We find this arbitrary and therefore unconvincing.

\subsection{Comparison with other clusters: M\,35, M\,50, and Blanco\,1}

%
\begin{figure}
	\includegraphics[width=\columnwidth]{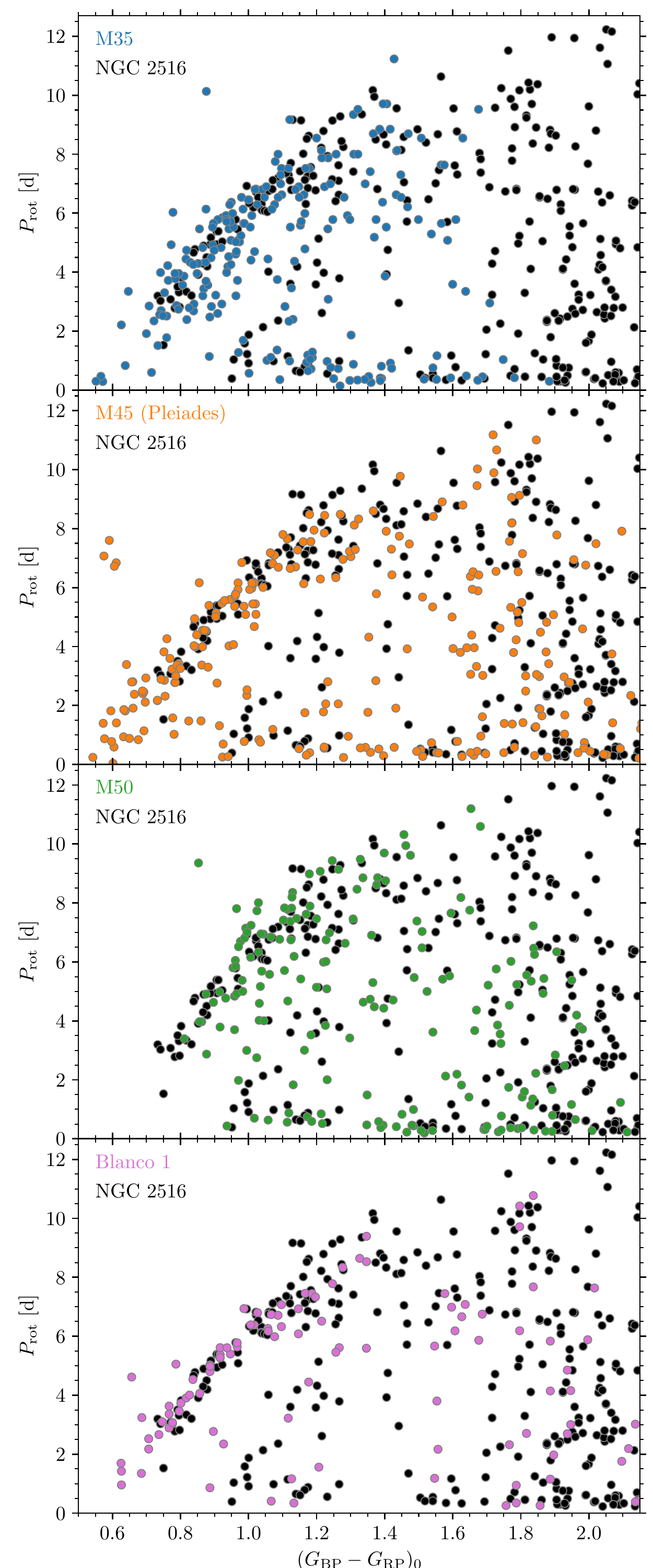}
	\caption{Colour-period diagrams for the four similarly-aged open clusters (varying colours) available to date, compared with the measured one for NGC\,2516 (black in each panel). 
    The clusters are (from top to bottom) M\,35, M\,45 (Pleiades), M\,50, and Blanco\,1, with $\mathrm{E}_{(B-V)}$ values taken from the respective publications, and scaled for \emph{Gaia} colour. 
    Note that their rotational distributions are visually essentially indistinguishable from that for NGC\,2516 to within the sensitivities of the individual studies.
			(See Fig.~\ref{fig:OCcompareBV} in the Appendix for the equivalent figure in $B-V$ colour, and the text for further details.)
            }
	\label{fig:compareOCs}
\end{figure}

Beyond the very well-studied Pleiades cluster, there are three others that have measured rotation period distributions and ages close enough to make them suitable for comparison, which we do in turn below.
The open clusters M\,35 and M\,50 are 150\,Myr \citep{2009ApJ...695..679M} and 130\,Myr old \citep{2009MNRAS.392.1456I}, respectively (each with an uncertainty of ${\sim}20$\,Myr).
Given the age uncertainties we consider both to be coeval with NGC\,2516, and now compare their rotation periods with those in NGC\,2516 (see Fig.~\ref{fig:compareOCs}).
Blanco\,1 is believed to be $115\pm10$\,Myr old \citep{2014ApJ...795..143J}, marginally younger than the others, but close enough to the FGK star ZAMS.
M\,35 and M\,50 are both roughly 1\,kpc distant, while Blanco\,1 is much closer, at 240\,pc \citep{2018A&A...616A..10G} distance.

\subsubsection{M\,35}

For M\,35, \cite{2009ApJ...695..679M} include not only long-baseline rotation period measurements (especially important in discovering long rotation periods), but also detailed membership and binarity information from extensive time-series radial velocity data presented soon thereafter in \cite{2010AJ....139.1383G}.
We have updated their membership list with the current \emph{Gaia}~DR2 astrometric measurements in a manner similar to our work on NGC\,2516.

Comparing the M\,35 rotational distribution (first panel in Fig.~\ref{fig:compareOCs}) with that of NGC\,2516 reveals them to be visually indistinguishable in all the regions of the CPD where both have derived periods.
In particular, both the fast- and slow rotator sequences lie one over the other, except that the M\,35 study is sensitive to somewhat warmer stars, and consequently samples the late-F fast-rotator region of the slow rotator sequence.
M\,35 also has visibly more scatter on its slow rotator sequence than NGC\,2516, one likely caused by the relatively large (and also differential) reddening towards this cluster.
In Fig.~\ref{fig:compareOCs}, following prior work, we have dereddened the photometry with a single value for all stars.
The fast rotator sequence (where colour differences are largely irrelevant) is essentially indistinguishable from that for NGC\,2516.
The stars between the sequences in M\,35, the so-called 'gap' stars, are equally sparse in both clusters, so essentially similar. 
The only truly visible difference between the distributions is that the low-mass (M\,star) rotation periods are missing in M\,35, perfectly understandable given its much-larger distance.
We therefore conclude that M\,35 and NGC\,2516 have essentially identical rotation period distributions.

\subsubsection{M\,50}

The visual impression in the comparison for M\,50 (third panel in Fig.~\ref{fig:compareOCs}) is again that its rotational distribution \citep{2009MNRAS.392.1456I} is equally indistinguishable from that for NGC\,2516.
The slow- and fast rotator sequences are both positioned in identical locations, although it must be admitted that the slow rotator sequence in M\,50 has somewhat greater scatter.
A caveat needs to be mentioned here - no radial velocity membership is available for this cluster, so we have had to use the \emph{Gaia} data exclusively for membership information.
The cluster's distance of 1\,kpc results in the retrieval of relatively few members from among the M\,50 M\,dwarfs listed in \cite{2009MNRAS.392.1456I}.
This is because they are typically fainter than $G=18$, which renders the \emph{Gaia} astrometry too imprecise for effective use.

There is also another issue.
We find a larger proportion of stars in the rotational gap between the fast- and slow rotators.
We suspect that this effect might originate in the sampling pattern of the observations, one optimized for finding exoplanets, and not rotation periods.
Consequently, a certain number of the stars in the gap region could actually be half-periods, in analogy with the observations of I07 of NGC\,2516 which were obtained in the same observing programme.
We are obviously not in a position to resolve this issue without additional observations.
However, given the excellent agreement in the locations of the primary rotational sequences, we consider M\,50 and NGC\,2516 to have equivalent rotational distributions.

\subsubsection{Blanco\,1}

While we were finalizing this work, \cite{2020MNRAS.492.1008G} published new rotation periods for the 115\,Myr-old open cluster Blanco\,1.
They found the rotation period distribution to be identical to that for the Pleiades.
To improve our comparison with NGC\,2516, we have supplemented those data with rotation periods from \cite{2014ApJ...782...29C} for the most complete picture of rotation in Blanco\,1, displayed in the final panel of Fig.~\ref{fig:compareOCs}. We observe that the distributions of both clusters follow each other closely, along both the fast- and slow rotator sequences.
The only little difference we notice is that NGC\,2516 hosts several stars with slower rotation at $(G_\mathrm{BP}-G_\mathrm{RP})_0\approx 1.2$.
We presume that this can be attributed to Blanco\,1 being of somewhat younger age, as already noted by \cite{2014ApJ...782...29C}.
The other areas of the CPD are too sparsely populated to show any obvious difference.

\subsubsection{No evidence for cluster-to-cluster variations}

In summary, we are able to find no compelling evidence for cluster-to-cluster variations.
All the (minor) differences between the Pleiades and NGC\,2516 rotational distributions can be explained plausibly, without questioning the universal validity of stellar physics and angular momentum evolution.
Additional comparisons with the M\,35, M\,50, and Blanco\,1 distributions also reveal no palpable differences.

In a prior comparison of the NGC\,2516 and Pleiades $v \sin i$ distributions, \cite{2002ApJ...576..950T} also found no significant differences. (This is unsurprising because of the relative insensitivity of $v \sin i$ data.)
Such issues also led to \cite{1998MNRAS.300..550J} finding no fast rotators ($v\sin i > 20$\,km\,s$^{-1}$) and consequently claiming that there are stars with faster rotation in the Pleiades than in NGC\,2516.
Our stance is that photometric rotation period comparisons are a far more unambiguous tool for answering those questions.

It should be noted that earlier studies of NGC\,2516 assumed a much lower photometric metallicity than is obtained from modern (high-resolution) spectroscopy.
The low photometric metallicity from \cite{1985A&A...147...39C} [Fe/H]$=-0.4$ is in strong contrast to the nearly solar values obtained by \cite{2018MNRAS.475.1609B} ([Fe/H]$=-0.08\pm0.01$) and \cite{2002ApJ...576..950T} ([Fe/H]$=0.05\pm0.06$).
This is potentially important because metallicity could be a driver of cluster-to-cluster variations in the rotational evolution.
A non-solar metallicity would imply different convection zone properties, and therefore somewhat different spindown on the main sequence \citep{1997MNRAS.287..350J}.
However, the measured near-solar metallicity of NGC\,2516 makes that argument moot.

Some evidence of cluster-to-cluster variations in the rotation periods for certain clusters have been proposed before by \cite{2016ApJ...833..122C}.
Their conclusion was that the cluster environment could influence the rotational evolution.
However, \cite{2016ApJ...833..122C} compared h~Per (13\,Myr) to the older Pleiades (125\,Myr) and therefore had to make large (and in our view, unacceptable) assumptions about the rotational evolution to arrive at their conclusion.

\subsection{A representative ZAMS rotational distribution}
\label{sec:reprZAMS}

We find the rotational distribution of the five zero-age main sequence open clusters to be nearly indistinguishable, at least to the extent allowed by characteristics of the relevant studies.
The comparison of NGC\,2516 with the Pleiades, M\,35, M\,50, and Blanco\,1 is in fact the strongest evidence to date against the existence of cluster-to-cluster variations.

To highlight this cluster-to-cluster similarity and emphasize the existence of this representative ZAMS rotation period distribution we display the full range of the measured rotation period distributions in both \emph{Gaia} $(G_\mathrm{BP}-G_\mathrm{RP})_0$ and $(B-V)_0$ colours\footnote{The $(B-V)_0$ values are calculated from $(G_\mathrm{BP}-G_\mathrm{RP})_0$ using a relationship derived in Gruner \& Barnes (in prep.).} in Fig.~\ref{fig:componepanel}.
The separate component distributions in $(B-V)_0$ are shown in the Appendix~\ref{app:CPD}.
The very close correspondence between all the component distributions is obvious\footnote{Note that the individual studies contributing to this distribution have their own selection effects, in brightness, rotation period sensitivity, etc., so that while the locations of stars in the CPD are likely robust, the relative numbers of stars in different locations of these CPDs should be treated with caution.}.
Incidentally, the slow and fast rotation limits of this representative distribution can be reproduced simply in any desired colour by the use of the approximating functions $P = \tau_c/8$ and $P = 45/\tau_c$, as we show in the Fig~\ref{fig:simpleModel}.
We also display the critical Rossby Number line $P = \tau_c/\sqrt{k_I k_C} = 0.06\,\tau_c$, which serves as an approximate dividing line for fast- and slow rotators, especially in the F,\,G,\,K spectral range.

%
\begin{figure*}
	\includegraphics[width=\textwidth]{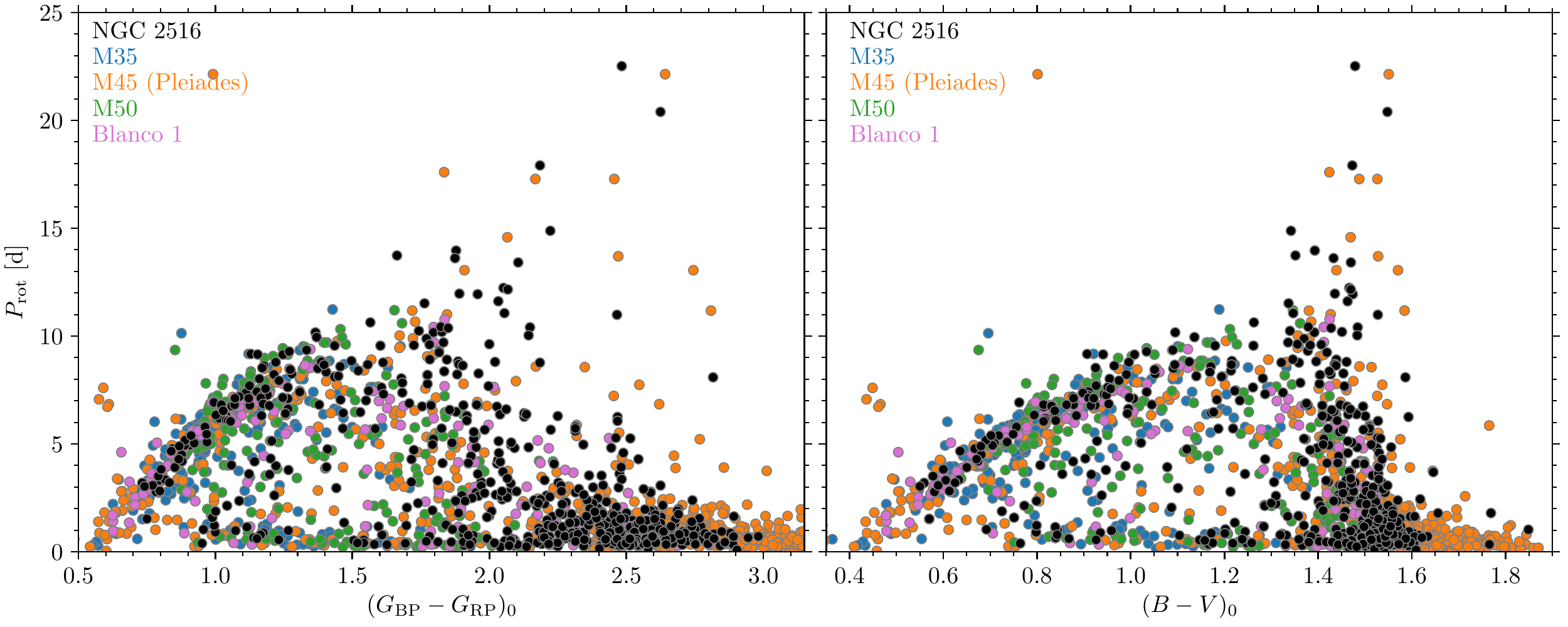}
	\caption{Representative ZAMS rotation period distribution shown in combined colour-period diagrams of all considered open clusters in both \emph{Gaia} $(G_\mathrm{BP}-G_\mathrm{RP})_0$ (\emph{left}) and $(B-V)_0$ colours (\emph{right}).
    The fast and slow sequences, including the extended slow rotator sequence are evident, as is the less-densely occupied region between the sequences.
     The display colours of the individual open clusters are the same as in Fig.~\ref{fig:compareOCs}, and Fig.~\ref{fig:OCcompareBV} provides separate comparison panels for the individual clusters in $B-V$ colour.
     See Fig.~\ref{fig:simpleModel} for an approximate way to represent this distribution in a colour-independent way using elementary functions of the convective turnover timescale.}
	\label{fig:componepanel}
\end{figure*}

%
%
\section{X-ray activity of the rotators}
\label{sec:xray}

We now turn to the X-ray activity of NGC\,2516, where our rotation periods permit the first X-ray activity vs. Rossby Number diagram for the cluster to be constructed.
We discuss this with particular reference to the corresponding diagrams for the Pleiades and Blanco\,1.

Fortunately, a substantial amount of X-ray data is available for NGC\,2516 because this open cluster was among the plate-scale calibration targets for both the XMM-Newton \citep{2000SPIE.4140....1G} and \emph{Chandra} missions \citep{2004ApJ...606..466W}. It is also being used in the recently-launched eROSITA mission \citep{2010SPIE.7732E..0UP, 2012SPIE.8448E..0YF}.

The first X-ray studies with \emph{ROSAT} data had already found a large number of X-ray active members \citep{1996A&A...312..818D, 1997MNRAS.287..350J, 2000A&A...357..909M}. The calibration data from \emph{Chandra} \citep{2001ApJ...547L.141H} and XMM-Newton \citep{2001A&A...365L.259S} identified many additional members.

For this study, we use the two deep surveys from \cite{2003ApJ...588.1009D} (\emph{Chandra}) and \cite{2006A&A...450..993P} (XMM-Newton).
Because of the deeper detection threshold, we prefer the XMM-Newton data over the \emph{Chandra} data for stars with multiple measurements\footnote{The spatial resolution, however, is significantly better with \emph{Chandra}.}.
In general these measurements are in good agreement.
\cite{2003ApJ...588.1009D} have also published a list of upper limits for photometric cluster members which we add for completeness.
Among the members with detected rotation periods, we identify 191 objects with a measured X-ray luminosity or an upper limit.

\subsection{Rotation-activity diagram for NGC\,2516}

%
\begin{figure}
	\includegraphics[width=\columnwidth]{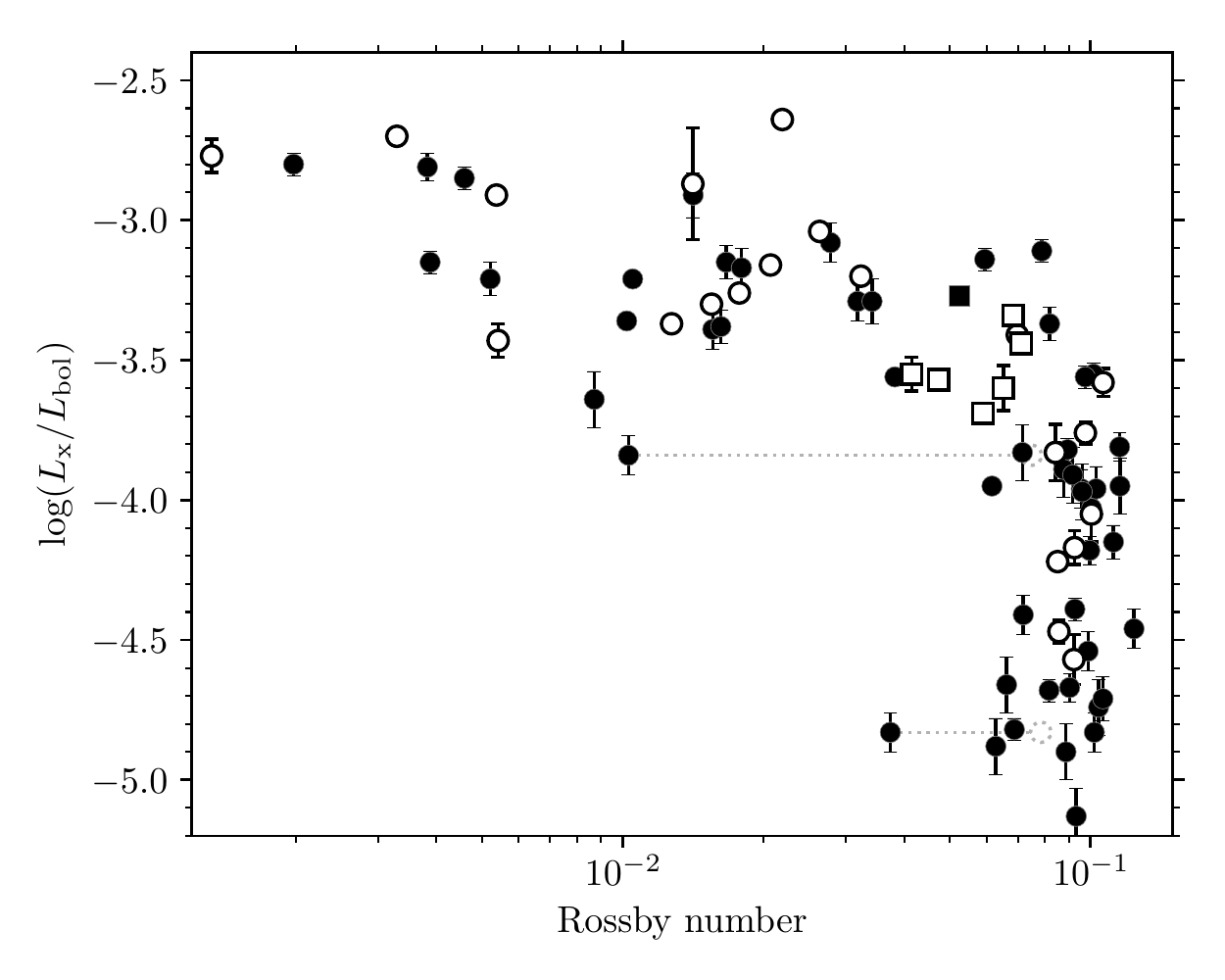}
	\caption{X-ray activity against Rossby number $Ro = P/\tau_c$ for NGC\,2516. 
          Single stars are shown with filled symbols, while known binaries (of all separations) are displayed with open symbols.
          Square symbols indicate the gap stars ($Ro \approx 0.06$) that are in transition from fast- to slow rotation. The dotted line indicates the alternative position of the two possible alias periods in our sample.}
	\label{fig:Roactivity}
\end{figure}

%
\begin{figure}
	\includegraphics[width=\columnwidth]{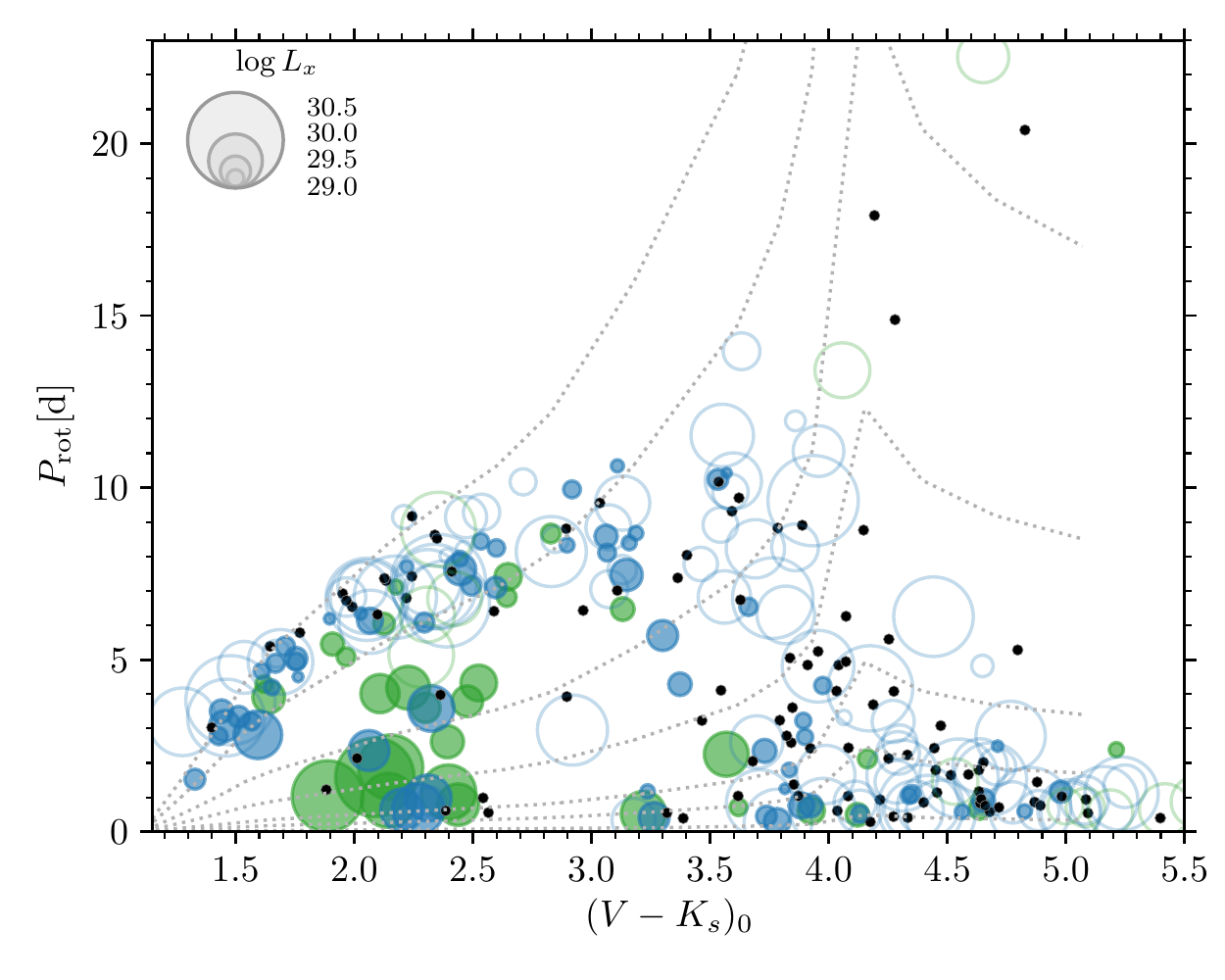}
	\caption{Colour-period diagram for NGC\,2516 encoded with the X-ray luminosities of the rotators in NGC\,2516 (data from both \emph{Chandra} and XMM-Newton).
          Green symbols indicate radial velocity and photometric binaries, while all other stars are in blue.
          Filled circles are detected members, while open circles denote upper limits from the \emph{Chandra} observations.
          Black points show the locations of all stars without measured X-ray luminosities.
          The dotted lines trace (from top to bottom) $Ro=\{0.15, 0.1, 0.05, 0.025, 0.01, 0.005, 0.001\}$.}
	\label{fig:actrot}
\end{figure}

Together with the newly-measured rotation periods for NGC\,2516, the X-ray data mentioned above allow us to construct the first rotation-activity diagram\footnote{We note, in fairness, that \cite{2010MNRAS.407..465J} have also provided a similar diagram, although restricted to M-type stars, because they were only able to use the low-mass rotation periods derived in I07.}
for FGKM stars in NGC\,2516 shown in Fig.~\ref{fig:Roactivity}. The general shape of our rotation-activity diagram is familiar, and resembles those from other open clusters, with a saturated branch for $Ro\lesssim 0.06$ and an unsaturated branch for higher Rossby numbers, the latter delineating stars that are believed to obey the rotation-activity relationship.
The so-called gap stars denoting those in the colour-period diagram that are in transition from fast to slow rotation are located near the knee of the rotation-activity diagram ($Ro \sim 0.06$), emphasising their transitional nature \citep{2003ApJ...586L.145B}.

The Rossby number is defined as usual as $Ro = P_\mathrm{rot}/\tau_c$ with $\tau_c$ as the convective turn-over timescale.
To estimate it, we use the theoretical values tabulated in \cite{2010ApJ...721..675B} and calculate $\tau_c$ via the effective temperature $T_\mathrm{eff}$, which itself was estimated for most stars in NGC\,2516 from the \emph{Gaia} data and is included in the second data release \citep{2018A&A...616A...8A}\footnote{The \emph{Gaia} $T_{\rm eff}$ values are adequate for our purposes, considering that possible systematic errors and uncertainties in the calculated convective turnover timescales are likely to be of greater significance.}.
Values of $\log L_\mathrm{X}/L_\mathrm{bol}$ were taken directly from \cite{2006A&A...450..993P} and \cite{2003ApJ...588.1009D}.

There also appear to be a certain number of outliers with lower $\log L_\mathrm{X}/L_\mathrm{bol}$ value than the majority of the stars in the diagram.
We find a secondary peak at longer rotation period in the periodogram for two of them.
(Star \emph{2i1383}, with $P_\mathrm{rot} = 1.17$\,d ($Ro=0.010$), has a secondary peak in the power spectrum at $P_\mathrm{rot}\approx8.5$\,d, and
Star \emph{1v1504}/\emph{4v79}, with $P_\mathrm{rot} = 1.52$\,d ($Ro=0.037$), has a secondary peak at $P_\mathrm{rot}\approx3.2$\,d).
The secondary periods would lie in perfectly reasonable regions of the CPD. 
Hence, their Rossby numbers could potentially be greater, putting both back onto the sequence in the X-rays vs. $Ro$ plot in Fig.~\ref{fig:Roactivity} in line with the other data points here.
However, we consider it inadvisable to alter our rotation period measurements for such cosmetic purposes, and instead quote the rotation periods as derived, and interpret the ratio $2/89 = 2\%$ as indicating our rotation period alias rate. (There could be four additional such stars in our 308-star sample.)

The usage of the Rossby number as a proxy for stellar rotation entails the loss of the direct connection to the colour-period diagram.
In order to reconnect the two we display the X-ray data in a CPD in Fig.~\ref{fig:actrot}. The sizes of the data points are proportional to the measured X-ray luminosities themselves ($\log L_\mathrm{x}$), rather than normalized to the bolometric luminosities.
Additionally, we indicate the lines of equal Rossby number for easier comparison with Fig.~\ref{fig:Roactivity}. The largest luminosities correspond to the warmest fast rotators.
However, most of these stars are in binary systems, which are believed to be somewhat X-ray bright. Corresponding, all but one of our likely binaries has an X-ray activity measurement.
On the other hand, as clearly visible in Fig.~\ref{fig:Roactivity}, the single- and binary stars clearly populate the same regions of the diagram at the current level of precision.

\subsection{Comparison with the Pleiades and Blanco\,1}

%
\begin{figure}
	\includegraphics[width=\columnwidth]{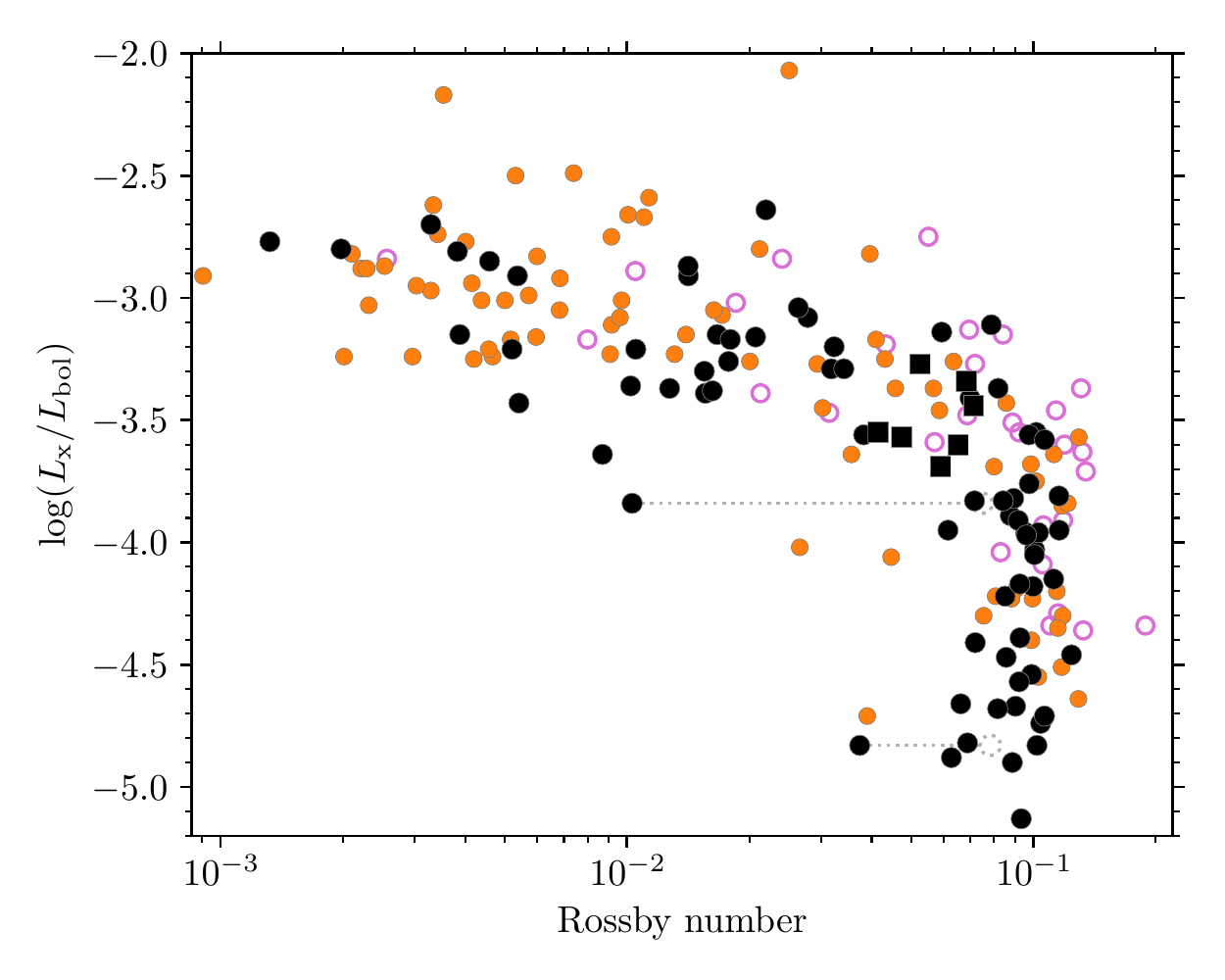}
	\caption{X-ray luminosity against Rossby number $Ro$ for NGC\,2516 (black), the Pleiades (orange), and Blanco\,1 (unfilled purple). The gap stars in transition from fast to slow rotation in NGC\,2516 are marked using squares and the alternative position of possible alias periods with the dotted lines. The distributions are almost indistinguishable.}
	\label{fig:RotActPl}
\end{figure}

Prior studies comparing the X-ray activity of the Pleiades and NGC\,2516 have found the latter to be less X-ray luminous for G and K stars (e.g. \citealt{2000A&A...357..909M, 2001ApJ...547L.141H}).
A possible explanation was sought in the differing rotation rates \citep{2000A&A...357..909M, 2006A&A...450..993P}, ascribed to the age difference believed to exist at that time between the clusters \citep{2002ApJ...576..950T}, or to non-member contamination \citep{2001ApJ...547L.141H}.
The latter has of course been mostly removed in our study.
We now investigate this claim in light of the clean and large samples of rotation periods available now for both open clusters.

We display the prior Pleiades X-ray activity from \cite{1999A&A...341..751M} in Fig.~\ref{fig:RotActPl} together with the new data for NGC\,2516.
We have restricted the Pleiades data to the same mass range as our NGC\,2516 data and recalculated the Rossby numbers in the same manner as for NGC\,2516, to enable a fair comparison.
We observe the overall X-ray vs. $Ro$ distributions of both clusters to be essentially identical.
Notably, we see no sign for any under-luminosity of NGC\,2516 with respect to the Pleiades in our rotationally selected sample.
The few stars in the Pleiades above the saturated level in Fig.~\ref{fig:RotActPl} have large uncertainties in their ROSAT-based count rates, as tabulated in \cite{1999A&A...341..751M}.
Consequently, we ascribe any corresponding scatter above the NGC\,2516 saturated stars to measurement uncertainties or non-member contamination.

We noted in Sect.~\ref{sec:comPleiades} that the one small difference visible in the colour-period diagrams of the two clusters was a small excess of stars in the rotational gap between the fast- and slow rotator regimes. No corresponding stars are found in the Pleiades CPD.
The different X-ray luminosities might originate from this.
Indeed, the gap stars in NGC\,2516, as marked in Fig.~\ref{fig:Roactivity}, are arguably visually slightly less X-ray luminous than the overall population of stars with the same Rossby number in the Pleiades.

It is not feasible to construct the X-ray luminosity function with and without the gap stars with the standard tool of X-ray astronomy, the Kaplan-Meier estimator. When removing a certain group of stars from the sample, we no longer have randomly censored data and the survival analysis with the Kaplan-Meier estimator would be biased \citep{1985ApJ...293..192F}. 

In addition to the gap stars, we note that the unsaturated sequence extends to lower $L_\mathrm{X}/L_\mathrm{bol}$ by about 0.4\,dex. This is an observational effect because the XMM-Newton observations for NGC\,2516 are much deeper than the \emph{ROSAT} data. The proximity of the Pleiades cannot compensate fully for this effect.

The new rotation periods provided in \cite{2020MNRAS.492.1008G} (joined with \cite{2014ApJ...782...29C}) enable us to also include Blanco\,1 in the rotation-activity diagram in Fig.~\ref{fig:RotActPl}.
Previously, \cite{2009AJ....137.3230C} compared the X-ray luminosities of stars in Blanco\,1 and the Pleiades and found the Pleiades to be more luminous.
In the rotationally selected sample we find the Blanco\,1 distribution also to be in agreement with that of NGC\,2516. The similarities of the rotation-activity diagrams of all three open clusters is additional evidence against cluster-to-cluster variations.

\section{Conclusions}
\label{sec:conclusion}
Our study of the southern open cluster NGC\,2516 confirms and extends prior findings that it is very similar to the well-studied Pleiades cluster in the north in age, richness, and other properties.
As such it contributes to the establishment of a Southern Pleiades-like benchmark cluster.

Our work is particularly aimed at elucidating the rotation periods and activity of the cool stars in the cluster.
This is leveraged off a 42\,d-long time-series photometric campaign at CTIO in the $V$ and $I_c$ filters of ${\sim}1$ sq. degree centred on NGC\,2516.

We used all available information to construct a membership list for NGC\,2516 over our survey area.
This list identifies 844 single and binary members of NGC\,2516 among the $\sim$14\,000 distinct sources for which we obtained light curves.
This identification uses the photometric membership from various sources, radial velocities from the literature and several large surveys, and also proper motions and parallaxes from \emph{Gaia}~DR2. \emph{Gaia} DR2 is subsequently used as a check, and to hunt down outliers.

Using a combination of several period-search methods applied to our time-series photometry, supplemented by \emph{manual verification of all light curves and periodicity}, we find 308 rotation periods among the members of NGC\,2516.
Evaluation of the membership status of stars observed by I07 yields 247 additional rotation periods, mostly for M\,dwarfs, and enlarges the sample to 555 stars.
This rich dataset allows us to construct rich, high-fidelity colour-period diagrams (CPDs) for NGC\,2516, enabling detailed comparisons with equivalent work in the Pleiades and other similarly young open clusters.

NGC\,2516 and the Pleiades appear to have near-identical rotation period distributions across the entire rotation period and colour range, confirming the similar ages ascribed to both clusters in the literature. We also identify about a dozen unusually slowly-rotating ($P_\mathrm{rot}{\sim}10 - 23$\,d) late-K and M~dwarfs among our NGC\,2516 members, thereby also confirming their counterparts in the Pleiades.
These stars appear to prolong the slow rotator sequence in the CPD to lower masses and far longer periods than previously thought possible for ZAMS open clusters.
We call this branch the `extended slow rotator sequence'.

We compare the NGC\,2516 rotation period distribution with those of the three other similarly-aged open clusters M\,35, M\,50, and Blanco\,1. For various reasons, these comparisons are not as authoritative as the one with the Pleiades.
Nevertheless, we can discern no substantive differences between them, leading us to conclude that they are all intrinsically similar and likely a natural outcome of pre-main sequence evolution, and perhaps even the star formation process preceding it.
We provide a simple colour-independent way to mark the boundaries of this representative ZAMS rotational distribution approximately.

We also compare the NGC\,2516 rotation data with a number of models constructed over the past decade. 
The models are summarized in relation to one another and their strengths and weaknesses described.
In each case, we find significant detailed issues in describing the measured NGC\,2516 rotation period distribution. 
Consequently, we tend to favour empirical inter-comparisons of the relevant cluster rotational distributions.

Our new rotation periods also enable us to construct the first rotation--X-ray activity diagram for the FGKM stars in NGC\,2516 and to compare it with those for the Pleiades and Blanco\,1.
As expected from prior work in the Pleiades and elsewhere, we see saturated behaviour in X-rays for $Ro<0.06$, and a sharp drop beyond for the unsaturated stars.
The X-ray-saturated stars in NGC\,2516 are on the same level as stars in the Pleiades and Blanco\,1,
suggesting that the similarity in their rotational distributions extends to that in their activity distributions.

In summary, we find that the very rich rotation period distributions for NGC\,2516 and the Pleiades are mostly indistinguishable, as are their stellar X-ray activity distributions.
Thus, we conclude that the Pleiades and NGC\,2516 are not only comparable in age, richness, and metallicity, but also in their rotation-activity properties, so that they may be considered as northern and southern benchmark counterparts.
The similarity in rotation and X-ray distributions also extends to the other available young open clusters M\,35, M\,50, and Blanco\,1.
More generally, this work provides the strongest evidence to date against the existence of cluster-to-cluster variations in rotation and suggests that the star formation process in different cluster environments is sufficiently similar to result in statistically identical outcomes for the rotational distributions of their cool stars.

\begin{acknowledgements}
We are grateful to the anonymous referee for a helpful and detailed report, and to Dr. Richard Jackson for prompt pre-publication access to related data in Jackson et al. (2020).
SAB and DJJ thank the staff of Cerro Tololo Inter-American Observatory (CTIO) for their valuable support during the observations.
A.~Schwope is thanked for useful comments on the manuscript.
SAB acknowledges support from the Deutsche Forschungs Gemeinschaft (DFG) through project number STR645/7-1.
D.J.J. acknowledges support from the National Science Foundation (AST-1440254) and from a John Templeton Foundation award to Harvard University's Black Hole Initiative. 
This work is based in part on observations at Cerro Tololo Inter-American Observatory, National Optical Astronomy Observatory (2008A-0476; S.~Barnes) and the SMARTS consortium through Vanderbilt University (D.~James), operated by the Association of Universities for Research in Astronomy (AURA) under a cooperative agreement with the National Science Foundation.
This research has made use of NASA's Astrophysics Data System Bibliographic Services.
This research has made use of the SIMBAD database and the VizieR catalogue access tool, operated at CDS, Strasbourg, France.
This paper includes data that has been provided by AAO Data Central (datacentral.aao.gov.au).
This publication makes use of the RAVE survey. Funding for RAVE has been provided by: the Australian Astronomical Observatory; the Leibniz-Institut für Astrophysik Potsdam (AIP); the Australian National University; the Australian Research Council; the French National Research Agency; the German Research Foundation (SPP 1177 and SFB 881); the European Research Council (ERC-StG 240271 Galactica); the Istituto Nazionale di Astrofisica at Padova; The Johns Hopkins University; the National Science Foundation of the USA (AST-0908326); the W. M. Keck foundation; the Macquarie University; the Netherlands Research School for Astronomy; the Natural Sciences and Engineering Research Council of Canada; the Slovenian Research Agency; the Swiss National Science Foundation; the Science \& Technology Facilities Council of the UK; Opticon; Strasbourg Observatory; and the Universities of Groningen, Heidelberg and Sydney. The RAVE web site is at https://www.rave-survey.org.
Based on observations made with ESO Telescopes at the Paranal Observatories under programme ID 188.B-3002 (\emph{Gaia}-ESO survey) and programme ID 179.A-2010 (VHS).
This work has made use of data from the European Space Agency (ESA) mission {\it Gaia} (\url{https://www.cosmos.esa.int/gaia}), processed by the {\it Gaia} Data Processing and Analysis Consortium (DPAC, \url{https://www.cosmos.esa.int/web/gaia/dpac/consortium}). Funding for the DPAC has been provided by national institutions, in particular the institutions participating in the {\it Gaia} Multilateral Agreement.
The GALAH survey is based on observations made at the Australian Astronomical Observatory, under programmes A/2013B/13, A/2014A/25, A/2015A/19, A/2017A/18. We acknowledge the traditional owners of the land on which the AAT stands, the Gamilaraay people, and pay our respects to elders past and present.
This publication makes use of data products from the Two Micron All Sky Survey, which is a joint project of the University of Massachusetts and the Infrared Processing and Analysis Center/California Institute of Technology, funded by the National Aeronautics and Space Administration and the National Science Foundation.
The Digitized Sky Survey was produced at the Space Telescope Science Institute under U.S. Government grant NAG W-2166. The images of the Digitized Sky Survey are based on photographic data obtained using the Oschin Schmidt Telescope on Palomar Mountain and the UK Schmidt Telescope.
\newline
\textbf{Software:}
\textsc{DaoPhot~II} was kindly provided by Peter. B. Stetson.
This research made use of \textsc{Astropy}, a community-developed core Python package for Astronomy \citep{2013A&A...558A..33A}.
This work made use of \textsc{Topcat} \citep{2005ASPC..347...29T} and \textsc{Astrometry.net} \citep{astrometry}.
This research made use of the following \textsc{Python} packages:
\textsc{Pandas} \citep{pandas};
\textsc{NumPy} \citep{numpy};
\textsc{MatPlotLib} \citep{Hunter:2007};
\textsc{IPython} \citep{ipython};
\textsc{SciPy} \citep{scipy};
\textsc{Scikit-learn} \citep{scikit-learn}
\end{acknowledgements}

\bibliographystyle{aa} 

\bibliography{N2516} 

\begin{appendix}
	\section{Light curves}
	\label{app:LCs}

	\begin{figure*}
	\includegraphics[width=\textwidth]{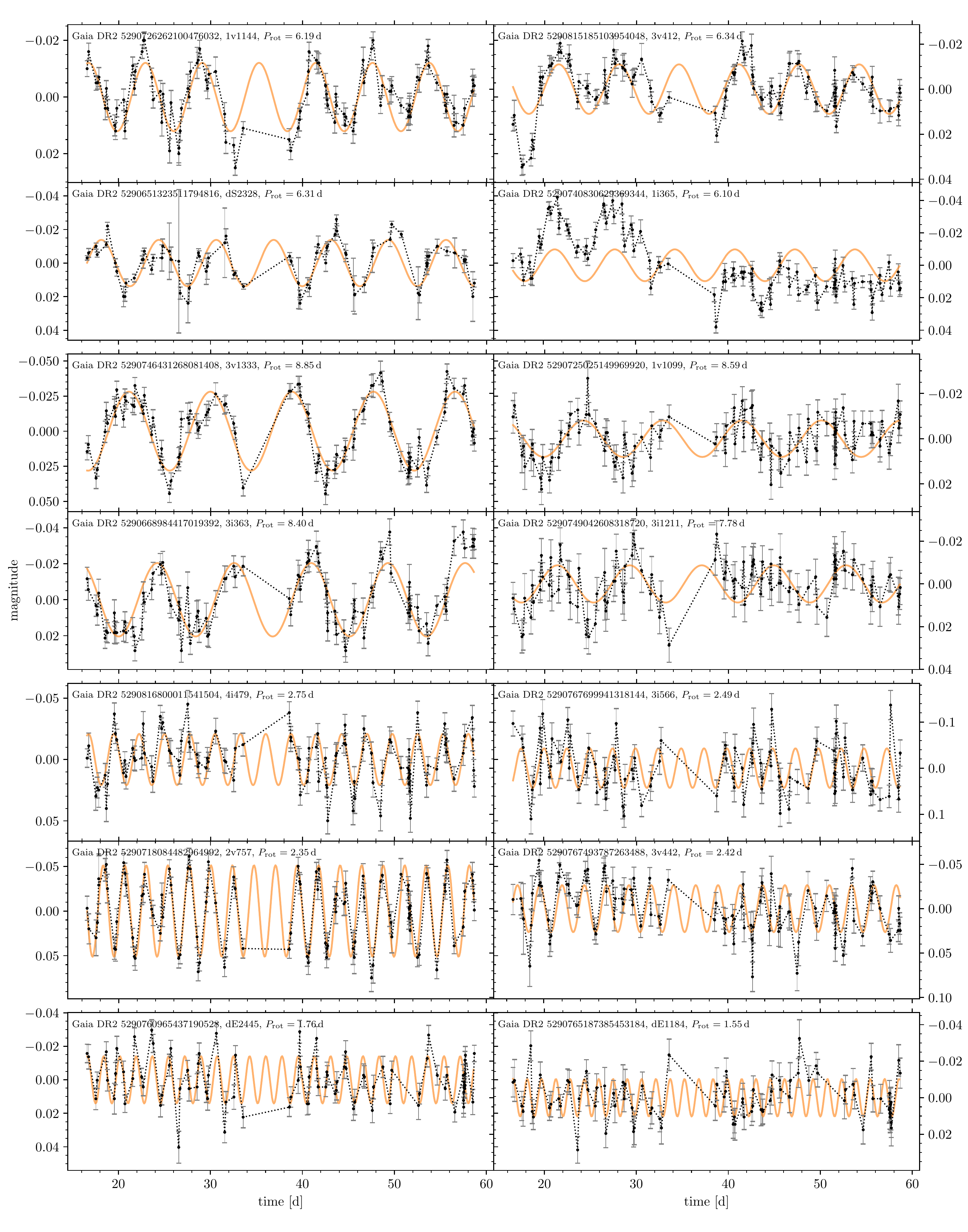}
	\caption{Light curves for selected stars (marked in Fig.\,12) displaying their differing characteristics. In the \emph{left} column, we show light curves of stars with first class (obvious) rotation periods, and in the \emph{right} column those of second class (algorithmic). Each figure block contains light curves of stars of similar colour and period, arranged from early (top) to later-type stars (bottom), as marked in Fig.~\ref{fig:CPD}.}
	\label{fig:exampleLCs}
	\end{figure*}

	\begin{figure*}
		\includegraphics[width=\textwidth]{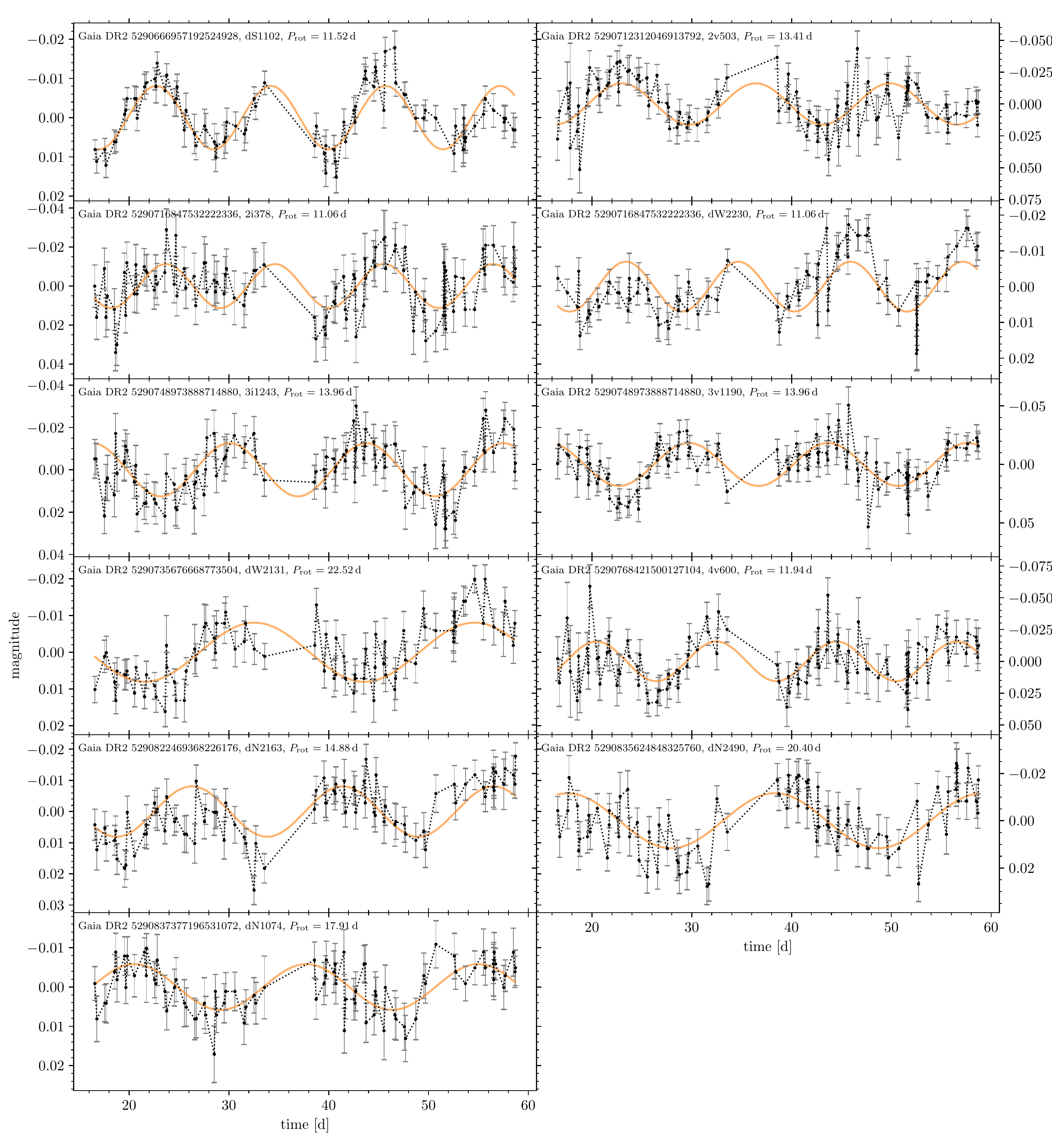}
		\caption{Light curves for the slowly rotating early M dwarfs. We show the data in black, along with the photometric uncertainties. The dashed line simply connects the data points and the orange line is a sinusoidal fit to the data with our determined period. In the second and third row both columns show light curves of the same star in two different bands.}
		\label{fig:lcslowM}
	\end{figure*}

\section{Additional colour-period diagrams} 
\label{app:CPD}

\begin{figure}
\includegraphics[width=\columnwidth]{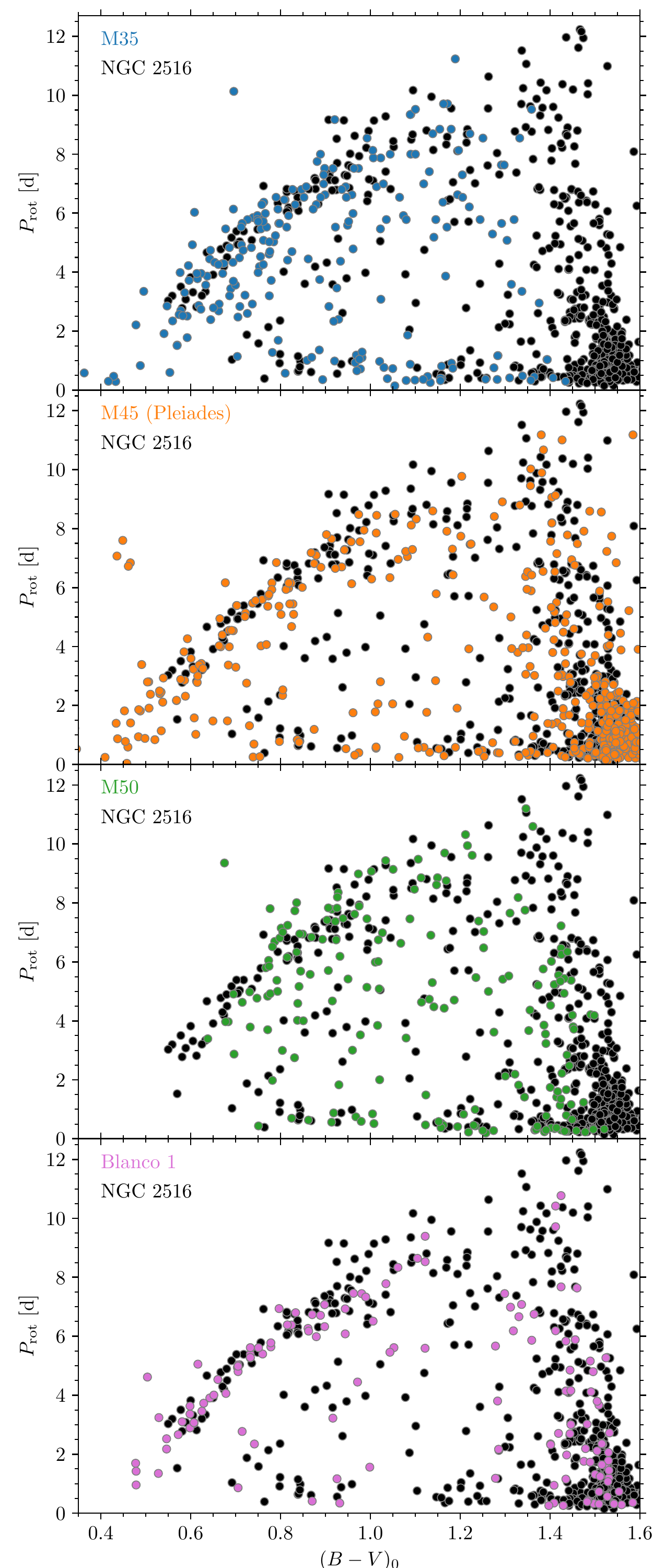}
\caption{Colour-period diagrams in $(B-V)_0$ colour, equivalent to those with \emph{Gaia} colour in Fig.~\ref{fig:compareOCs}, comparing NGC\,2516 (black in all panels) with the four open clusters M\,35, Pleiades, M\,50, and Blanco\,1.
The $(B-V)_0$ colours were obtained by transformation from \emph{Gaia} $(G_\mathrm{BP}-G_\mathrm{RP})_0$ photometry (Gruner \& Barnes, in prep.) except for NGC\,2516, where the observed $B-V$ values were used when available.
}
\label{fig:OCcompareBV}
\end{figure}

\begin{figure*}
\includegraphics[width=\textwidth]{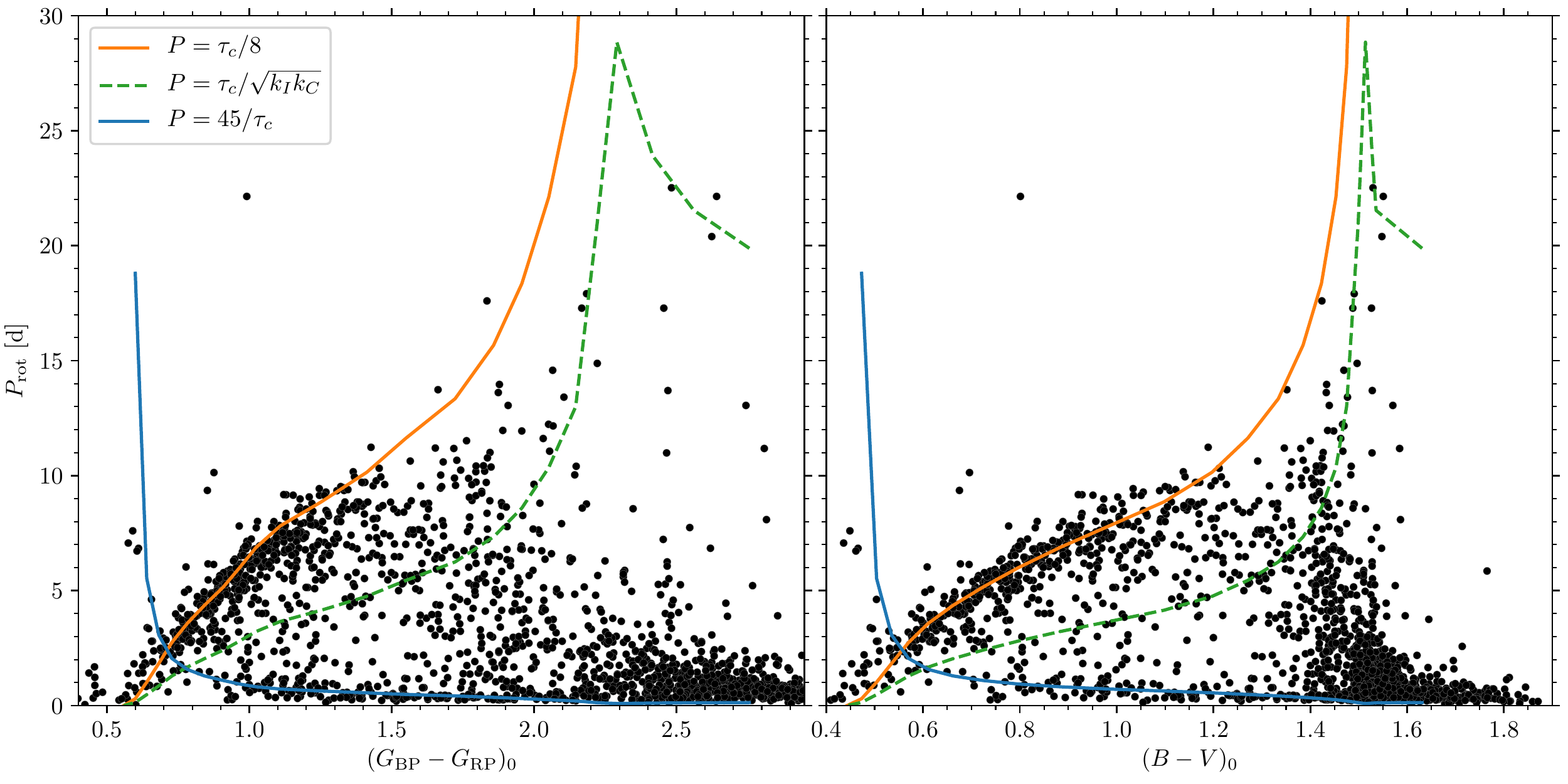}
\caption{Simple colour-independent way to reproduce the ZAMS rotational distribution. Colour-period diagrams in \emph{Gaia} $G_{BP}-G_{RP}$ and $B-V$ colours for the combined representative ZAMS distribution (data as in Fig.~\ref{fig:componepanel}), showing how the elementary functions $P = 45/\tau_c$ and $P = \tau_c/8$, of the convective turnover timescale $\tau_c$ may be used to specify its fast and slow rotation limits. 
The dashed line $P = \tau_c / \sqrt{k_I k_C}$ separates the fast and slow rotator regions in the \cite{2010ApJ...721..675B} model and approximately describes the less-densely occupied region between the fast- and slow rotator regions.
This model is not relevant for the M\,dwarfs that are still on the pre-main sequence.
See text in Sect.~\ref{sec:reprZAMS}.}
\label{fig:simpleModel}
\end{figure*}

\section{Potential non-members}
\label{app:nonmembers}
After our analysis was completed, we learned about a new membership study of stars in open clusters (including NGC\,2516) by \cite{2020arXiv200609423J}. 
Their membership mostly confirms our own independent analysis, but
10 of these 555 rotators are classified as non-members in their work.
However, all of these appear to be radial velocity members of the cluster, and none of these is an outlier in our rotational distribution. 
Apart from three of them lying in the `gap' region between the fast- and slow sequences (thereby lowering the fraction of gap-type stars), they are otherwise normally located, and none of them is on the `extended slow rotator sequence'.
Their presence or absence does not alter any of our conclusions, but we favour retaining them in our rotational distribution.
However, for the convenience of readers, we include the IDs of the potential non-members in \cite{2020arXiv200609423J} in Table~\ref{tab:nonmembers}.

\begin{table}
	\caption{List of stars in our sample of rotators that have recently been classified as non-members in \cite{2020arXiv200609423J}.}
	\label{tab:nonmembers}
	\begin{tabular}{llll}
	\hline\hline
	ID & \emph{Gaia} DR2 designation & J01ID & I07ID\\
	\hline
	\dots & 5290003638143523968 & \dots & 2-5-2476\\
	\dots & 5290646719306810880 & \dots & 1-2-1929\\
	1i1014, 1v939 & 5290738047490662016 & 5887 & \dots\\
	1i1524, 1v1402 & 5290820648301994112 & 7596 & \dots\\
	2v421 & 5290713136680635136 & 4089 & \dots\\
	deepS2147 & 5290649811683301888 & 9790 & 1-7-105\\
	1i1128, 1v1045 & 5290725849783612288 & 6268 & \dots\\
	4v950 & 5290769933328623616 & 11485 & \dots\\
	deepW1367 & 5290734027401142400 & 2277 & \dots\\
	3v260 & 5290672660908938752 & 8645 & \dots\\
	\hline
\end{tabular}
\end{table}

\end{appendix}

\end{document}